**И. З. ШКУРЧЕНКО**

# ЭНЕРГЕТИЧЕСКАЯ СТРУКТУРА СОЛНЦА И ПЛАНЕТ СОЛНЕЧНОЙ СИСТЕМЫ С ТОЧКИ ЗРЕНИЯ МЕХАНИКИ БЕЗЫНЕРТНОЙ МАССЫ[i]

## ***От автора[1]***

Мною уже была написана работа на тему исследования планет солнечной системы под названием «Строение Солнца и планет солнечной системы с точки зрения механики безынертной массы» (1974). Она была посвящена моим детям и носила определённый характер. Мы в общих чертах рассмотрели строение Солнца и планет солнечной системы и получили, что Солнце и планеты имеют одинаковое строение. Различия выражаются только в специфике температурного режима. Одинаковое строение Солнца и планет солнечной системы определило нам, что все планеты в прошлом являлись частями Солнца, которые оно отстрелило на определённом этапе своего существования. В общем, все планеты имеют солнечное происхождение. Происхождение самого Солнца для нас пока остаётся загадкой.

В отличие от прочих наук механика безынертной массы как механика жидкости и газа даёт возможность заглянуть внутрь планет и самого Солнца. В работе «Строение Солнца и планет солнечной системы с точки зрения механики безынертной массы» [2] мы дали описание внутреннего строения планет и Солнца. Теперь настало время подвергнуть материал этой работы математической проверке, чтобы придать конкретность строению планет и Солнца. Именно этим мы займёмся в данной работе.

Главной теоретической базой по нашему вопросу остаётся механика безынертной массы, или механика жидкости и газа. В данной работе мы будем пользоваться положениями механики безынертной массы как известными для вас пояснениями.

Механика безынертной массы является общей теорией, а в данном случае нам придётся иметь дело с планетами и звёздами. По этой причине сначала нам предстоит получить теорию, касающуюся непосредственно планет и звёзд, которая бы учитывала специфику их механического состояния как прикладная теория.

## **Прикладная теория к энергетической структуре планет и Солнца**

Начнём с самого простого и хорошо всем известного.

Все знают, что планеты и само Солнце имеют шарообразную форму объёма, в котором заключена их масса. Масса каждой такой планеты находится под силовым воздействием радиального силового поля, которое своим силовым воздействием стремится притянуть всю массу планеты к её центру. Вот это положение является *общим* для всех планет и Солнца.

Во всём остальном они имеют те или иные специфичные различия, которые выражаются в том, что одни планеты – горячие, другие - холодные, одни планеты имеют свою массу в жидком и газообразном состоянии, другие – в твёрдом состоянии. Одни планеты имеют идеальную шарообразную форму, другие основательно сплющены. Если руководствоваться этими признаками, мы не найдём двух одинаковых планет в нашей солнечной системе, но вы, наверное, не сомневаетесь в том, что все эти различия определяются их специфическим энергетическим состоянием.

Для планет также существуют различия химического характера и по плотности. Различия химического состава не относятся к компетенции механики безынертной массы, но в то же время мы и здесь должны будем вмешаться в определённой степени.

Как видите, мы сталкиваемся с большим перечнем вопросов, на которые, кажется, почти невозможно ответить. В то же время мы занялись данной работой, чтобы получить ответы на все эти вопросы. Притом мы должны получить не просто ответы, а ответы, выраженные с помощью математического аппарата в количественной форме.

Начнём по порядку, с общего для всех планет и Солнца принципа (отметим сразу, что всё сказанное в одинаковой мере относятся к спутникам планет).

Коль все планеты находятся под воздействием гравитационного поля и имеют шарообразную форму объёма, то мы будем должны изобразить их именно в таком виде на рисунке 1.

На рис. 1 мы видим идеализированную, или общую планету, которая имеет определённый радиус (*R*) своей сферической поверхности. Шаровая поверхность ограничивает объём планеты (*V*), в котором содержится масса величиной (*M*). Полагаем, что массу планеты составляет жидкость с постоянной плотностью ($\rho$). В центре планеты сосредоточивается гравитационное поле, которое своим силовым воздействием охватывает всю массу планеты. Силовое воздействие гравитационного поля на жидкую массу мы называем давлением ($P$). Все эти обозначения показаны на рис. 1. Вот с этой идеальной планеты нам предстоит начать работу.

---

[1] Игорь Захарович Шкурченко (1935 – 2003), об авторе см. Предисловие редактора к работе «I.Механика жидкости и газа, или механика безынертной массы».

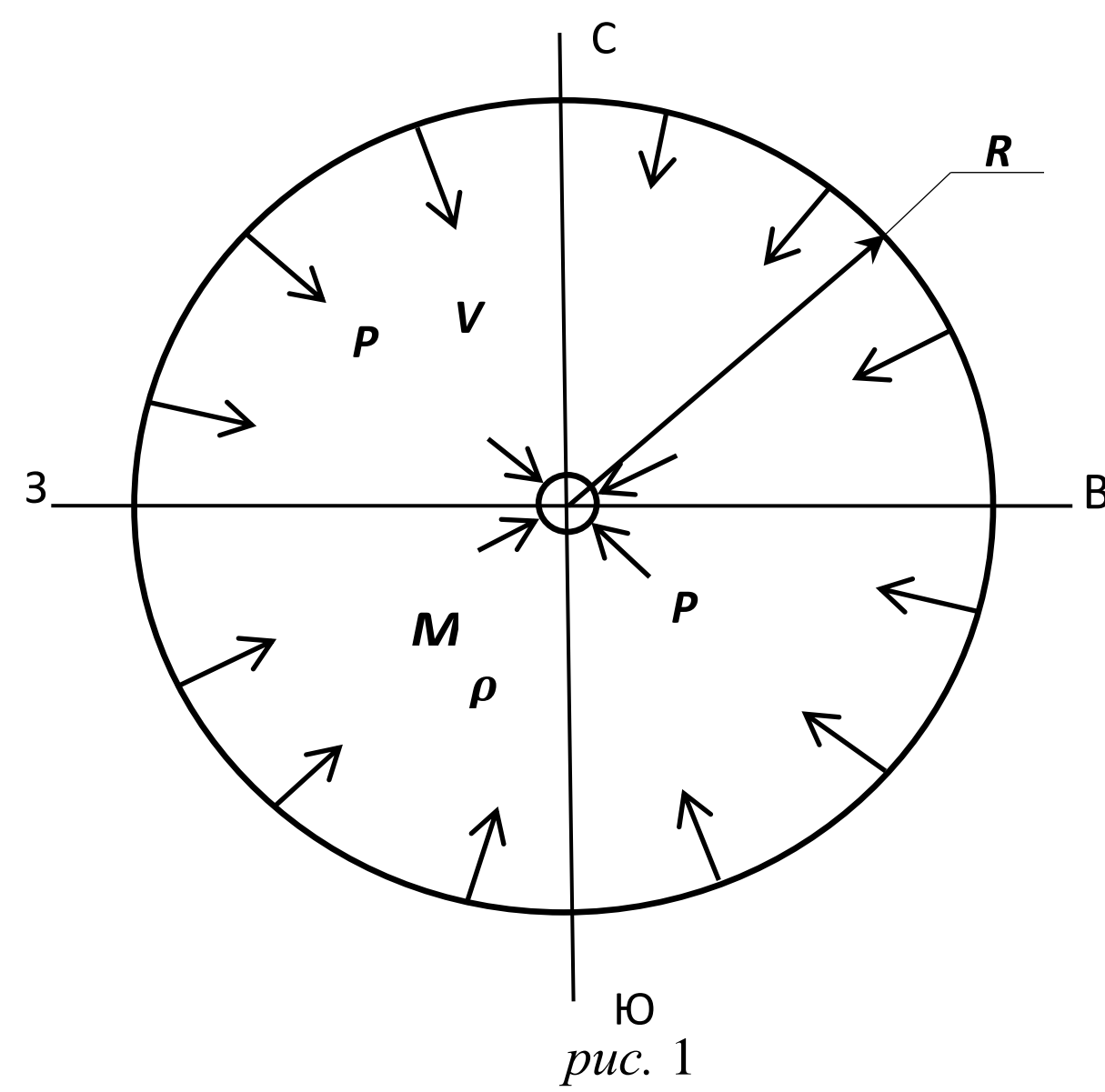

*рис.* 1

Природа поселила людей на поверхности земного шара. Отсюда у них сложилась привычка принимать поверхность всех планет при их исследовании за начало отсчёта. Мы тоже останемся верными традиционной привычке людей и за начало своих отсчётов примем поверхность нашей идеальной планеты. Все люди знают, что они находятся на шаровой поверхности, но в своих исследованиях они принимают её за плоскость. Мы тоже поступим аналогичным образом.

Отметим, что единицы измерения, которые будут встречаться в данной работе, относятся к метрической системе единиц (МКС) в силу моего консерватизма. Это значит, что единицей длины является метр, сила измеряется в килограммах, время – в секундах, работа и энергия - в килограммометрах и т.д.

### а) Исследование энергетического состояния идеальной планеты относительно её поверхности, где плоскость служит началом отсчёта.

Не думайте, что в данном названии скрыто что-то особенное. Здесь надо понять, что для исследования принимается столб жидкости в виде цилиндра или прямоугольного столба, высота которого измеряется от поверхности планеты. Всё это покажем на рис. 2.

Мы видим, что на рис. 2 выделен столб жидкости высотой ($h$). Высота этого цилиндра может быть любой в зависимости от нашего желания. Всё это вам хорошо известно, ибо вы подобным пользуетесь при вычислении давления и энергии в глубинах океана.

Вам также хорошо известно, что лежит ли камень на поверхности Земли или движется в свободном падении, он всё равно имеет некоторое ускорение в соответствии с законами Ньютона. На нашей планете это ускорение равно $g$ = 9,81 м/сек$^2$. Масса, умноженная на ускорение ($g$), определяет нам вес, или силу, с которой гравитационное поле Земли действует на тот или иной предмет. Нам понятно, что тело в свободном падении должно иметь постоянное ускорение, но нам непонятно, что и лежащий камень тоже должен иметь постоянное ускорение. Всё станет на свои места, если мы определим положение неподвижного тела (лежащего камня) как застывшее движение. Тогда материальным параметрам всегда будет соответствовать движение, которое в одном случае является непосредственно самим движением материального тела, в другом случае – застывшим движением. Застывшее движение материальных тел мы привыкли обозначать как состояние покоя материального тела.

В объёме идеальной планеты жидкость тоже неподвижна и находится под воздействием гравитационного поля. По этой причине мы тоже должны отнестись к жидкости в состоянии покоя как к жидкости находящейся в состоянии застывшего движения. Ведь жидкость под действием гравитационного поля стремится двигаться к центру планеты. В механике безынертной массы наглядная форма движения жидкости выражается потоком как объёмом движущейся жидкости. В нашем конкретном случае таким потоком является неподвижный цилиндрический столб жидкости высотой ($h$), который показан на рис. 2.

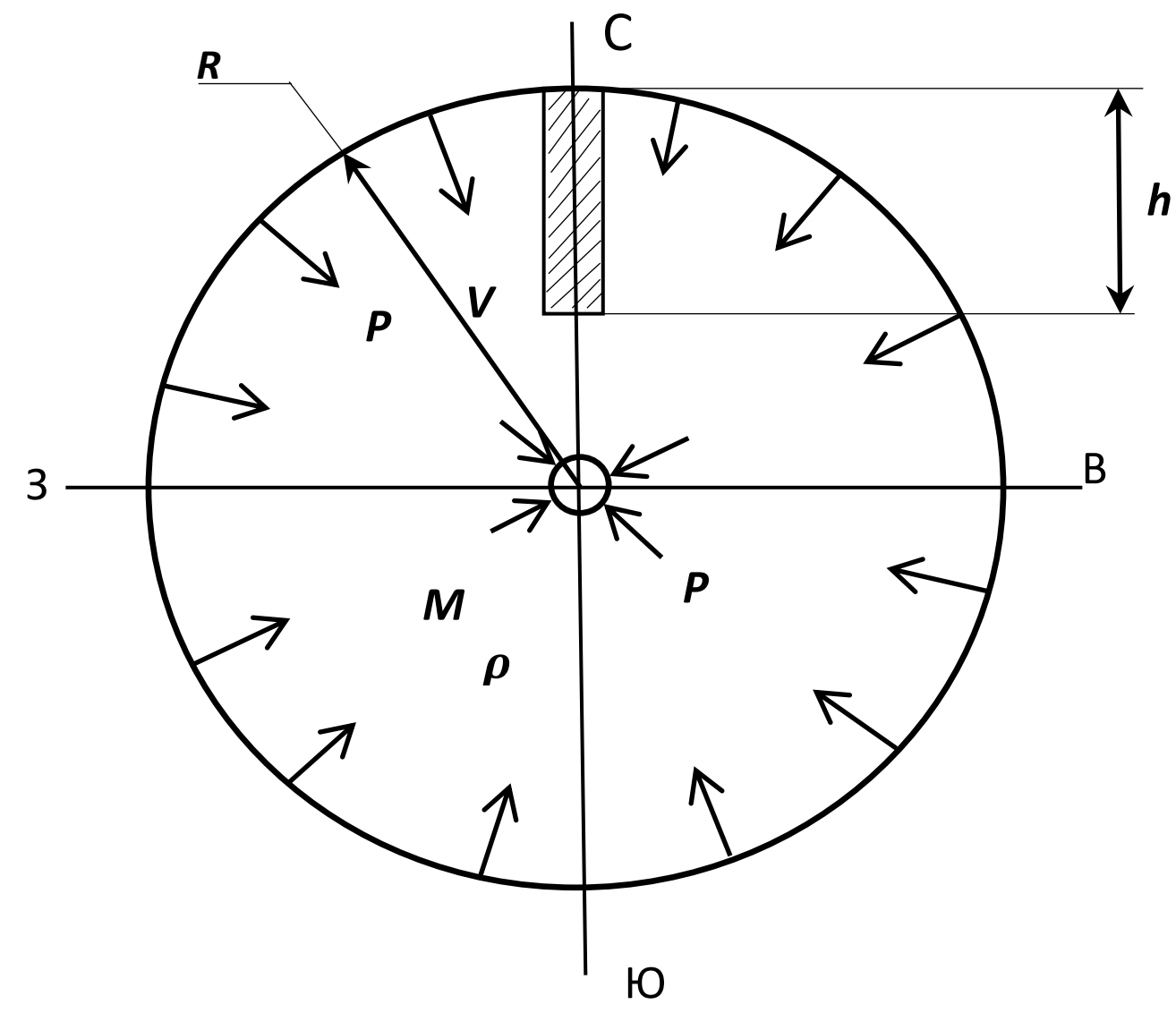

*Рис. 2*

В механике Ньютона для твёрдых тел сила выражается просто силой, величина которой определяется вторым законом Ньютона. В механике безынертной массы сила определяется как давление, величина которого определяется вторым законом механики безынертной массы.

Второй закон механики безынертной массы гласит, что сила в жидкостях и газах равна произведению расхода массы в единицу времени на скорость, т.е. $P = \dot{M}W$

В поле земного тяготения твёрдые тела при свободном падении движутся с постоянным ускорением $g$ = 9,81м/сек$^2$, а жидкости и газы – с постоянной скоростью $w$ = 3, 132 м/сек$^2$. По этой причине люди, спускающиеся на парашютах, не разбиваются, а приземляются плавно. Величина постоянной скорости ($w$) находится в такой зависимости с величиной постоянного ускорения свободного падения твёрдых тел, как $w^2 = g$, или $w = \sqrt{g}$ ([1]).

Мы выяснили, что выделенный в идеальной планете столб жидкости высотой ($h$) представляет собой поток жидкости, находящейся в застывшем движении. В механике безынертной массы такого рода поток называется установившимся движением жидкости.

Всякое движение жидкости характеризуется двумя уравнениями: это уравнение сил и уравнение движения.

Уравнение сил для застывшего потока идеальной планеты будет иметь такой вид:

$$P = \dot{M}w \qquad (1),$$

где $P$ - сила давления,

$\dot{M}$ – расход массы в единицу времени,

$w$ – постоянная скорость, которая соответствует воздействию гравитационного поля идеальной планеты, т.е. $w = \sqrt{g}$,

а уравнение движения – такой:

$$\dot{M} = \rho W F \qquad (2),$$

где $F$ – площадь сечения потока.

В нашем случае площадь сечения потока равна единице площади, т.к. мы имеем дело с давлением, т.е. $F = 1$.

$W$ – пока неизвестная для нас скорость движения жидкости.

Величину этой скорости мы должны получить. Скорость – скоростью, но не забывайте, что мы имеем дело с застывшим движением. По этой причине жидкость не движется, а находится в состоянии покоя.

Принятая нами высота ($h$) жидкостного столба определяет нам то сечение потока, в котором мы желаем знать его характеристики. Это значит, что жидкость в данном сечении потока может двигаться только под действием гравитационного поля, а гравитационным силам соответствует вполне определённая и постоянная скорость ($w$). Следовательно, скорость жидкости ($W$) в уравнении (2) по величине равна постоянной скорости ($w$), т.е. $W = w$. В тоже время при такой логике мы получаем неверный результат для скорости уравнения (2), т.е. $W \neq w$. Вопреки логике мы должны записать скорость ($W$) уравнения (2) равной

$$W = hw \qquad (3).$$

В этом выражается особенность застывшего движения. Подробный вывод и определение этого описания смотри в работе [1]. Тогда уравнение (2) примет вид:

$$\dot{M} = \rho hwF \qquad (4).$$

Вот такой вид принимает уравнение движения для застывшего движения.

Уравнение (4) мы должны разделить на площадь сечения потока ($F$) т.к. мы стремимся получить давление. После такого преобразования подставим уравнение (4) в уравнение (1), получим:

$$P = \rho hw^2 \qquad (5).$$

Таким путём мы получили нужное нам давление сил как давление на соответствующей глубине ($h$).

В механике безынертной массы работа и энергия определены общей зависимостью как равные произведению объёма на давление, т.е.

$$Э = VP \qquad (6).$$

Здесь имеется в виду потенциальная энергия. Для *кинетической энергии* существует иная зависимость.

Давление ($P$) нам известно из уравнения (5). Объём ($V$) жидкости здесь равен тому объёму, который жидкость занимает на заданной глубине ($h$). Это значит, что на различной глубине потенциальная энергия жидкости будет различной. Если мы подставим в уравнение (6) уравнение (5), то получим зависимость для энергии в таком виде:

$$Э = V\rho hw^2 \qquad (7).$$

Будем считать, что мы получили уравнение энергий (7) для нашей идеальной планеты.

Вы не увидели почти ничего нового для себя и скажете, что мы здесь просто заменили постоянное ускорение сил планетного тяготения ($g$) на квадрат постоянной скорости ($w^2$) – только и всего.

Уравнение (5) хорошо известно всем, если в нём заменить квадрат скорости ($w^2$) на ускорение ($g$). В этом случае величина давления не изменяется, но понимание сущности самого явления меняется в корне.

При застывшем движении мы как бы не замечали этого различия, но когда мы имеем дело с натуральным движением, то этим различием уже нельзя пренебрегать. До настоящего времени уравнением (5) в видоизменённой форме обозначали одновременно и давление и энергию. Таким образом, в количественном отношении здесь не делалось ошибки, но сущность была тоже неправильной.

Дело в том, что в уравнении (7) объём ($V$) фактически количественно определяет объём поверхности, расположенной на глубине ($h$), который имеет одинаковое давление ($P$). Как говорится, объём поверхности представляет собой нечто неопределённое, которое может быть каким угодно малым. По этой причине поступают проще – относят энергию к единице объёма, т.е. делят уравнение (7) на величину объёма ($V$), тогда оно примет вид:

$$Э = \rho h w^2 \quad (8).$$

В таком виде уравнение (8) количественно равно уравнению (5). Количественно они равны, но размерность имеют разную. Ибо уравнение (7) мы делим не на объём ($V$) с его размерностью, а на число объёмных единиц, которое количественно соответствует объёму ($V$), а сама объёмная размерность при этом сохраняется.

В общем, уравнение (8) служит для практических целей людей, которым они должны пользоваться для определения потенциальной энергии на глубине ($h$).

Будем считать, что мы покончили с нашим первым теоретическим положением.

Если теперь мы снова посмотрим на полученные нами выкладки и на нашу идеальную планету, то увидим неувязки в наших теоретических положениях. Ведь наша идеальная планета имеет шарообразный объём. Её масса приближается к центру планеты гравитационным полем, а мы для своих исследований взяли цилиндрический столб жидкости. Но, как теперь сами видите, с движением к центру планеты шаровые поверхности всё время уменьшается. Говоря другими словами, с увеличением глубины нижняя поверхность столба непрерывно уменьшается. Это значит, что давление будет расти не только в зависимости от глубины, но и в зависимости от уменьшения поверхности с ростом глубины.

Теперь мы  должны будем учесть ещё влияние уменьшения поверхности с ростом глубины. Только в этом случае мы сможем правильно учесть изменение давления и энергии в объёме идеальной планеты. По этой причине мы должны будем взять жидкостной столб не в виде цилиндра, а в виде конуса, вершина которого располагается в центре идеальной планеты, а основание – на её шаровой поверхности. Покажем всё это на рисунке 3.

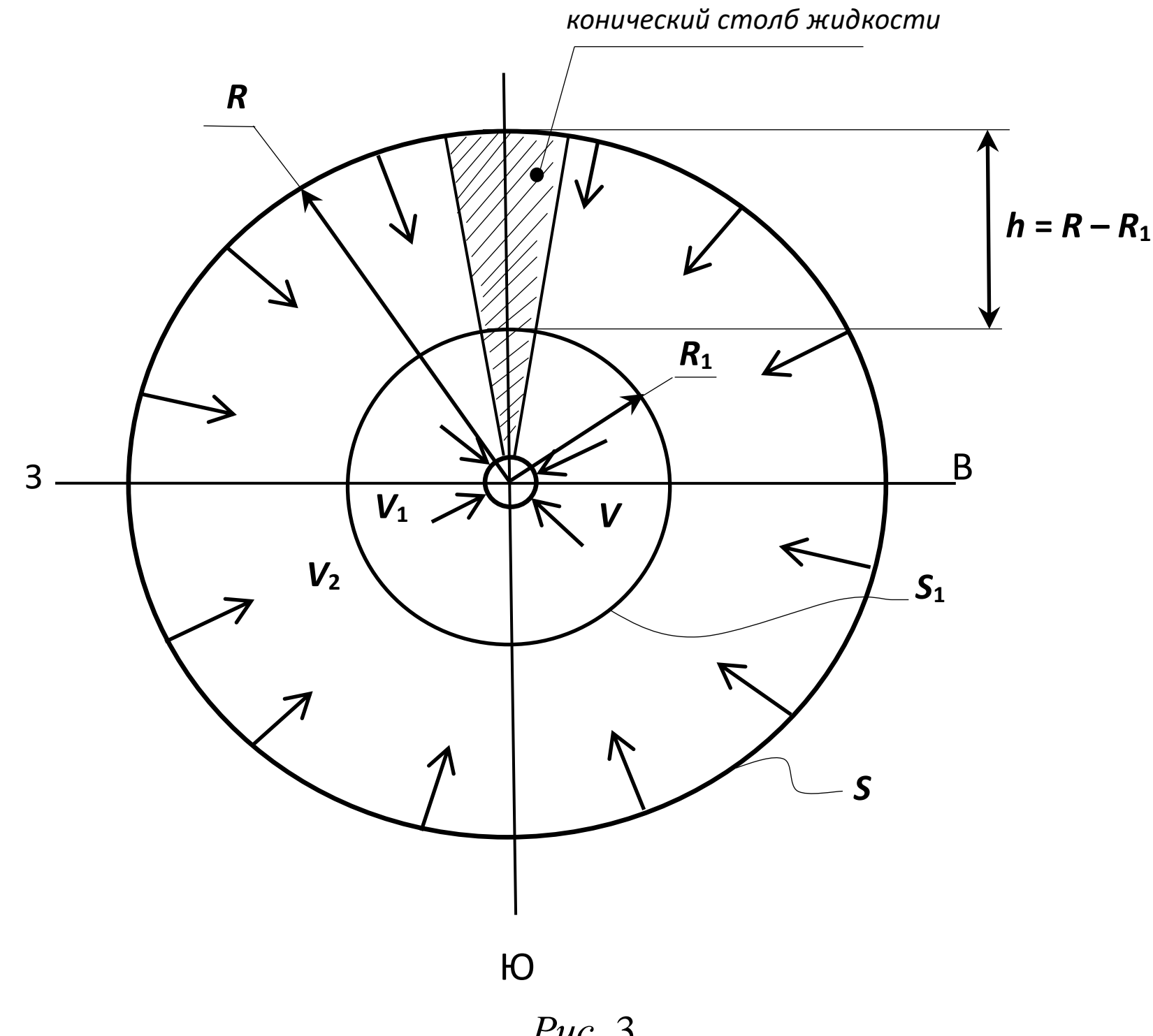

*Рис.* 3

Если при исследовании цилиндрического столба жидкости мы могли обходиться без остальной геометрии идеальной планеты, то при конусообразном столбе жидкости нам уже не обойтись без неё. Поэтому нам предстоит вспомнить геометрию шара.

Площадь шаровой поверхности ($S$):

$$S = 4\pi R^2 \quad (9).$$

Объём шара ($V$): $$V = \frac{3}{4}\pi R^3 \qquad (10),$$

где $\pi = 3{,}14$ – является постоянной величиной,
$R$ – радиус шара.

Коническая форма принятого нами столба жидкости не изменяет нам сущности застывшего движения, которое при цилиндрическом столбе жидкости было установившимся. По этой причине в этом случае мы можем использовать те же уравнения движения (2) и сил (1), которые мы использовали для цилиндрического потока установившегося застывшего движения.

Теперь приступим к получению уравнений для сил, или давлений.

Хотя мы здесь имеем дело с коническим жидкостным столбом, но для удобства расчётов мы тоже будем приводить его к цилиндрическому жидкостному столбу.

На рис. 3 для наших исследований мы выделили заданной нами высотой ($h$) соответствующий объём ($V_2$). Определим его с помощью уравнения (10).

$$V_2 = V - V_1 = \frac{4}{3}\pi R^3 - \frac{4}{3}\pi R_1^3 = \frac{4}{3}\pi (R^3 - R_1^3) \qquad (11).$$

Плотность нашей идеальной планеты нам известна, т.к. мы ею задались. Она равна у нас ($\rho$). С её помощью мы можем определить общую массу, заключённую в выделенном объёме ($V_2$):

$$M_2 = \rho V_2 \qquad (12).$$

Вот эта масса ($M_2$) опирается, или располагается на поверхности ($S_1$), которая является поверхностью шара с радиусом ($R_1$). Поэтому принимаем поверхность шара ($S_1$) за основание цилиндра. Поверхность ($S_1$) определяем по уравнению (9). Получим:

$$S_1 = 4\pi R_1^2 \qquad (13).$$

Основание цилиндра ($S_1$) мы определили по уравнению (13). Высота цилиндра ($h$) нам известна, т.к. мы ею задаёмся по своему усмотрению. Тогда цилиндрический объём ($V_{цил}$) будет иметь такую зависимость:

$$V_{цил} = S_1 h \qquad (14).$$

В тоже время объём цилиндра ($V_{цил}$) будет меньше выделенного объёма шара ($V_2$), т.е. $V_2 > V_{цил}$. По этой причине, чтобы сохранить условие силового равновесия на поверхности ($S_1$), мы должны будем всю массу ($M_2$) этого объёма шара ($V_2$) разместить в цилиндрический объём ($V_{цил}$), т.е. мы должны записать, что

$$\rho V_2 = V_{цил}\, \rho_{цил} \qquad (15).$$

Из уравнения (15) мы определим плотность ($\rho_{цил}$) жидкости в цилиндрическом объёме, т.е.

$$\rho_{цил} = \frac{V_2}{V_{цил}}\rho \qquad (16).$$

Вы понимаете, что плотность ($\rho_{цил}$) будет больше плотности ($\rho$) нашей идеальной планеты, но такое условие преобразования плотности даёт возможность преобразовать конический столб жидкости в цилиндрический. Следовательно, вы должны понимать, что плотность цилиндрического объёма ($\rho_{цил}$) не соответствует действительной плотности ($\rho$) нашей идеальной планеты, т.е. она не является реальной величиной, но необходима нам для преобразований.

После того, как мы таким путём превратили конический столб жидкости в цилиндрический, мы можем воспользоваться зависимостями для цилиндрического столба (5) и (7), которые мы получили выше.

Тогда уравнение сил, или давлений, примет вид:

$$P = \rho_{цил} \cdot h \cdot w^2 = \rho_{цил}(R - R_1) \cdot w^2 \qquad (17).$$

Уравнение энергии примет такой вид:

$$Э = V\rho_{цил} \cdot h \cdot w^2 = V\rho_{цил}(R - R_1) \cdot w^2 \qquad (18).$$

Теперь проведём исследование уравнений (17) и (18). Полагаем, что ($R_1$) стремится к нулю, т.е. $R_1 \to 0$, или $h \to R$, тогда в уравнении (17) поверхность ($S_1$) тоже будет стремиться к нулю, т.е. $S_1 \to 0$. В этом случае давление ($P$) будет стремиться к бесконечности, т.е.

$$P \to \infty \quad \text{при} \quad S_1 \to 0, \quad h \to R \qquad (19).$$

Вспомним, что энергия равна произведению объёма ($V$) на давление ($P$). В этом случае для энергии уравнения (18), т.е. когда радиус ($R_1$) стремится к нулю, ($R_1 \to 0$, или $h \to R$), то и объём ($V$) стремится к нулю ($V \to 0$), а давление ($P$) стремится к бесконечности ( $P \to \infty$), мы будем иметь:

$$Э = 0 \cdot \infty \quad \text{при} \quad S_1 \to 0, \quad h \to R \qquad (20).$$

Из этих исследований мы видим, что в центре нашей идеальной планеты мы имеем полную неопределённость относительно энергетических характеристик. Ибо давление ($P$) в центре планеты стремится к бесконечности ($\infty$), а энергия представляет собой неопределённость как произведение нуля на бесконечность.

Отсюда также следует общий вывод, что, какой бы радиус ни имела наша планета, большой или маленький, в центре её мы всегда будем иметь бесконечно большое давление и неопределённость относительно энергии. Ибо центр планеты, как всякая точка, по поверхности и по объёму в пределе стремится к нулю, или в пределе имеет ноль.

С помощью математики мы получили в центре любой планеты бесконечно большую величину давления и неопределённость относительно энергии, но та же математика не даёт нам возможность представить их себе материально. Наше воображение не в силах представить себе в предметном, или материальном, виде количественную величину бесконечно большого числа и неопределённости. Следовательно, ответ на этот вопрос надо искать в самих материальных предметах, которые несут в себе, или содержат, бесконечность и неопределённость. Пока такими материальными предметами для нас являются планеты, их спутники и само Солнце. Всё это так же означает, что специфические различия между планетами определяются их центром, в котором мы имеем бесконечно большую величину давления и неопределённость относительно энергии. По этой причине мы должны будем обращаться за дальнейшими разъяснениями непосредственно к самим планетам солнечной системы и к самому Солнцу.

Отметим, что вы теперь сами видите необходимость учитывать в зависимостях для застывшего движения различие между давлением и энергией, которое вы старались не замечать. Теперь вы видите, что существует коренное различие между энергией и давлением.

Полученные зависимости для вычисления давления и энергии планет очень просты. Они могли быть получены ещё Архимедом. В настоящее время даже ученики хвастают тем, что они знают больше Архимеда, но никто из учёных до сих пор не удосужился найти зависимости для планетных давлений и энергий. <…>

Главным отличительным признаком планет является то, что их масса притягивается гравитационным полем к их центру. Астрономы делят небесные тела на звёзды, планеты, спутники. С точки зрения механики безынертной массы все отличительные признаки типа блеска, массивности, цвета и т.д. не имеют существенного значения, т.к. мы руководствуемся в этом плане пока одним-единственным общим для них признаком - что масса притягивается к центру. В этом случае подобное тело мы будем называть планетой, независимо от того, называют ли астрономы это тело звездой или спутником планеты. В общем, мы ещё раз уточнили, что мы имеем в виду под названием «планета». Надеемся, что астрономы будут снисходительны к нам и простят нам подобные вольности, например то, что мы ставим Солнце в одинаковое положение с нашей планетой Земля или даже с её спутником – Луной.

Изолированное небесное тело типа планеты имеет чётко выраженную шарообразную форму объёма. После исследования планет вы теперь не сомневаетесь, что шарообразная форма планет является единственной для них приемлемой формой, которая позволяет с ростом глубины увеличивать, прежде всего, давление и производить таким путём наращивание энергии до определённой степени.

В центре планет мы имеем бесконечно большую величину давления. Следовательно, шарообразная форма объёма позволяет иметь планетам максимальную величину давления. Это значит, что любая другая форма планеты, связанная с любым отклонением от шарообразной формы, будет приводить к большему или меньшему снижению давления внутри планеты. Данное положение очевидно и не требует доказательств.

Из этого положения в качестве следствия следует, что снизить давление в планете можно только за счёт изменения формы её объёма, отличной от шарообразной формы. Увеличить же давление в планете невозможно, если она имеет шарообразную форму объёма. Ведь бесконечно большое давление нельзя увеличить. Бесконечно большая величина всегда будет оставаться бесконечно большой величиной, сколько бы к ней не добавляли. Это значит, что для планет можно только снижать давление, но не увеличивать его. В общем, данное положение можно считать законом, который определяет энергетическую структуру планет.

В переводе на обыкновенный язык это положение означает, что увеличение давления в планете мы не можем наблюдать из-за того, что шарообразная форма объёма планеты при этом не изменяется. Уменьшение величины давления в планете мы всегда можем наблюдать по изменению формы её объёма, отличной от шарообразной.

Это положение, или, если хотите, даже закон, даёт возможность качественно и количественно оценить результаты силового воздействия на планету по изменению её формы.

Сами силовые воздействия на планеты мы не будем выдумывать, а возьмём лишь те, которые нам известны из наблюдений за планетами солнечной системы, и проведём для них исследование.

Фактически мы можем сказать, что в определённой степени нам известно три силовых воздействия на планеты. Перечислим их.

1) Силовое влияние одной планеты на другую. Примером подобного силового влияния могут служить взаимное влияние Земли и Луны. По этой причине мы наблюдаем приливы и отливы в морях и океанах нашей планеты.
2) Силовое воздействие на планеты, которое зависит от внутреннего состояния планет. Мы знаем, что большинство планет имеет сплюснутую форму из-за того, что они вращаются вокруг своей оси.
3) Соударение планет с другими планетами и с другими материальными телами.

Начнём разбираться по порядку со всеми неопределённостями и силовыми воздействиями планет. Первыми у нас на очереди стоят бесконечная величина давления и неопределённость энергии

## ПЛАНЕТЫ СОЛНЕЧНОЙ СИСТЕМЫ

### а) Материальное выражение бесконечной величины давления и неопределённости энергии в планетах солнечной системы

Планеты солнечной системы делятся на планеты-гиганты, планеты земной группы, на спутники планет и само Солнце. Распределяя небесные тела солнечной системы таким образом, астрономы вкладывали в них свою специфику. Для нас подобное распределение планет солнечной системы даёт возможность распределить планеты по величине их массы, как планеты с большей или меньшей массой. Соответственно и объёмы этих планет являются большими или меньшими.

Если мы теперь посмотрим на планеты солнечной системы, то увидим, например, что сравнительно небольшие по массе и объёму планеты такие, как Луна, Меркурий, имеют шарообразную форму, и представляют собой остывшие материальные тела. Солнце, как самая большая планета, имеет самую высокую температуру и излучает большое количество тепловой и световой энергии. Если мы все планеты солнечной системы расположим между Солнцем и остывшими планетами, то заметим, что, чем больше масса и объём планеты, тем большее количество тепловой энергии она излучает.

Отметим, что в настоящее время основная масса людей считает, что планеты-гиганты тоже представляют собой остывшие тела с очень низкой температурой порядка минус 200˚. Это мнение в корне не верно. Все планеты-гиганты имеют толщину атмосферы в несколько тысяч километров. Следовательно, давление на поверхности этих планет достигает нескольких тысяч атмосфер. При таких давлениях и температурах любой нам известный газ переходит в жидкое и даже твёрдое состояние. Это значит, что мощные атмосферы планет-гигантов содержатся за счёт тепла, которое излучают планеты-гиганты. Отсюда следует, что поверхности этих планет раскалены до тысяч градусов по Цельсию. Это означает, что масса планет настолько горячая, что находится в жидком и газообразном состоянии. Если сравнить атмосферы планет-гигантов с нашей земной атмосферой, то мы увидим, что планеты-гиганты излучают в сотни раз больше тепловой энергии, чем планета Земля.

Исходя из всех этих конкретных условий существования планет, мы можем сделать лишь один вывод, что с ростом глубины, или высоты ($h$), давление непрерывно растёт, но не до бесконечности, а до определённой величины, хотя и очень большой.

Когда давление на определённой глубине планеты достигнет своей определённой критической величины, то за этой критической величиной давления масса планеты не может существовать в том материальном виде, в котором мы её привыкли видеть, а переходит полностью в энергию типа тепловой, электрической и им подобной. Хотя мы пользуемся многими видами подобной энергии в повседневной практике, но мы почти ничего не знаем о них. Одно мы можем сказать определённо, что энергии не могут создавать давление в объёме.

Если в объеме отсутствует масса в летучем состоянии, то мы считаем, что в данном месте мы имеем дело с вакуумом, хотя в это же время в нём может находиться и световая и тепловая, и электрическая энергия. Возможно, существуют ещё какие-то виды энергии, о которых мы ничего не знаем, но всё равно для них должно быть общим свойством, что эти неизвестные нам виды энергии, если такие существуют, не могут создавать давления в объёме. Мы, например, знаем, что световая энергия, воздействуя непосредственно на поверхность, создаёт на нём давление, но в объёме она всё равно не создаёт никакого давления. С помощью законов механики безынертной массы мы больше не можем установить в этом плане. По этой причине в дальнейшем мы будем руководствоваться этими условиями, вернее, исходить из них при исследовании энергетического состояния реально существующих планет. Покажем такую планету на рисунке 4.

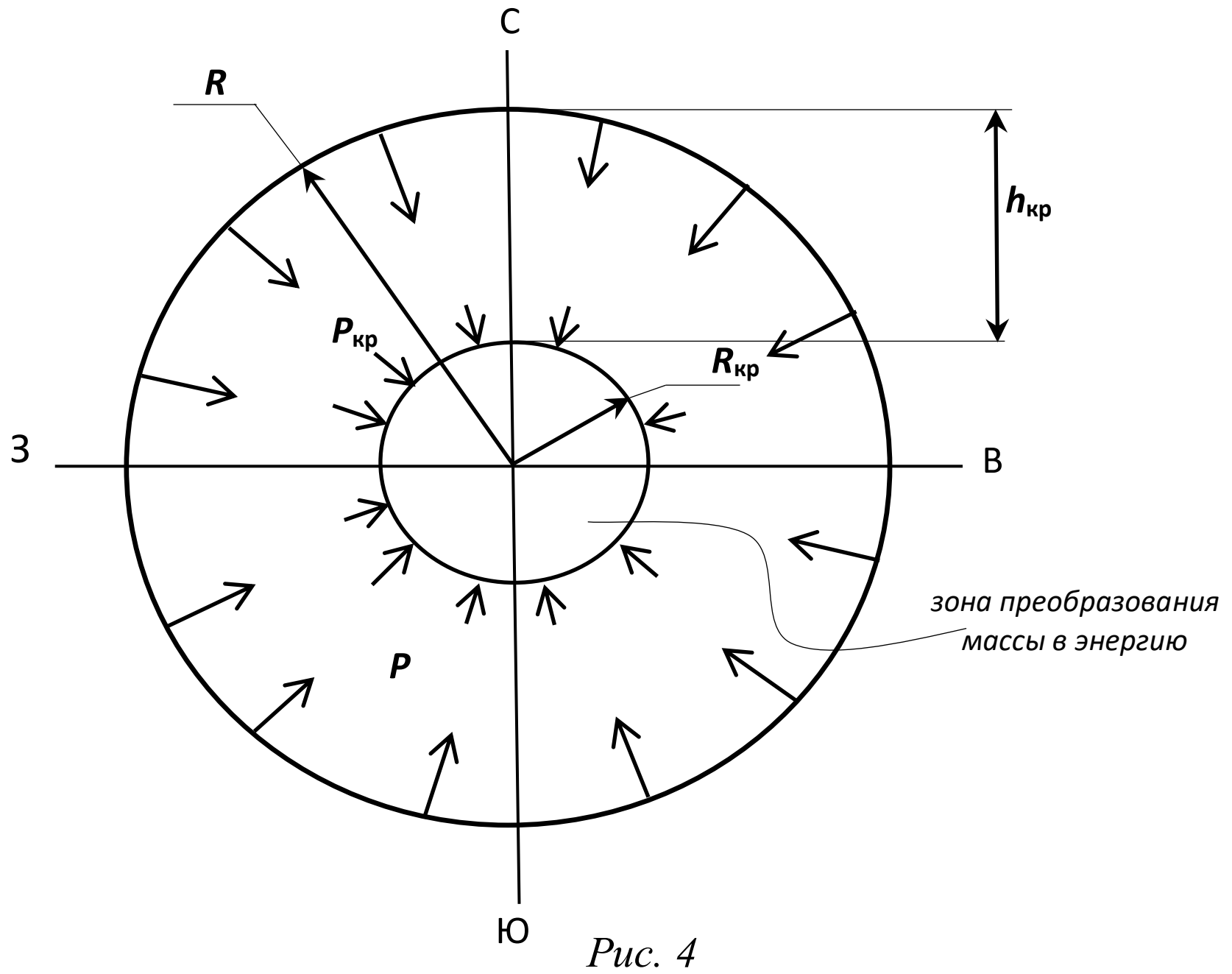

*Рис. 4*

На рисунке 4 показана планета с зоной преобразования массы планеты в энергию. Коль энергия не создаёт давления в объёме, то масса планеты будет поступать в эту зону беспрепятственно. В этом случае её масса могла бы выгореть, т.е. превратиться в энергию, за сравнительно короткий срок. Коль этого не происходит, то это означает, что энергия из зоны преобразования снова поступает в массу планеты и преобразуется там опять в её массу.

Одновременно таким путём производится непрерывный подогрев массы планеты, ибо в планете существует бесконечный источник энергии. Одновременно зона преобразования массы в энергию является источником циркуляции массы планеты, которая определяет одинаковость плотности во всём её объёме. По причине циркуляции мы можем считать, что плотность вещества в объёме реальных планет имеет постоянную величину.

Таким конкретным образом в планетах осуществляется неопределённость энергии, как ноль, умноженный на бесконечность, в определённом материальном виде. Просто в этой области образуется зона преобразования массы планеты в энергию.

Коль механика безынертной массы не может дать ответа на такой вопрос, какое количество энергии выделяется при полном переходе массы в энергию, то будем считать, что на всех планетах единица массы преобразуется в одинаковое количество энергии.

Мы знаем, что плотности различных планет отличаются друг от друга. Как нам известно, на плотность влияют два фактора – химический состав массы и температура. В то же время реально существующие планеты солнечной системы подтверждают положение, что большой ошибки не будет, если принять, что единица массы превращается в постоянную величину энергии независимо от её химического состава и температурного, или теплового режима. В этом случае величина, или объём, зоны преобразования массы в материю каждой планеты будет находиться в прямой зависимости от общей величины массы каждой конкретной планеты. Коль из единицы массы во всех случаях, например, выделяется одинаковое количество тепловой энергии, то общее количество тепловой энергии, выделяемой планетой, будет прямо зависеть от величины, или объёма, зоны преобразования массы в энергию планеты. Используя этот принцип, в основе которого лежит величина массы планеты, мы можем сопоставить планеты, например, по количеству тепловой энергии, выделяемой каждой планетой конкретно.

Солнце обладает самой большой массой. Оно выделяет самое большое количество тепловой энергии из всех планет солнечной системы. Это значит, оно имеет самый большой объём зоны преобразования массы в энергию, или объём своей печки. Следовательно, оно сжигает большее количество массы по сравнению с другими планетами. По этой причине оно получает большее количество тепловой энергии на единицу своей массы, чем другие планеты. Поэтому его называют даже звездой, а все остальные тела – просто планетами и даже спутниками.

Планеты земной группы имеют минимальную массу по сравнению с другими планетами. Минимальной массой обладают и спутники планет.

Среди планет земной группы и спутников мы видим как горячие планеты, так и полностью остывшие. То, что планеты должны быть горячими, вас теперь не удивляет, а то, что существуют холодные планеты типа Луны, должно вас удивлять.

Подобное не трудно объяснить. Мы с вами живём на планете Земля, которая имеет земную кору, или литосферу. Она представляет собой остывшую часть массы планеты. Земная кора подобно льду на реке или в океане, покрывает расплавленную массу планеты и плавает в ней благодаря выталкивающей силе Архимеда. Наличие литосферы на нашей планете указывает на то, что мощность теплового источника недостаточна, чтобы перевести всю массу Земли в жидкое состояние.

Отметим, что мощности теплового источника нашей планеты вполне достаточно, чтобы, если не расплавить полностью земную кору, то нагреть её до 500° - 600° С. Подобного не происходит лишь потому, что большая часть тепловой энергии затрачивается на вращение нашей планеты вокруг своей оси. Вы ведь все хорошо понимаете, что шаровой объём, заполненный жидкостью, не может вращаться по инерции, как твёрдое тело, на его вращение необходимо затрачивать энергию. Пока мы вам дали эти сведенья только для справки. Более подробно на этом вопросе мы остановимся ниже.

Планеты с массой меньшей, чем масса Земли, имеют ещё меньший объём своей печки, которая располагается в их центре. Её тепловой энергии не хватило, чтобы содержать всю планетную массу в жидком состоянии. В своё время все планеты находились в жидком состоянии. Сначала на их поверхностях образовалась первичная твёрдая планетная кора. Из-за того, что мощность теплового источника самой планеты была достаточно низкой, масса планеты продолжала остывать. Плавающая твёрдая кора планеты продолжала утолщаться и в один прекрасный момент она перестала быть плавающей и образовала твёрдую несущую поверхность шаровой формы.

Образовавшаяся твёрдая шаровая оболочка, которая была способна нести свой вес, больше не оказывала давления на оставшуюся жидкую часть планеты. В результате чего получилось так, что жидкая часть планеты стала представлять собой как бы меньшую по объёму и по массе планету. По этой причине объём планетной печки уменьшился, и её мощность уменьшилась соответственно. Планета продолжала остывать дальше, образовалась новая твёрдая несущая сфера, которая разгрузила оставшуюся часть жидкой массы планеты. Подобное происходило до тех пор, пока планета полностью не остывала и превращалась в окаменевшее тело типа Луны и ей подобных планет.

Прочитав подобное объяснение, вы скажите, что с нас довольно всяких гипотез, их и так развелось предостаточно, одной больше, одной – меньше, что от этого изменится. В данном случае вы имеете дело не с гипотезой, а с самой настоящей реальностью, которую вы можете проверить практически.

Дело в том, что если даже планеты с малой массой остыли и превратились в каменную глыбу, но при этом не утратили своего радиального гравитационного поля, которое притягивает их массу к центру планеты, то в этом случае планета не утрачивает свойств планеты с жидкой массой. Она просто переходит в состояние застывшего движения – только и всего. По этой причине, если остывшую планету встряхнуть так, чтобы её сферические оболочки утратили свою несущую способность (что можно сделать с помощью взрыва или столкновением с достаточно большим метеоритным телом), то в этом случае твёрдая масса планеты получит возможность двигаться и создаст в области её центра критическое давление. В результате чего произойдёт мгновенный переход массы планеты в зоне её преобразования в энергию. Коль этот переход будет мгновенным, то планета не разрушится, а просто взорвётся. По этой причине остывшие планеты фактически представляют собой бомбы, с которыми необходимо обращаться очень осторожно. Это значит, что Луна и ей подобные планеты представляют собой бомбы. При современной технике вы легко сможете убедиться в том, что остывшие планеты могут взрываться. Не следует начинать подобный эксперимент с Луны. Почему не следует, вы об этом узнаете ниже.

Можно сказать, что однажды сама природа провела подобный эксперимент. Между планетами Марс и Венера существует пояс астероидов. В начале прошлого века Ольберсом была выдвинута гипотеза, что астероиды являются обломками крупной планеты. Назвали погибшую планету Фаэтоном. Теперь это положение можно считать уже почти не гипотезой. Мы со своей стороны должны добавить, что погибшая планета Фаэтон по величине своей массы была несколько меньше Марса, коль она смогла остыть.

Гипотезу О. Ю. Шмидта о том, что планеты солнечной системы образовались путём конденсации космической пыли, а затем они медленно разогревались в результате большого давления, надо считать не состоявшейся. Ибо в этом случае планеты не смогли бы разогреться, т.к. они бы просто взорвались и снова пришли в исходное состояние.

Вы можете сказать, что этих доказательств вам не достаточно. Положение здесь осложняется тем, что мы не можем только с помощью законов механики безынертной массы определить величину критического давления, когда масса начинает переходить полностью в энергию. Но в то же время мы можем замерить размеры зоны преобразования массы в энергию, или планетной печки. На нашей планете Земля замеры её сделаны. Вы знаете, что замерами установлено, что наша планета имеет два ядра: большое и малое. На большом ядре мы остановимся ниже, а малое ядро является непосредственно планетной печкой. Замеренный радиус ($R_{кр}$) этой печки равен 500 километров. Зная этот радиус, мы можем определить величину критического давления ($P_{кр}$) для массы.

Отметим, что в настоящее время бытует мнение, что малое ядро нашей планеты составляют металлы типа железа и никеля. Это мнение ошибочно, что тоже подтверждается практическим или опытным путём. Это практическое подтверждение дано в работе [2]. Ниже мы остановимся ещё раз на этом положении.

Современные технические средства и условия планет позволяют произвести замеры планетных печек на Венере и Марсе. Большой интерес представляют собой размеры планетной печки Венеры. Ибо Венера является такой планетой без всяких искажений, которую мы показали на рис. 4 как идеальную планету.

На планетные печки Земли и Марса некоторое искажение накладывает плоский установившийся поток их массы, который вращает и Землю, и Марс и все остальные планеты. На

Венере такого же потока нет. Вот этим она интересна. Полученные таким путём данные трёх планет дадут возможность с большой точностью определить размеры планетных печек других планет и, в том числе, самого Солнца.

Отметим, что в настоящее время известно самое большое выделение энергии из массы, которое происходит при ядерных реакциях. По этой причине считают, что Солнце получает свою энергию путём термоядерной реакции, при которой водород переходит в гелий. В то же время практические наблюдения и исследования Солнца не подтверждают того, что оно черпает свою тепловую энергию из ядерных реакций. Всё равно, так называемые, ведущие учёные утверждают, что на Солнце происходят термоядерные реакции вопреки всяким фактам. Все их утверждения основываются на том, что они знают лишь о ядерных реакциях и потому ничего другого не может быть. <…>

Все мы живём на поверхности планеты Земля. Также нам известно, что толщина материков составляет всего 50-80 км, а дно мирового океана и того меньше – 5-10 км. Всю остальную массу нашей планеты составляет расплавленная масса, находящаяся под действием высоких температур. Всем хорошо известно, что необходимо большое количество тепловой энергии, чтобы содержать массу планеты в расплавленном состоянии. Начёт тепловой энергии Земли, как и других планет, учёные ведут себя скромно. Негласно считают, что масса Земли разогревается от большого давления. Полагают, что этого достаточно, т.к. они не находят радиоактивности в материалах изверженных пород. В целом же считают, что вулканическая деятельность является стихийным бедствием. <…> Иначе вулканологи заметили бы, что в вулканической деятельности поражает своей необычностью не вулканические бомбы и даже ни сама лава, а то, каким образом в многокилометровой толще земной коры проплавляется выход самой лавы на поверхность. Ведь проплавленный канал в земной толще выглядит тонкой ниточкой по сравнению с её толщиной. Также всем хорошо известно, чтобы вести проплавление в любом материальном теле, необходимо подводить энергию в зону проплавления. Из-за небольшого диаметра зоны проплавления подвод энергии в неё невозможен. Циркуляция в зоне проплавления расплавленной магмы тоже невозможна. Как видите, тепловую энергию невозможно подвести в зону проплавления, в то же время проплавление – налицо. Например, если вы смоделируете земную кору из свинца и попытаетесь сделать в нём подобное проплавление с помощью, например, расплавленного железа, то если вам даже и удастся проплавить свинец, железо из свинца будет выходить в твёрдом состоянии. Вот как интересно получается.

Опять же, расплавленная магма движется снизу вверх вопреки законам физики. Ведь материки представляют собой «айсберги» в расплавленной массе Земли. Это явление природы указывает нам на то, что в зонах вулканической деятельности имеет место выход энергии, которая поступает из планетной печки.

На это же указывают кимберлитовые, или алмазоносные, трубки. Исследования этих трубок показывает, что проплавления в них происходили быстро, со скоростью несколько сот метров в секунду. Притом происходило даже не проплавление, а поглощение твёрдой массы земной коры с очень большой скоростью. Ведь кимберлитовые трубки заполнены не расплавленной массой, а осадочными породами – кимберлитом, или синей глиной. Это говорит о том, что неизвестная нам энергия поглощала в объёме трубки массу, а её место заняла синяя глина, которая была вытеснена в эту пустоту с каких-то глубинных слоёв земной коры. Отсюда мы видим, что никакая ядерная энергия не может сравниться по мощности с энергией планетной печки. Притом эта глубинная энергия имеет необычную форму упаковки в пространстве. Притом эта энергия не взрывает, а поглощает массу на пути своего движения.

Вот эти факты прямо указывает на существование планетной печки, где масса при большом давлении полностью переходит в энергию.

Если вы посмотрите на карту океанических течений, то увидите, что она во многом схожа с картиной движения облаков, например, на Юпитере и Сатурне. В океанах нет перепадов по высоте, а течения всё равно существуют. Движение жидкости может происходить только под действием силового поля. Это говорит о том, что в природе существуют неизвестные нам силовые поля. Например, можно прочитать сообщения, что в Тихом океане наблюдались волны, высота которых достигала 25 метров. Шли они непрерывно через каждые 20 секунд. Вы знаете, что земная атмосфера создаёт давление на поверхности океана равное одной атмосфере, что эквивалентно десятиметровому столбу жидкости, а волны достигают 25 метров. В сводках погоды иногда говорят, что волны в таких-то местах достигают 9 или 14 метров. Подобное считается заурядным явлением. Никому даже в голову не приходит, что энергии атмосферы может хватить на создание волн высотой не более пяти метров. Такая наглядная картина движения океанических вод указывает на то, что расплавленная масса Земли под поверхностью земной коры существует не спокойно, а совершает бурные движения.

Вот это бурное движение магмы, или расплавленной массы Земли, в определённой степени отражают океанические воды в своём движении. В большой степени оказывает своё влияние движение расплавленной массы Земли на движение воздушных масс в атмосфере Земли.

Механика безынертной массы даёт возможность установить факт существования неизвестных энергий и силовых полей. Она даже даёт возможность по движению океанических вод определить величину воздействия этих силовых полей, но физическую сущность этих силовых полей и энергий могут определить другие науки, вернее, люди, занимающиеся другими науками.

Ещё одним интересным явлением природы солнечной системы являются кометы. Наличие хвостов не является их обязательным признаком, но зато для кометы обязательным является твёрдое ядро диаметром несколько километров и газовая оболочка типа атмосферы. Вот эти внешние признаки говорят о том, что какие-то вещества, которые составляют твёрдую часть кометы, не способны создавать твёрдых самодостаточных сферических оболочек, которые бы удерживали свой вес и тем самым снижали бы давление в центре кометы. По этой причине в центре кометы должна существовать печка по преобразованию массы в энергию. Ибо только это условие способно обеспечить существование газовой оболочки, или кометной атмосферы.

Надо полагать, что под Тунгусским метеоритом мы должны понимать столкновение кометы с нашей планетой.

Мы привыкли к тому, что всякое тело, обладающее массой, имеет гравитационное поле. Притом мы здесь не делаем никакого различия. В то же время оно существует. Ибо одни тела имеют радиальное гравитационное поле, которое действует на массу в радиальном направлении и стремится образовать шарообразный объём из массы, а другие тела просто притягиваются друг к другу, как намагниченные тела. Например, кометы обладают радиальным гравитационным полем, а астероиды – нет. Вот с этой точки зрения кометное твёрдое вещество представляет для нас очень большой интерес. Добыть его очень необходимо.

Будем считать, что мы определились количественно с неопределённостями, связанными с центральными областями планет. Ибо мы установили, что величину, или размеры, зоны преобразования массы в энергию можно пока установить непосредственными замерами. По размерам планетной печки мы затем определяем критические давления и таким путём получаем материальное выражение для неопределённостей, связанных с бесконечностью давления и неопределённостью энергии.

## б) СИЛОВОЕ ВЛИЯНИЕ ОДНОЙ ПЛАНЕТЫ НА ДРУГУЮ

Здесь мы тоже будем иметь дело с планетами, которые имеют радиальное гравитационное поле и массу в жидком состоянии. Для удобства выявления силового воздействия одной планеты на другую, одну планету мы будем рассматривать полностью, а вторую – условно, в виде её силового воздействия.

Когда речь заходит о силовом воздействии, то в вашей памяти оно сразу же ассоциируется с векторами. Математический векторный анализ доведён до виртуозности. Если вы имеете дело с твёрдым телом, то они вам предлагают большие стрелы, которые называются векторами. Если вы имеете дело с жидкостями, то они вам предлагают эти векторы мелкой россыпью, которая называется градиентами. По этому способу у них даже расписана теория поля. Когда возьмёшь в руки математику, то создаётся такое впечатление, что всё уже изучено, и изучать больше нечего. Когда же переходишь к действительности, то получается всё - наоборот: везде сталкиваешься с неизвестным, а математикой делать нечего. Современная механика жидкости и газа существует сама по себе, а реальная практика людей – сама по себе. Чтобы как-то прикрыть ненужность подобной теории, придумали ещё теорию подобия, сущность которой сводится к тому, что она даёт возможность пересчитывать с уже известного, как большего с меньшего или наоборот. Подобное положение оправдывают тем, что считают, что жидкости и газы подчиняются законам

Ньютона, которые он нашёл для твёрдых тел. Вместо того, чтобы найти законы механики для жидкости и газов, люди продолжают городить одну нелепость хуже другой. Современная теория механики жидкости и газа представляет собой сплошной векторный анализ, в основе которого лежат законы Ньютона. Постарайтесь забыть всю эту векторную муру, тогда вы сможете воспринимать всё, как оно должно быть.

Твёрдое тело воспринимает силовое воздействие в виде непосредственного контакта двух тел, а жидкости – через изменение давления в объёме. По этой причине на рис. 5 стрелками у нас показаны направления действия гравитационных полей планет.

Наша исследуемая планета, которая подвергается силовому воздействию другой планеты, имеет радиальное гравитационное поле. Вот это условие мы показали стрелками. Планета, которая воздействует на нашу исследуемую планету, находится на значительном расстоянии от неё. По этой причине её гравитационное поле будет воздействовать на исследуемую планету своими силовыми линиями, которые располагаются параллельно линии, соединяющей центры планет. Вот эти линии тоже показаны на рис. 5 стрелками. Во всех случаях стрелки на рис. 5, как и на всех остальных рисунках, показывают, в каком направлении притягивается масса – только и всего, без всякой величины.

Теперь остаётся получить зависимости для силового воздействия. Для этой цели покажем на рис. 5 плоскость ($S$), которая проходит через центр исследуемой планеты и располагается перпендикулярно действию силовых полей, воздействующих на нашу планету. Вот эта плоскость ($S$) будет для нас плоскостью начала отсчётов, а для радиального гравитационного поля планеты началом отсчёта так и останется её поверхность с радиусом ($R$). После этого мы уже можем рассматривать изменение давления в планете. Для этого выделим, как обычно, столб жидкости в центре планеты величиной ($l$), как показано на рис. 5.

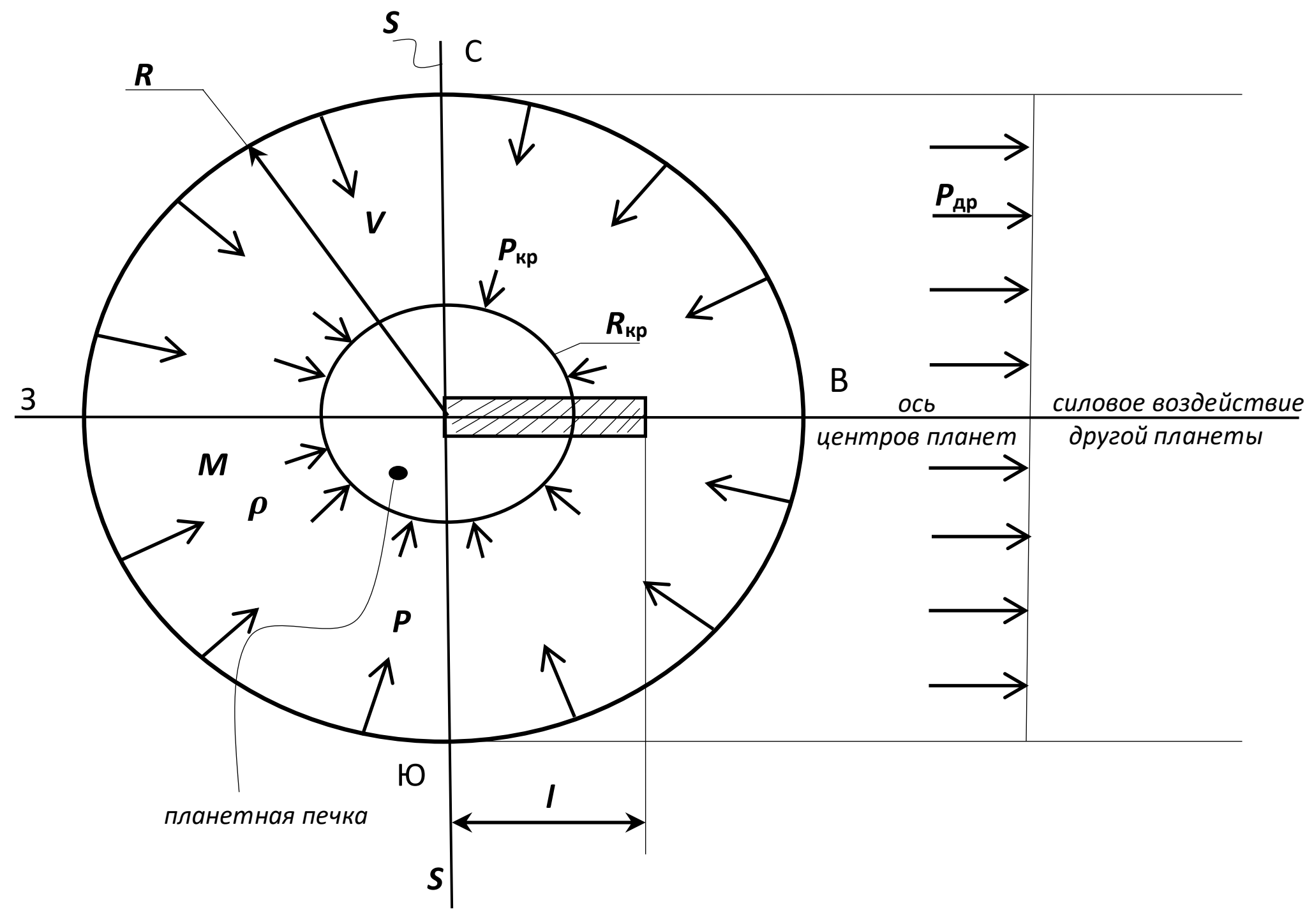

*Рис. 5*

Здесь может последовать такой вопрос, почему плоскость ($S$) начала наших отсчётов расположена на расстоянии радиуса ($R$) планеты, а не на расстоянии диаметра планеты? Ответ здесь будет таков, что мы исходим из реальных условий силового воздействия, которое определяется изменением давления в планете и законами механики безынертной массы. Если бы мы плоскость ($S$) начала отсчётов расположили на расстоянии её диаметра или за планетой, то в этом случае мы могли бы рассматривать притяжение всей массы планеты, которое уже определено законами всемирного тяготения Ньютона. Мы не собираемся повторяться. Нас интересует то силовое воздействие, которое изменяет давление планеты с одновременным изменением её внутренней энергии и которое предстаёт перед нами в наглядном виде через изменённую форму объёма планеты. Это условие мы определили выше. Изменение формы планеты мы связываем с уменьшением давления и энергии потому, что при подобных силовых воздействиях в реальных условиях невозможно увеличивать давление и энергию планеты, т.к. они имеют предельно большие величины.

Для своих исследований мы будем использовать принцип независимости действия силовых полей. Этот принцип является общим для любых силовых взаимодействий тел. По этой причине мы можем рассматривать раздельно силовое воздействие на нашу планету её собственно радиального гравитационного поля и гравитационного поля воздействия другой планеты. Действие радиального гравитационного поля мы рассмотрели выше и получили для него необходимые зависимости.

Для получения зависимостей мы будем пользоваться теми же принципами, которые мы использовали раньше, т.е. дело мы будем иметь с застывшим движением и соответственно для него запишем уравнения сил и движения. Как это делается, вы теперь знаете. Поэтому мы сразу запишем уравнение давлений по аналогии уравнения сил для цилиндрического столба жидкости (5) с соответствующими изменениями в обозначениях.

$$P_{\text{др}} = \rho l w_{\text{др}}^2 \qquad (21).$$

Мы получили уравнение (21), которое является зависимостью для давления силового воздействия от другой планеты. Но мы не можем воспользоваться им, т.к. мы имеем здесь два неизвестных. Нам неизвестна величина давления ($P_{\text{др}}$) и величина постоянной скорости ($w_{\text{др}}$) другой планеты. Плотность ($\rho$) нам известна, т.к. характеристики исследуемой планеты мы знаем. Высота жидкостного столба ($l$) нам тоже известна, т.к. мы задаём её сами. Чтобы мы могли пользоваться уравнением (21), нам необходимо получить ещё одну зависимость. Вот эту вторую зависимость мы можем получить по изменению формы шарового объёма нашей исследуемой планеты.

Уравнение (21) нам показывает, что с увеличением высоты жидкостного столба ($l$) от центра шара к его периферии давление ($P_{\text{др}}$) растёт пропорционально высоте ($l$). Когда высота ($l$) станет равной радиусу исследуемой планеты ($R$), то величина давления ($P_{\text{др}}$) силового воздействия другой планеты достигнет максимальной величины. Сила воздействия другой планеты стремится переместить массу от центра исследуемой планеты к её периферии, или к наружной шаровой поверхности. Наружная поверхность планеты не может быть препятствием для силового воздействия. По этой причине жидкость должна будет начать перемещаться от центра планеты к её периферии, коль для неё нет препятствия. В этом случае жидкость из одной планеты должна была бы начать перемещаться к другой планете. Чего не происходит в действительности. Это значит, что воздействующая сила уравновешивается силами планеты. Материально это выражается в том, что уравновешивание сил воздействия должно производиться соответствующим столбом жидкости.

На рис. 6 мы показали силы воздействия и противодействующие им силы. На исследуемый цилиндрический столб жидкости высотой ($l$) действует сила воздействия, величину которой мы определили уравнением (21). Максимальная величина высоты ($l$) может быть равна планетному радиусу ($R$). За планетным радиусом уже будет располагаться уравновешивающий столб жидкости высотой ($h_{\text{пр}}$). Будем считать, что уравновешивающий столб жидкости находится только под действием сил гравитационного силового поля нашей исследуемой планеты. Тогда для определения противодействующего давления ($P_{\text{пр}}$) мы можем воспользоваться уравнением (5). Запишем его ещё раз:

$$P_{\text{пр}} = \rho h_{\text{пр}} w^2 \qquad (22).$$

Различие между уравнением (5) и уравнением (22) заключается лишь в том, что в уравнении (5) мы сами задаёмся высотой ($h$) жидкостного столба, а в уравнении (22) высота ($h_{\text{пр}}$) получается сама, как высота, компенсирующая силовые воздействия другой планеты. Мы считаем, что находимся на нашей исследуемой планете. Поэтому высоту ($h_{\text{пр}}$) в уравнении (22) мы можем получить путём прямого замера. В этом случае высота ($h_{\text{пр}}$) станет известной, а плотность ($\rho$) и постоянная скорость ($w$) нам были известны. Следовательно, по уравнению (22) мы можем найти величину противодействующего давления ($P_{пр}$).

Мы получили необходимые зависимости для действующих независимо друг от друга сил. Теперь нам необходимо рассмотреть их при совместном действии на исследуемую планету.

Коль другая планета воздействует на нашу планету не радиальным, а параллельным гравитационным полем, то силы давления гравитационного поля исследуемой планеты мы должны будем взять определёнными по типу цилиндрического столба жидкости, т.е. уравнение (5). Величину давления для взаимодействующих сил другой планеты мы определим уравнением (21).

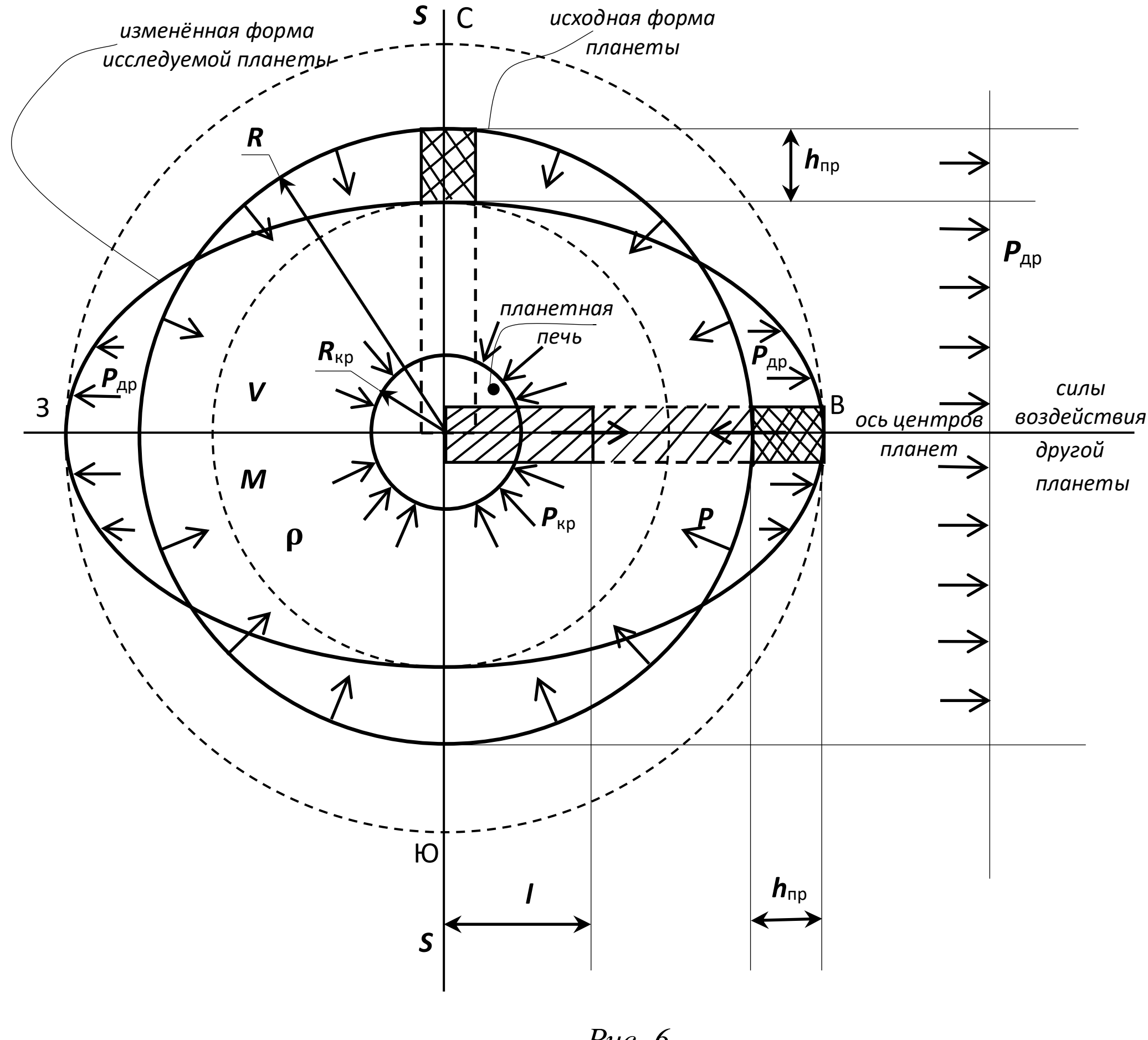

*Рис. 6*

Согласно уравнению (5) величина давления растёт от поверхности планеты к её центру пропорционально высоте жидкостного столба ($h$). Для сил воздействия другой планеты согласно уравнению (21) величина давления растёт от центра планеты к её наружной поверхности пропорционально высоте жидкостного столба ($l$).

Если следовать векторному анализу, то мы должны были бы изобразить эпюры давлений. Затем сложить их и получить нужный нам ответ. Если мы так поступим, то не получим искомого результата, хотя полученные выше зависимости являются правильными, т.е. соответствующими действительной сути механического состояния жидкости.

Чтобы получить решение, мы должны проникнуть в суть самого явления. Здесь суть явления такова, что в направлении действия гравитационного поля другой планеты масса исследуемой планеты теряет часть своего веса. В направлении перпендикулярном действию гравитационного поля другой планеты масса исследуемой планеты не теряет своего веса и остаётся неизменной. Воспользуемся этими условиями и запишем их в математическом выражении. Для чего воспользуемся цилиндрическим столбом жидкости, но коль теперь мы будем иметь дело с весом, то представим себе, что он вдруг стал твёрдым, но всё равно находится в жидкой массе планеты.

Запишем вес такого столба:

$$G = Fh\rho g \qquad (23),$$

где $G$ – вес столба,

$F$ - площадь поперечного сечения столба,

$\rho$ - плотность,

$g$ - постоянное ускорение гравитационного поля планеты. Для нашей планеты Земля $g = 9{,}81$ м/сек$^2$.

Мы в своих жидкостных столбах принимаем площадь сечения ($F$) равной единице. Что удобно для записи давления ($P$). Для отвердевших столбов мы принимаем площадь их сечения тоже равной единице, т.е. $F = 1$. Тогда уравнение (23) примет вид:

$$G = h\rho g \quad (24).$$

Возьмём столб отвердевшей жидкости, расположенной по оси центров планет (см. рис. 6) и запишем для него условия равновесия. Принимаем высоту столба ($h$) равной радиусу планеты ($R$), т.е. $h = R$ . Тогда исследуемый столб отвердевшей жидкости будет иметь такой вес:

$$G = R\rho g \quad (25).$$

Такой вес ($G$) столб имел до воздействия гравитационного поля другой планеты. Когда же стало действовать гравитационное поле другой планеты, вес этого столба уменьшился, т.к. уменьшилась величина постоянного ускорения гравитационного поля и стала равна ($g_1$), т.е. $g > g_1$. Запишем изменённый вес столба жидкости:

$$G_1 = R\rho g_1 \quad (26).$$

В результате мы будем иметь, что $G > G_1$. По этой причине столб жидкости весом ($G_1$) будет всплывать. Чтобы этого не происходило, мы должны сверху или со стороны силового поля другой планеты добавить «кусочек» жидкостного столба высотой ($h_{\text{пр}}$), плотностью ($\rho$) и постоянным ускорением ($g$) исследуемой планеты, т.е.

$$G_{\text{пр}} = h_{\text{пр}}\rho g \quad (27).$$

Тогда условие равновесия запишется в таком виде:

$$G = G_1 + G_{\text{пр}} \quad (28).$$

Запишем уравнение (28) через параметрическое выражение весов, т.е. по уравнениям (25), (26) и (27), тогда получим:

$$R\rho g = R\rho g_1 + h_{\text{пр}}\rho g \quad (29).$$

В уравнении (29) сократим плотность ($\rho$) и распишем его относительно изменённого постоянного ускорения ($g_1$), т.е.

$$g_1 = \frac{R - h_{\text{пр}}}{R} g \quad (30).$$

В уравнении (30) нам неизвестны постоянное ускорение ($g_1$) и высота принятого дополнительного столба жидкости ($h_{\text{пр}}$). Следовательно, для решения уравнения (30) нам необходима ещё одна зависимость. Мы теперь можем её получить.

В направлении действия гравитационного поля другой планеты вес массы уменьшается, а в перпендикулярном ему направлении он остаётся неизменным. Это значит, что давление внутри планеты упадёт на определённую величину.

На уменьшение планетного давления отреагирует столб жидкости, расположенный в направлении плоскости ($S$), или полюсов планеты (см. рис. 6). При действии гравитационного поля другой планеты для этого столба жидкости характеристики остаются неизменными, т.е., ни плотность ($\rho$), ни постоянное ускорение ($g$) не изменяются. Следовательно, при изменении давления внутри планеты может измениться только его высота. По этой причине уменьшения давления внутри планеты высота жидкостного столба равная величине ($R$) уменьшится на величину ($h_{\text{пр}}$).

Как мы отмечали выше, эту высоту ($h_{пр}$) мы можем замерить, т.к. мы находимся на поверхности исследуемой планеты.

Коль во всём объёме исследуемой планеты в направлении перпендикулярном направлению действия гравитационного поля другой планеты характеристики массы будут оставаться неизменными, т.е. плотность ($\rho$) и постоянное ускорение ($g$) не изменятся, то высота компенсирующего столба жидкости, которую мы записали уравнением (27) и применили в уравнении равновесия (29), будет равна по величине величине уменьшения высоты жидкостного столба в плоскости ($S$) (см. рис. 6). Это значит, что высоту ($h_{пр}$) мы можем определить путём прямого измерения на планете.

Найдя таким путём высоту ($h_{пр}$) компенсирующего столба жидкости, мы можем с помощью уравнения (30) вычислить величину постоянного ускорения ($g_1$).

Величина постоянного ускорения ($g_1$) является изменённой величиной постоянного ускорения ($g$) нашей исследуемой планеты. Изменение здесь выражается в том, что гравитационное поле другой планеты уменьшит своим действием величину постоянного ускорения исследуемой планеты. Величина уменьшения постоянного ускорения является в этом случае величиной постоянного ускорения гравитационного поля другой планеты, которое воздействует на нашу планету. Величину постоянного ускорения ($g_{др}$) другой планеты мы можем получить, вычтя из величины постоянного ускорения ($g$) нашей планеты величину изменённого ускорения ($g_1$), т.е.

$$g_{др} = g - g_1 \qquad (31).$$

Затем уже по величине постоянного ускорения ($g_{др}$) другой планеты мы можем определить постоянную скорость ($w_{др}$) для поля другой планеты, которым она воздействует на нашу планету. Вы знаете, что величина постоянной скорости равна корню квадратному из величины постоянного ускорения, т.е.

$$w_{др} = \sqrt{g_{др}} = \sqrt{g - g_1} \qquad (32).$$

Вот таким путём мы найдём величину возмущения, которая возникает от действия гравитационного поля другой планеты.

Отметим, что вам такие выводы могут показаться несколько искусственными. Мы всегда оперировали с давлением, а здесь перешли на вес и на выталкивающую силу, т.е. использовали законы Ньютона и Архимеда. Вот такого рода положения на начальной стадии у нормальных людей порождают сомнения, с которыми они принимаются разбираться молча. Поэтому пока они считают себя не в праве высказываться ни за, ни против нового. В общем, это нормальная человеческая реакция. <…>

***

Вернёмся снова к своей теме.

Выше мы разобрались с давлением, а теперь остаётся разобраться с энергией.

В механике твёрдого тела потенциальная энергия, как и работа, равна произведению силы на путь. Поэтому, чем дальше твёрдое тело расположено от центра, тем больше его энергия, т.е. чем больше ($h$), тем больше энергия.

Для жидкостей и газов всё выглядит наоборот. Для них энергия равна произведению давления на объём. Давление в планете растёт с приближением к её центру и уменьшается при удалении от центра планеты. Следовательно, при всяком уменьшении энергии в планете жидкость должна удаляться от центра на соответствующее расстояние.

В нашем случае, когда одна планета воздействует на другую своим гравитационным полем, тоже происходит уменьшение энергии в планетах. Коль энергия равна произведению объёма на давление, то она не может уменьшиться просто только за счёт уменьшения давления. В этом случае с уменьшением давление должно в обязательном порядке произойти и перемещение объёма, т.е. изменение его формы. Вот это явление мы наблюдаем в планете при воздействии на неё гравитационным полем другой планеты. Рассмотрим изменение объёма в планете на рис. 7.

До приложения гравитационного поля другой планеты наша исследуемая планета имела шаровой объём радиусом ($R$) и определённую величину энергии. Вот этот шаровой объём и энергию мы принимаем за базовые величины, что показано на рис. 7. Масса планеты представлена в виде несжимаемой жидкости. Следовательно, при уменьшении энергии в планете происходит отток части массы планеты в одном месте и приток её в другом месте планеты. На рис. 7 показан отток массы в виде объёма ($V_1$) и её переток в объём ($V_2$).

Коль мы берём действительную картину перераспределения жидкости в объёме планеты при воздействии на неё гравитационного поля другой планеты, то мы убеждаемся, что такое перераспределение жидкости в объёме планеты соответствует уравнению потенциальной энергии жидкостей и газов.

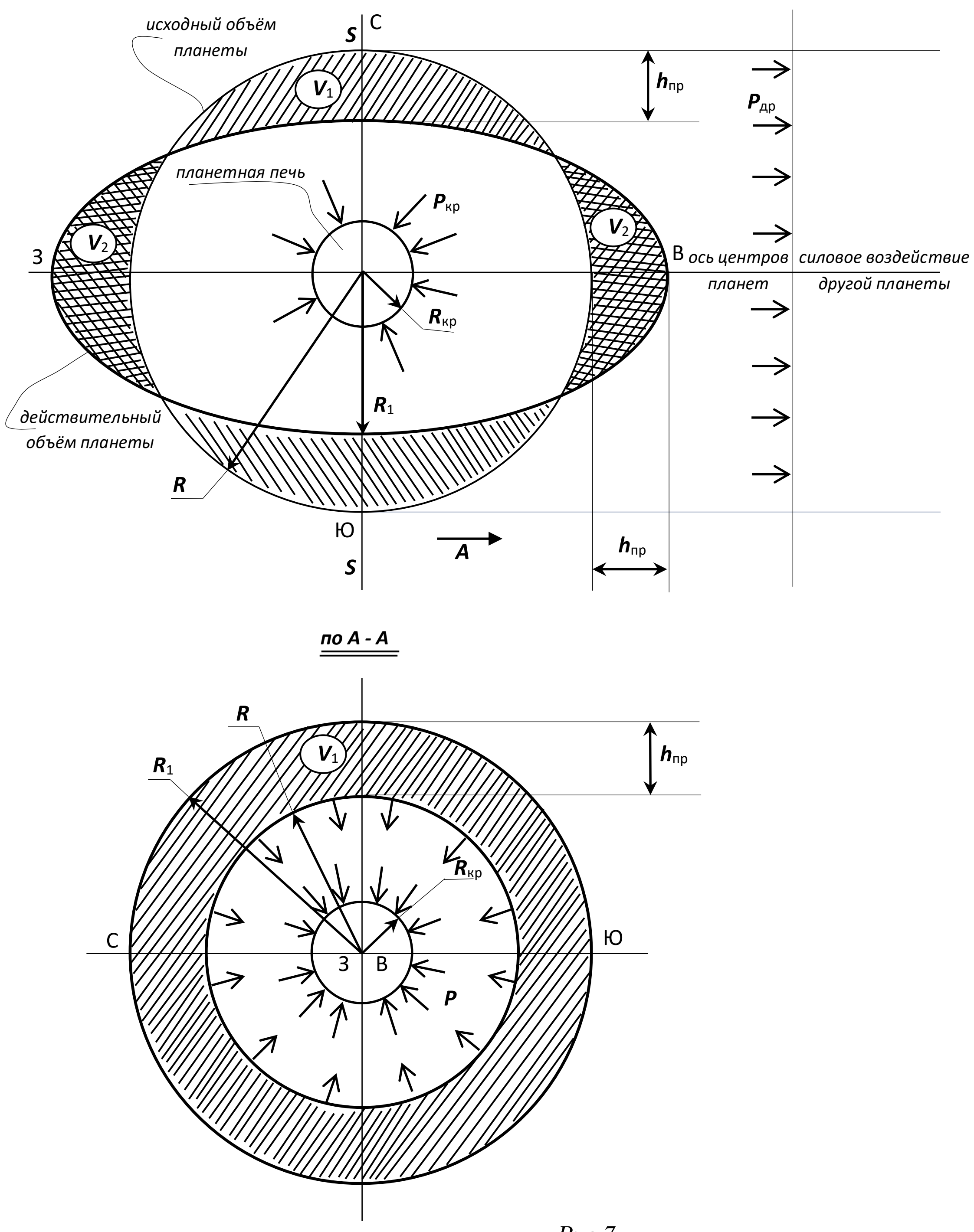

*Рис.7*

При воздействии силового поля другой планеты масса вытекает из объёма ($V_1$) исследуемой планеты, расположенного перпендикулярно направлению действия силового поля другой планеты. Вытекшая из объёма планеты масса располагается в двух симметрично расположенных объёмах ($V_2$) в направлении действия силового поля другой планеты. В целом исследуемая планета под воздействием силового поля другой планеты принимает форму подобную эллипсу. В то же время основная контурная линия изменившегося объёма планеты не является эллипсом, а является какой-то иной фигурой, для которой математика ещё не получила соответствующие зависимости.

Мы не будем искать этих зависимостей, чтобы не отбирать куска хлеба у математиков, желающих оставить своё имя в математической науке. Мы лишь показали большую нужность такой фигуры для практического исследования планет и определили все необходимые для неё характеристики.

В общем, изменившийся объём планеты имеет ось симметрии, которая соединяет центры взаимодействующих планет, и плоскость симметрии ($S$), которая располагается перпендикулярно направлению действия сил другой планеты. В конечном результате мы видим, что такая симметрия изменённого объёма планеты соответствует её уравновешенному состоянию. Ведь объём планет, заполненный жидкостью, не имеет жёстких материальных границ и сохраняется лишь за счёт того, что жидкость располагается в нём симметрично, уравновешивает сама себя. Это значит, что характер воздействия силового поля другой планеты, ни смотря на свою прямолинейность, носит более сложный характер взаимодействия, который определяется законами механики безынертной массы и симметричным самоуравновешиванием жидкости. По этой причине в исследовании планет в первую очередь мы должны исходить из законов природы и затем оформлять их соответственно количественными математическими зависимостями.

Вытекший объём ($V_1$) является ни просто объёмом жидкой массы, а он включает в себя вполне определённое количество энергии, которую мы можем записать в таком виде:

$$\Delta Э_1 = V_1 \cdot P_1 = V_1 \cdot \rho h_{пр} w^2 \qquad (33).$$

Ведь это количество энергии ($\Delta Э_1$) вместе с жидкой массой переместилось из объёма ($V_1$) в два объёма ($V_2$).

Согласно закону сохранения энергии количество энергии, заключённое в объёме ($V_1$), должно равняться количеству энергии, заключённому в объёме ($V_2$). В связи с тем, что мы имеем дело с несжимаемой жидкостью, то объём ($V_1$) и два объёма ($V_2$) должны быть равны по величине друг другу, т.е. $V_1 = V_2 + V_2$.

Следовательно, чтобы сохранялось равенство энергий в этих объёмах, как

$$\Delta Э_1 = \Delta Э_2 \qquad (34),$$

должно автоматически сохраняться равенство давлений в этих объёмах. В свою очередь давления находятся в прямой зависимости от глубины жидкостного столба ($h_{пр}$). Это значит, что в объёмах ($V_1$ и $V_2$) должно в обязательном порядке сохранятся равенство высот ($h_{пр}$) и постоянство скорости планетного тяготения ($w$).

В целом же мы видим, что уменьшение энергии в исследуемой планете произошло за счёт удаления части её массы от центра планеты. Что находится в полном соответствии с зависимостью количественного распределения энергии в объёме жидкости.

Распределение энергии в объёме планеты полностью подтвердило нам правильность наших приёмов при составлении зависимостей для распределения давлений в объёме планеты или, то, что мы пользовались не искусственными приёмами, а единственно правильными приёмами, которые единственно приемлемы в данном случае.

Теперь нам остаётся разобраться непосредственно в самих вычислениях изменения энергий в планете, т.е. в конкретном вычислении объёмов и давлений.

Что касается давлений, то они находятся в прямой зависимости от высоты ($h$), или глубины. По этой причине мы должны были бы составить для них дифференциальные уравнения. Затем, уже с помощью интегрирования, получить необходимые величины давлений.

Если для шарового объёма составление таких зависимостей не составляет большого труда, то для других объёмов, отличных от шарового объёма, составление дифференциальных и интегральных зависимостей в большинстве случаев является невозможным делом. Но всё равно в таких случаях невозможно отказаться от дифференциального и интегрального исчисления. В этом случае приходится пользоваться их приближёнными методами. К разряду приближённых методов относится графическое дифференцирование и интегрирование. В нашем случае для количественного вычисления энергии наиболее приемлемым является графический метод.

Разъясним свою мысль примером. Энергию объёма ($V_1$) мы записали уравнением (33) в арифметическом виде. Для количественного вычисления энергии в объёме ($V_1$) арифметической формой уравнения (33) пользоваться нельзя, т.к. получим ошибочные результаты. Получается, что мы такой записью сделали ошибку. Ибо правильная запись уравнения (33) должна производиться не в арифметической, а в интегральной форме записи. В данном случае мы умышленно пошли на такую ошибку. Сделано это здесь потому, что почти каждый человек поймёт зависимость энергии, записанной в арифметической форме, а если бы она была записана в интегральной форме, то понять её смог бы не каждый.

Специалист же в области механики жидкости и газа знает, что в таких случаях энергия записывается в интегральной форме. Когда ему потребуется решить какой-то конкретный пример, то он без всякой подсказки преобразует зависимость (33) в интегральную форму и воспользуется ею, в общем, настоящий специалист не сочтёт нашу умышленную ошибку за ошибку, а примет её как удобную для всех форму записи энергии.

*Всё же данное положение требует более подробного разъяснения. Сделаем его.*

*В настоящее время бытует негласное мнение, что математика является наукой наук. Это мнение связано с неверным пониманием назначения математики.*

*Математика является наукой о вычислении количеств, которые существуют для неё абстрактно, или отвлечённо. Вот это назначение математики является главной её сущностью. С этой меркой надо подходить к математике. По этой мерке она не может быть наукой наук, т.к. не связывает своё количественное вычисление с явлениями природы. Просто мы можем пользоваться теми или иными количественными математическими вычислениями для анализа исследований соответствующих явлений природы – только и всего. Это значит, что математикой можно пользоваться только при обработке результатов исследования, а не наоборот. В настоящее время навязали эту обратную неправильную тенденцию, что математика является главным методом в исследовании явлений природы.*

*В неправильности такой тенденции вы могли убедиться на наших примерах. Если бы мы прибегли к абстрактной математической форме теории силового поля другой планеты, то мы бы не смогли количественно определить воздействие силового поля другой планеты на исследуемую планету. Если бы мы пользовались абстрактной математической теорией поля, то мы в своих исследованиях конкретного явления природы должны были бы отказаться от законов природы, каковыми являются закон сохранения энергии и законы механики безынертной массы, а просто пользоваться математическими зависимостями, которые в математике предписаны для теории поля. Если бы мы поступила так, то мы бы не смогли исследовать явление природы.*

*Далее, математику делят на два больших раздела: элементарную математику и высшую математику. Таким путём математике придают неверное назначение, т.е. неверный смысл, как будто элементарная математика имеет какие-то ограничения в количественных вычислениях, а высшая математика их не имеет.*

*Как вам известно, элементарная математика включает в себя арифметику, геометрию, алгебру, тригонометрию и сами числа. Все эти разделы включает в себя и высшая математика. Различие между ними заключается лишь в том, что, так называемая, высшая математика в своих вычислениях разделяет любые количественные величины на предельно малые составляющие их части, которые там называются дифференциалом, а так называемая элементарная математика вычисляет своими методами всякие количественные величины в целом. Это значит, что все ограничения количественных вычислений в элементарной математике в одинаковой мере присущи и высшей математике, т.е. элементарная математика определяет ограничения для высшей математики, а не наоборот. По этой причине, так называемая, элементарная математика находится в высшем положении по отношению к так называемой высшей математике. Как видите, даже в самой математике всё переиначено наоборот.*

*Если провести образное сравнение между теперешними элементарной математикой и высшей математикой по их действительному назначению для конкретных исследований, то оно будет выражаться в таком виде: элементарная математика является нормальным человеческим зрением, которое является общей нормой для каждого человека, а высшая математика является микроскопом или телескопом  в зависимости от того, какой предмет желает детально рассмотреть тот или иной человек. Ибо теперешняя высшая математика даёт возможность рассмотреть более мелкие детали в общем количестве, как детальную структуру самого количества, а общее количество определяется зависимостями элементарной математики. Ведь в жизненном обиходе мы пользуемся своим обычным зрением и с его помощью рассматриваем окружающие нас предметы, даже их детальное построение. Когда не хватает*

*нашего зрения, чтобы детально рассмотреть тот или иной предмет, тогда мы прибегаем к помощи телескопа или микроскопа.*

*Элементарная математика находится в точно такой же зависимости, как наши глаза с микроскопом и телескопом. По этой причине количественные величины любой или всякой научной работы в любом случае должны быть определены в первую очередь зависимостями элементарной математики, а зависимости высшей математики должны применяться лишь в необходимых случаях, когда возникла в этом необходимость. Если даже научная работа основывается на детальном исследовании количеств, которые сразу требуют зависимостей высшей математики, то всё равно, для общих количеств в первую очередь должны быть записаны зависимости в форме элементарной математики, а затем уже должен идти детальный анализ в форме зависимостей высшей математики. Отсутствие такой последовательности в математических записях научных работ любого назначения приведёт к тому, что многие, даже специалисты, не могут разобраться в работах своих же коллег, а большинству людей они просто становятся не доступными для понимания.*

*Вот такая неразбериха в современной математике нужна лишь опять же всё тем же мошенникам в науке. Ведь высшая математика даёт возможность, как и всякий телескоп или микроскоп, видеть в мухе слона, а в слоне – муху. Этот эффект так же даёт возможность выдавать хорошо известное старое за новое. Мошенники даже очень хорошо знают об этом эффекте и неразберихе в математике и ловко, даже очень ловко, пользуются этим в своих корытных интересах. Они используют этот эффект в первую очередь для добывания научных степеней. По этой причине существуют учёные степени с приставкой «физико-математических наук», а не просто, скажем, кандидата физических наук или доктор математических наук. Ведь математика и физика являются абсолютно разными науками, не совместимыми ни в какой степени. Приставка «физико-математических наук» даёт возможность подменять законы природы в физике абстрактными математическими зависимостями. Чего делать нельзя, но мошенники делают это совершенно спокойно. Ведь подобные фальшивки не преследуются законом человеческого общества. Хотя мы в этом случае имеем дело с самыми настоящими мошенниками по сути своей одинаковыми с фальшивомонетчиками. Разница между ними заключается лишь в том, что для одних есть статьи в уголовном кодексе, для других их нет. Мошенники в науке наносят человеческому обществу во много раз больший ущерб, чем все фальшивомонетчики и прочие мошенники вместе взятые, если бы их даже не преследовали законом, а дали бы действовать свободно.*

*Зависимости теперешней элементарной математики являются исходными зависимостями для так называемой высшей математики, вернее, зависимости элементарной математики являются базовыми зависимостями для высшей математики. Элементарная математика может существовать без высшей математики, а высшая не может без элементарной. Выходит, что элементарная математика является высшей по отношению к так называемой высшей математике. Просто так называемая высшая математика образовалась на основе развития приближённых методов вычисления в элементарной математике – только и всего.*

*По этой причине и в силу вышеизложенных причин не должно быть деления математики на высшую и элементарную. Просто вся математика так и должна называться математикой, а её разделы, которые относятся к теперешней высшей математике, должны просто называться анализом в соответствии со своим прямым назначением. В этом случае от физика будут требовать знание и поиск законов природы, а от математика – знание и поиск правил математики и решения её зависимостей. Каждый человек в школе изучает и законы природы, и математику. По этой причине большинство людей будут свободно разбираться в научных работах любых специалистов и оценивать их достоинства. В современных условиях людей от науки отпугивает математическая неразбериха. Теперь же каждый должен знать, что если в любой научной работе нет объяснений, связанных с законами природы, и нет количественных зависимостей, относящихся к теперешней элементарной математике, то такая работа не является научной. В ней может быть написана только разнотипная чепуха, замаскированная математическими зависимостями, которые призваны сделать недоступными для понимания научного смысла или сущности подобной «научной» работы.*

*Вы скажите, что учёные не знали об этой неразберихе в математике и поэтому писали свои работы, как хотели. Это не так. Каждый учёный, если он является действительно учёным, прежде всего, знает назначение всего того, чем он пользуется в своих исследованиях, в том числе и математики. Ведь ему приходится следить за тем, что выходит в его исследованиях за рамки*

*ему неизвестного. По этой причине, хоть бессознательно, он всё равно относится к высшей математике как к анализу, а не как к высшей. Не подчёркивали они этих границ лишь потому, что они даже не подозревали, а большинство людей в современных условиях даже не догадывается, что в науке могут существовать мошенники, хотя в настоящее время наука полностью находится в руках мошенников. Ведь все считают, что мошенничать можно, например, в религии, астрологии, в различных гаданиях и знахарствах, в общем, там, где нет законов природы и математики, но только не в науке. Вот этой убеждённостью людей и пользуются мошенники в науке. Если вы знаете законы природы и теперешнюю элементарную математику, то вы всегда сможете установить, что за автор скрывается за той или иной научной работой – действительно учёный или мошенник, или просто заблудившийся.*

*Снова вы видите лирические отступления. Но в то же время от них невозможно отказаться, т.к. они тоже по своей сути являются необходимыми пояснениями и разъяснениями. Ведь всё равно на эти тему у вас будут возникать ваши «почему». Одно из первых ваших «почему» будет такого характера: почему механика Ньютона была принята без всякого сопротивления, с восторгом триста лет тому назад, когда фактически не было техники, и наука играла незначительную роль в жизни людей? В таких случаях говорят, мол, люди изменились. Ничего подобного. Просто это говорит о том, что люди чего-то недопонимают. Ведь в современных условиях наука и техника играют главную роль в жизни людей, как их основная энергетическая база, которая выполняет за них очень большую работу по обеспечению жизни людей. Теперь можно разобраться с тем, чего люди не понимают в науке.*

*Механика безынертной массы так же проста, как и механика Ньютона. Законы природы не выдумываются людьми, а находятся. Затем они проверяются экспериментом и принимаются в практическую деятельность людей. Экспериментальная проверка законов механики безынертной массы так же проста, как и механики Ньютона. Законы механики безынертной массы дают возможность увеличить вдвое коэффициент полезного действия жидкостных и газовых двигателей, винтов, лопастных насосов и т.д. Повышение вдвое коэффициента полезного действия энергетических установок, которыми пользуется современное общество людей, означает в буквальном смысле ликвидацию энергетического кризиса во всех странах и одновременно, как бы даёт возможность увеличить топливные ресурсы нашей планеты тоже вдвое. Ведь сейчас человечество обеспокоено тем, что природные запасы топлива невелики и что они за несколько десятков лет могут быть исчерпаны полностью. Что в конечном итоге означает гибель человеческой цивилизации.*

*Также вы видите, что без данной работы современные практические исследования планет с помощью ракет и автоматических станций могут носить лишь бессмысленный характер, т.е. подобные исследования можно рассматривать как очень дорогую забаву для людей – только и всего.*

*Законы механики безынертной массы найдены мною в 1969 году. Сейчас идёт 1977 год, а они продолжают оставаться не внедрёнными в практику людей. Отсюда уже получается однозначный ответ, что механику безынертной массы в практику людей могут не допускать только мошенники и что современная наука находится полностью в руках мошенников. По этой причине она не приносит людям пользы, а только вред. Если люди не могут разобраться с мошенничеством в науке, то они всё равно получают его результаты. Отсюда уже вытекает их безрадостное, порой даже враждебное отношение к науке. Так что не люди стали другими, а наука стала другой. Коль она находится в руках мошенников, то это означает, что науки нет совсем. От неё осталась лишь оболочка, изнутри заполненная паразитами.*

*Как видите, научная работа получается с «картинками».*

Снова вернёмся к рисунку 7 и полюбуемся на него. Мы видим, что наша исследуемая планета от воздействия силового поля другой планеты изменила свой шарообразный объём на иную форму подобную эллипсу. Выше мы установили, что шарообразная форма объёма планет определяет максимальную величину их энергии, а всякая иная изменённая форма объёмов планет свидетельствует об уменьшении их энергии. Мы установили изменённую форму объёма планеты при воздействии на неё гравитационным полем другой планеты и определили её соответствующими количественными, или математическими, зависимостями. Теперь нам остаётся определить количественными зависимостями величину, на которую уменьшается энергия в планете при воздействии гравитационного поля другой планеты.

Как вы понимаете, энергия равна произведения объёма на давление. Из-за того, что планета изменила шарообразную форму своего объёма на иную, величина объёма планеты от этого не изменилась. Просто мы получили объём типа эллипсоида равновеликий по объёму шаровому объёму планеты. Ведь мы имеем дело с несжимаемой жидкостью, которая составляет объём планеты.

Это положение даёт нам возможность обойтись без знания математических зависимостей новой объёмной фигуры при вычислении её объёма. Объём новой фигуры мы всегда сможем вычислить как шаровой объём, зависимости для которого нам известны. Следовательно, мы не будем испытывать никаких затруднений при вычислении объёмов планет с изменённой формой.

С объёмами мы разобрались. Переходим к давлениям. Чтобы разобраться с давлениями, обратимся к рисунку 6. Вспомним, что радиальный столб жидкости цилиндрической формы, расположенный в направлении плоскости ($S$), как плоскости, расположенной перпендикулярно направлению действия силового поля другой планеты, не испытывает силового воздействия другой планеты и ведёт себя точно так же, как если бы не было этого силового воздействия. В то же время при силовом воздействии другой планеты максимальная его высота, равная радиусу планеты ($R$), уменьшилась на высоту ($h_{пр}$). Коль мы не досчитались высоты ($h_{пр}$) в жидкостном столбе, то всё это означает, что все величины давлений этого жидкостного столба, начиная с глубины ($R$ - $h_{пр}$) и кончая центром планеты, уменьшились на одну и ту же величину давления, которая вычисляется зависимостью (22) как

$$P_{пр} = \rho h_{пр} w^2.$$

Хотя столб жидкости уменьшился по высоте, но закон распределения давления в нём не нарушился. Величина давления как была прямо пропорциональна высоте, так и осталась. Коль этот закон не нарушается, то остались поверхности равного давления. При шарообразной форме объёма планеты поверхности равного давления имели шарообразную форму. Когда объём нашей планеты стал иметь форму подобную эллипсоиду, то поверхности равных давлений тоже приняли подобную форму. Это положение говорит о том, что давление уменьшилось на равную величину ($P_{пр}$) не только в нашем столбе жидкости, но и в объёме всей планеты. Это значит, что при воздействии силового поля другой планеты на нашу исследуемую планету, уменьшается давление во всём объёме планеты на величину равную ($P_{пр}$). Давление ($P_{пр}$) пропорционально ($h_{пр}$). Выше мы установили, что высота ($h_{пр}$) вычисляется непосредственными замерами, а затем по уравнению (22) вычисляется давление ($P_{пр}$). Таким путём мы установили уменьшение величины давления в объёме планеты.

Мы нашли необходимые величины для определения величины, на которую уменьшается энергия планеты при воздействии на неё силового поля другой планеты. Объёмом является объём всей планеты ($V$), а давлением – противодействующее давление ($P_{пр}$), которое выражается зависимостью (22). Тогда величину уменьшения энергии ($Э_{ум}$) мы можем записать такой зависимостью:

$$Э_{ум} = VP_{пр}, \text{ или } Э_{пр} = V\rho h_{пр} w^2 \qquad (35).$$

Уравнение (35) определяет нам величину уменьшения энергии исследуемой планеты при воздействии на неё гравитационного поля другой планеты.

Мы получили все необходимые зависимости для силового воздействия одной планеты на другую.

***

Мы рассмотрели случай, когда на одну планету воздействует другая планета, но в солнечной системе существует много планет, которые имеют несколько спутников. Как быть в данном случае? Очень просто. В этом случае мы должны будем для каждого спутника в отдельности вычислить его силовое воздействие на планету так же, как мы это делали при вычислении силового взаимодействия двух планет. Затем результаты воздействия для всех спутников планеты мы должны будем не складывать, а накладывать друг на друга. В результате мы получим картину суммарного воздействия всех её спутников на планету.

Теперь посмотрим на реальные вещи.

На нашей планете Земля существуют приливы и отливы. Естественно, что люди попытались проникнуть в сущность этого явления. Но, как вы теперь знаете, без знания законов механики безынертной массы разобраться с подобным явлением невозможно. Единственно, что они правильно установили, так это то, что причиной приливов является спутник нашей планеты – Луна. Всё же остальное является выдумкой. Ибо они считают, что приливы вызываются приливным ускорением, которое образуется в результате воздействия силового поля Луны на нашу планету. Теперь вы знаете, что для жидкостей и газов не существует ускорений, а есть только постоянные скорости ($w$), которые соответствуют силовым воздействиям гравитационного поля планеты.

Так же мы можем прочитать, что приливное ускорение изменяется обратно пропорционально кубу расстояния между планетами. Такое можно прочитать, но саму теорию с выкладками подобного мне ни разу не удалось посмотреть.

По закону всемирного тяготения Ньютона постоянное ускорение ($g$) меняется обратно пропорционально квадрату расстояния между планетами. Этот закон позволяет с большой степенью точности определить силовое воздействие одной планеты на другую. Теперь он проверен практикой космических полётов. Сомневаться в нём не приходится. Закон всемирного тяготения Ньютона в одинаковой степени пригоден для определения постоянной скорости гравитационного поля планет, т.к. квадрат постоянной скорости равен по величине ускорению, т.е. $g = w^2$.

В общем, здесь мы просто убедились, что закон всемирного тяготения Ньютона можно было бы также получить, если исходить из законов механики безынертной массы.

Вы можете сказать, что мы здесь старались напрасно, т.к. закон всемирного тяготения давно получен Ньютоном. Поэтому вы скажите, что в исследовании планет можно обойтись без постоянных скоростей, а пользоваться постоянным ускорением. Если вы так подумаете, то получите неверное представление о явлениях, связанных с силовым взаимодействием планеты. Сейчас мы постараемся разубедить вас в этом.

Для начала мы вам приведём такой факт, что на Земле существуют приливные явления, а на Луне их нет. Но в то же время люди продолжают сомневаться в том, какая из себя Луна внутри, то ли её основная масса является твёрдым застывшим телом, то ли находится в жидком состоянии. Теперь они не будут сомневаться на этот счёт. Коль на Луне нет приливных явлений, то она представляет собой твёрдое остывшее тело.

Здесь вы можете возразить, что об этом можно догадаться, если иметь представление о твёрдых и жидких телах. Догадаться, конечно, можно, но быть уверенным – нельзя. Мы не будем спорить с вами по таким мелочам, а снова рассмотрим силовое воздействие Луны на нашу планету Земля.

Мы знаем, что основную массу нашей планеты составляет жидкость. Ибо на её массу воздействует большое количество тепла, которое поступает из планетной печки. То, что мы называем земной корой, или литосферой, представляет собой отвердевшую массу Земли. Как плавающий лёд не изменяет механического состояния воды в бассейне, так и земная кора не изменяет механических характеристик жидкой массы, т.е. её уровень остаётся неизменным. Ибо на земную кору действуют архимедовы выталкивающие силы. В свою очередь, на твёрдой земной коре в большом количестве располагается жидкость, которую мы называем водой, или мировым океаном. Из-за того, что твёрдая земная кора плавает на жидкой массе Земли, а мировой океан располагается на твёрдой коре, то можно считать его изолированным от жидкой массы земли в механическом отношении.

Мировой океан занимает порядка 75% площади земной коры, а то, что мы называем материками, представляет собой лишь острова в мировом океане. Неважно, что они являются большими островами. Ведь они всё равно не влияют на механическое состояние мирового океана. Главное здесь заключается в том, что материки не разделяют мировой океан на отдельные изолированные водные бассейны, т.е. он представляет собой единый общий бассейн. Мировой океан имеет определенный уровень своей поверхности. Средняя глубина – порядка четырёх километров.

В силу того, что мировой океан располагается на плавающей твёрдой поверхности, то мы должны будем представить нашу планету как бы состоящей из двух планет. Одну планету должна составлять расплавленная масса Земли, какой она является на самом деле. Другую, воображаемую, планету должна полностью составлять вода. Радиус шарового объёма планеты, состоящей из воды, должен определяться современным уровнем мирового океана.

Воображаемую планету, состоящую из воды, мы могли бы взять даже в том случае, если бы на нашей планете существовал только один Атлантический океан или его часть, протяжённостью от полюса до экватора. Главным здесь является то, чтобы высота водного столба была бы не меньше радиуса принятой нами водяной планеты. Вам хорошо известно, что для жидкости не имеет значения, проходит ли столб жидкости через центр планеты или жидкость располагается по периферии твёрдого тела соответствующим слоем. Просто толщина жидкости должна быть такой, чтобы она не затрудняла приток и отток жидкости в область увеличения и уменьшения объёма планеты. Вот эти особенности жидкости позволяют нам занять водой весь объём нашей планеты.

Радиус первой планеты, состоящей из расплавленной массы Земли, тоже должен соответствовать действительному радиусу нашей планеты. Действительный радиус нашей планеты, как уровень расплавленной массы Земли, пока ещё никто не догадался точно определить. Что является большим пробелом в современной науке о Земле.

В общем, разница между радиусом водяной планеты и планеты, состоящей из расплавленной массы Земли, будет порядка 5 километров, если учесть, что средняя глубина мирового океана – 4 километра, а толщина океанического дна составляет 5 - 10 километров и что часть этой толщины дна погружена в расплавленную массу Земли, как плавающее тело. Если мы возьмём подобную приблизительную разницу в радиусах за расчётную, то она повлияет незначительно на точность расчётов, т.к. радиус Земли составляет порядка 6380 км.

Теперь следует вопрос, зачем нам нужна вся эта возня с разделением планет на две планеты? Отвечаем, Луна воздействует своим гравитационным полем как на ту, так и на другую планету. Следовательно, Луна затрачивает одинаковые величины энергии как на ту, так и на другую планету. В общем, она уменьшает энергию той и другой планеты на одну и ту же определённую величину при условии равенства их объемов. Мы видим, что радиус водяной планеты на 5 км больше. Замерив высоту ($h_{пр}$) той и другой планеты, мы затем сможем вычислить по уравнению (35) величину уменьшившийся энергии в планетах. Дело здесь заключается в том, что для водяной планеты мы знаем точно величину плотности воды, а о плотности расплавленной массы Земли мы имеем лишь смутное представление. Мы знаем, что средняя плотность нашей планеты равна 5,515 г/см$^3$. На этом исчерпываются все наши познания о плотности Земли. Разделение планеты на две планеты мы сделали для того, чтобы определить плотность расплавленной массы Земли.

К большому сожалению, мне известны лишь приблизительные величины амплитуды колебаний приливов и отливов той и другой планеты. Для водяной планеты она равна порядка 10 метров, а для планеты, состоящей из расплавленной массы Земли, она равна порядка одного метра. Средняя величина между максимальной и минимальной высотой подъёма жидкости даёт нам искомую величину ($h_{пр}$). В общем, для этого мы должны разделить на два амплитуды колебаний жидкостных столбов обоих планет. Тогда для водяной планеты мы будем иметь высоту ($h_{пр}$) равную $h_{пр}$ = 5м, а для планеты, состоящей из расплавленной массы Земли, высота ($h_{пр}$) будет равна $h_{пр}$ = 0,5м.

Выше мы установили, что величины уменьшения энергии в планетах от воздействия гравитационного поля Луны будут одинаковыми лишь в том случае, когда их объёмы будут равны между собой. У нас же водная планета имеет радиус на 5 км больше. Чтобы сохранить равенство энергии, мы должны будем сделать объёмы планет одинаковыми, т.е. либо увеличить радиус одной планеты на 5 км, либо уменьшить радиус другой на 5 км. Соответственно изменившемуся объёму планеты мы должны будем скорректировать высоту ($h_{пр}$). Корректировку высоты ($h_{пр}$) можно проделать путём интерполяции относительно изменившегося объёма планеты. Для чего мы должны будем составить такую пропорцию:

$$h_{пр} = h_{пр}\frac{V_{кор}}{V_{пр}} \qquad (36).$$

Объёмы планет мы можем записать через их радиусы, тогда зависимость (36) примет вид:

$$h_{кор} = h_{пр}\frac{\frac{4}{3}\pi R_{кор}^3}{\frac{4}{3}\pi R_{пр}^3} = h_{пр}\frac{R_{кор}^3}{R_{пр}^3} \qquad (37).$$

Если мы увеличим радиусы планет на 5 км, то отношение кубов радиусов будет иметь такую величину: 1,0003. Практически такая корректировка не влияет на точность результатов.

Чтобы определить плотность нашей жидкой планеты мы должны будем воспользоваться одинаковостью величин уменьшения энергии в планетах, т.е.

$$\Delta \text{Э}_{\text{в.пл}} = \Delta \text{Э}_{\text{ж.пл}} \qquad (38),$$

где $\text{Э}_{\text{в. пл}}$ – энергия водной планеты, а $\text{Э}_{\text{ж.пл}}$ - энергия жидкой планеты

В равенстве (38) заменим энергии зависимостями по уравнению (35), тогда получим:

$$V_{\text{в.пл}}\rho_{\text{в.пл}}w^2 h_{\text{в.пл}} = V_{\text{ж.пл}}\rho_{\text{ж.пл}}w^2 h_{\text{ж.пл}} \qquad (39).$$

В равенстве (39) мы можем сократить объёмы планет ($V$) и постоянные скорости ($w^2$). Затем запишем равенство относительно искомой плотности, т.е.

$$\rho_{\text{ж.пл}} = \rho_{\text{в.пл}} \cdot \frac{h_{\text{пр в.пл}}}{h_{\text{пр ж.пл}}} \qquad (40).$$

Если мы в равенство (40) подставим замеренные величины высот ($h_{\text{пр}}$) той и другой планеты, то получим, что плотность жидкой массы Земли в десять раз больше плотности воды. Это значит, что плотность расплавленной массы Земли равна 10 г/см$^3$, т.е. почти вдвое больше средней плотности нашей планеты, которая равна 5,515 г/см$^3$.

Мы получили сравнительно приблизительную величину плотности расплавленной массы Земли, т.к. не имеем точных замеров высот ($h_{\text{пр}}$). Надо полагать, что точные замеры не внесут большой корректировки в эту величину, т.е., возможно, действительная величина плотности будет равна либо 8 г/см$^3$, либо 11 г/см$^3$. В общем, она будет отличаться на большую величину от средней плотности планеты.

Действительную плотность расплавленной массы Земли можно было бы также установить с помощью выталкивающей силы Архимеда. Для этого необходимо установить действительный уровень поверхности расплавленной массы планеты. Затем в районе дна мирового океана произвести определение плотности расплавленной массы Земли. Дно мирового океана имеет сравнительно небольшую толщину. Поэтому погрешность в определении плотности будет меньшей. Несмотря на то, что Луна нам помогла определить плотность расплавленной массы Земли, всё равно необходимо сделать определение величины плотности по выталкивающей силе.

Хотя в настоящее время ничего не знают о величине действительной плотности расплавленной массы Земли, но всё равно на этот счёт существует определённое мнение как общепринятое. Считается, что плотность с глубиной растёт, а малое ядро нашей планеты имеет плотность порядка 11 - 12 г/см$^3$, мы же в своих измерениях получили, что расплавленная масса нашей планеты имеет почти такую же плотность. Такая величина плотности прямо указывает на то, что малое ядро нашей планеты является планетной печкой, где масса полностью переходит в энергию. Энергия не обладает ни массой, ни плотностью. Это значит, что в объёме планетной печки мы имеем вакуум относительно массы. По геодезическим замерам планетная печь имеет радиус в 500 километров. Такой объём является солидным даже для нашей планеты, который существенно может изменить среднюю плотность нашей планеты.

Получив такую величину плотности расплавленной массы Земли, мы таким путём получили вторичное подтверждение о существовании в нашей планете планетной печки, где масса полностью переходит в энергию.

Конечно, в таких случаях бывает обидно, когда ты проделаешь такую работу, а точных результатов не можешь получить. Выходит, что работа остаётся незавершённой. <…>

Возвращаемся снова к своим планетам. С взаимодействием Земли и Луны мы ещё не закончили. Мы определили плотность расплавленной массы Земли.

Помимо мирового океана наша планета имеет ещё воздушный океан в виде атмосферы. Но в этом океане не заметны приливные явления, связанные с воздействием гравитационного поля Луны. Это говорит о том, что газообразная масса атмосферы остаётся нейтральной к воздействию гравитационного поля Луны. Почему она так ведёт себя? На этот вопрос механика безынертной массы не в состоянии ответить. Будем надеяться, что в каком-то будущем физики дадут нам ответ на этот вопрос.

Выше мы установили, что если хорошенько встряхнуть Луну так, чтобы её панцирь разрушился, и вся лунная масса получила бы возможность двигаться к центру Луны, то в этом случае она бы просто взорвалась и прекратила своё существование, превратившись в пояс астероидов. Это бы означало, что гравитационное поле Луны больше бы не воздействовало на нашу планету. По этой причине давление во всей массе Земли увеличилось бы на пол атмосферы. Естественно, что это мизерная величина по сравнению с общим давлением в Земле. Увеличение давления должно в обязательном порядке вызвать увеличение объёма планетной печки. Ибо величина критического давления имеет однозначно определённую величину. По этой причине границы печки удалятся от центра к периферии, пусть даже на какие-то доли миллиметра, но всё равно удалятся. В свою очередь планетная печка станет превращать в энергию большее количество земной массы. По этой причине должна возрасти температура расплавленной массы Земли, предположительно, на один-два градуса. Всех последствий такого изменения, естественно, невозможно предусмотреть. В общем, подобное увеличение не должно сказаться на материках, т.к. они имеют очень большую толщу. Дно мирового океана имеет сравнительно небольшую толщину. Это значит, что температура мирового океана должна возрасти. Нельзя сказать, что от такого потепления мирового океана растают льды Антарктиды, но надо полагать, что льды Ледовитого океана должны обязательно растаять. Хотя современная техника позволяет разрушить Луну, но, в общем-то, остаётся неизвестным главный вопрос, получат ли люди от этого больше пользы для себя или больше вреда? Пока на этот вопрос люди не получат точного ответа, не стоит разрушать Луну. Что выше мы уже отметили.

Мы рассмотрели все те взаимодействия, которые касаются нашей планеты и её спутника – Луны.

## В) СИЛОВОЕ ВОЗДЕЙСТВИЕ НА ПЛАНЕТЫ, КОТОРОЕ ЗАВИСИТ ОТ ИХ ВНУТРЕННЕГО СОСТОЯНИЯ

Если мы посмотрим на планеты солнечной системы, то увидим, что большая группа планет, в том числе и наша планета, имеет форму отличную от шаровой. Вот эти планеты имеют вращение вокруг своей оси, которое называется суточным вращением.

Подобное изменение объёмов планет связывают с их вращением, т.е. причиной в этом случае выставляют вращение. В действительности же само вращение таких планет является лишь следствием, и причина здесь совсем другая. Ибо масса этих планет находится в жидком и газообразном состоянии. По этой причине эти планеты не могут вращаться по инерции, точно так же, как не может вращаться по инерции сырое яйцо. Чтобы планеты вращались, к ним необходимо подводить энергию. Энергия же преобразуется во вращательное движение в двигателях. Это значит, что в каждой такой планете существует двигатель, который потребляет определённую часть планетной энергии и преобразуют её в силу, которая реализуется в виде механического вращательного движения. Следовательно, двигатель является причиной изменения формы объёмов планет, а не вращательное движение, которое является лишь следствием работы двигателя.

В своей работе «Строение Солнца и планет солнечной системы с точки зрения механики безынертной массы» [2] мы дали принципиальную схему этого двигателя и полный расчёт к нему. В общем, там даны все необходимые математические зависимости для расчёта этого двигателя. Но в целом всё там носит описательный характер, т.к. мы тогда не знали зависимость изменения формы планет от её внутренней энергии. Теперь мы выяснили этот вопрос. Внутренняя энергия планет определяется её радиальным гравитационным полем. Максимум внутренней энергии планета имеет, когда её объём имеет шарообразную форму. Всякое отклонение от этой формы объёма связано с уменьшением энергии в ней. Вот это положение даёт нам возможность определить убыль энергии в планете качественно и количественно. В этом вы уже убедились.

Внутренне силовое воздействие в планетах мы рассмотрим на конкретном примере нашей планеты Земля. На рисунке 8 показана изменённая форма планеты и циркуляция её массы.

Двигателем нашей планеты, как и других подобных планет, является плоский установившийся поток насосного типа. Наглядное представление о плоском установившемся потоке вы имеете – это смерчи, тайфуны, турбины, центробежные насосы. Плоский установившийся поток имеет цилиндрическую форму. Внутри этого цилиндра располагается ещё один цилиндр. Жидкость, находящаяся между двумя этими цилиндрами, составляет тело самого плоского установившегося потока. Когда жидкость в поток подводится через стенки малого внутреннего цилиндра, а выходит она через стенки большого наружного цилиндра, то в этом случае мы имеем дело с плоским установившимся потоком насосного типа.

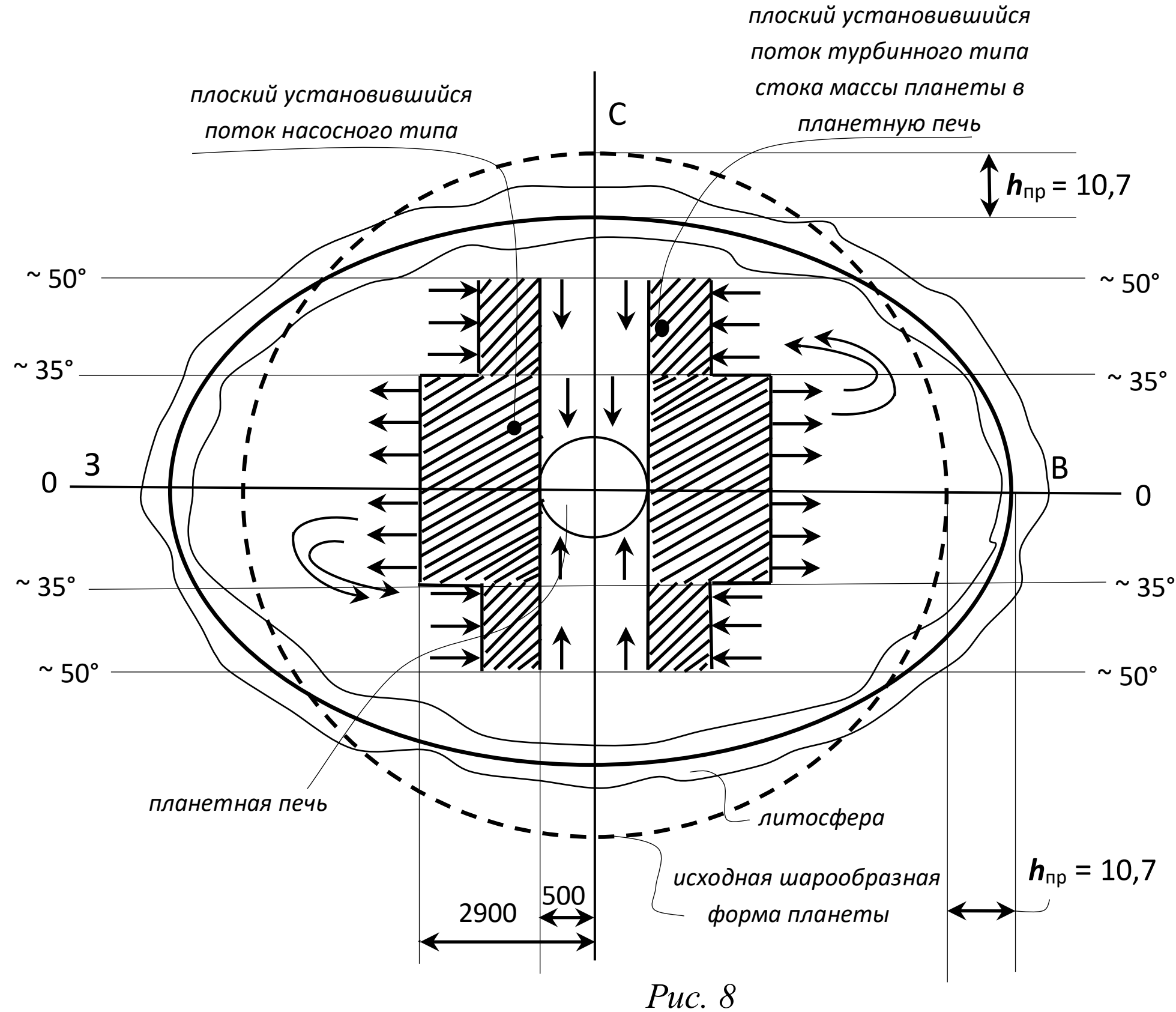

*Рис. 8*

*Схема циркуляции массы в планете Земля (размеры на рисунке даны в километрах)*

Если жидкость подводится в поток через стенки большого цилиндра и выходит из потока через стенки малого цилиндра, то в этом случае мы имеем дело с потоком турбинного типа. Как вы поняли, плоский установившийся поток своими цилиндрическими поверхностями разделяет области с различным энергетическим уровнем, или с различным давлением. Когда жидкость в потоке движется из низких давлений в область высоких давлений, то мы имеем дело с потоком насосного типа. Ибо в этом случае, чтобы жидкость могла двигаться из области низких давлений в область высоких давлений, к ней необходимо подводить энергию наподобие того, как это делается в центробежных насосах. В потоке турбинного типа, когда жидкость движется из области высоких давлений в область низкого давления, она наоборот, выделяет энергию, которую можно использовать по аналогии с тем, как это делается в турбинных агрегатах. Теорию и расчёт этих потоков смотри в работе [2].

Двигателем нашей планеты является поток насосного типа. В него подводится энергия из планетной печки. Эта энергия одновременно превращается в массу в плоском установившемся потоке. Затем уже масса из потока выходит через наружную большую цилиндрическую поверхность потока и устремляется в сторону полюсов нашей планеты. Затем жидкая масса планеты снова возвращается в планетную печку в малой цилиндрической поверхности потока через плоские установившиеся потоки турбинного типа, которые располагаются в районе полюсов нашей планеты и которые разделяют области высоких и низких давлений. То, что именно в таком направлении происходит циркуляция массы на нашей планете, свидетельствует движение материков к полюсам. Ибо материки могут дрейфовать к полюсам только под действием такой циркуляции. Здесь сомневаться не приходится. В свою очередь такая циркуляция массы планеты полностью подтверждает, что двигателем нашей планеты является плоский установившийся поток насосного типа, именно – насосного.

То, что турбина является двигателем, вас этим не удивишь, но то, что центробежный насос может быть двигателем, для вас является большой новостью. В то же время, что именно так обстоят дела на нашей планете и на других ей подобных, полностью подтверждает полностью её массы. Одновременно этот необычный факт подтверждает и то, что в планетной печке масса полностью переходит в энергию и не создаёт давления. Хотя мы привыкли к тому, что переход массы в энергию является взрывом, а взрыв - это, прежде всего, ударная волна и очень высокое давление. Таким путём мы снова сводим преобразование массы в энергию к массе. Ибо масса может создавать давление, а энергия – нет. Что полностью подтверждает то положение, что в планетной печке выделяется огромное количество энергии, и если бы она могла создавать давление в виде ударной волны, то от нашей планеты уже давно бы не осталось даже клочков. Вы имеете представление об атомной энергии, при которой происходит частичный переход массы в энергию, а в планетной печке – полностью.

В плоском установившемся потоке энергия снова превращается в массу и частично преобразуется в механическую энергию. Ведь в планетной печке давление стремится к нулю, а за пределами плоского установившегося пока оно достигает многих тысяч атмосфер. Вот таким путём поток насосного типа преобразуется в двигатель.

В малом цилиндрическом объёме этого потока располагается планетная печь как область низкого давления. По этой причине из области высокого давления через полюсные отверстия плоского установившегося потока турбинного потока в область низкого давления устремляется расплавленная масса планеты. Области низкого и высокого давления в жидкостях соединяются между собой через плоские установившиеся потоки турбинного типа. Они располагаются в полюсных областях нашей планеты, как показано на рис. 8.

В настоящее время принято считать, что наша планета имеет два ядра: большое и малое. Малое ядро имеет радиус в 500 километров, а большое – в 2900 километров. Эти размеры получены путём специальных измерений на натуре. Мы же с вами теперь должны понимать, что эти размеры относятся к двигателю нашей планеты, т.е. к плоскому установившемуся потоку насосного типа. Вот эти размеры проставлены на рис. 8 для этого потока. Они определяют нам размеры потока в экваториальной плоскости. Распределение этого потока в сторону полюсов нашей планеты нам неизвестно. Об этих размерах мы можем пока судить приблизительно. Сороковые широты на нашей планете называются ревущими. Это значит, что в области этих широт плоский установившийся поток насосного типа переходит в плоский установившийся поток турбинного типа. По этой причине мы можем сказать, что плоский установившийся поток насосного типа располагается до 35° – 40° широты. Что показано на рис. 8.

Наблюдениями отмечено, что в таких планетах, как Солнце, Юпитер, Сатурн, Уран и Нептун, экваториальная область вращается быстрее, чем их полюсные области. Скорость вращения планет вокруг своей оси определяется окружной, или тангенциальной, скоростью плоского установившегося потока. Каждая подобная планета имеет два типа таких потоков. Только по этой причине полярные и экваториальные области планет могут иметь разную скорость вращения. На нашей планете мы не видим подобного лишь потому, что она имеет земную кору, но всё равно этот раздел определяется сороковыми ревущими широтами. Отсюда мы также можем сделать вывод, что такие планеты как Юпитер, Сатурн, Уран и Нептун не имеют твёрдой планетной коры, а их масса находится в жидком и газообразном состоянии. По линиям раздела между областями планет с различной скоростью вращения мы можем определить высоту цилиндрической части плоского установившегося пока насосного типа этих планет или размеры их двигателей. Так же отсюда мы можем определить окружную скорость плоских установившихся потоков турбинного типа, расположенных в полюсных областях планет. Ясно здесь одно, что эти потоки вращаются в одну и ту же сторону, что и поток насосного типа, но имеют меньшую скорость вращения.

Теперь, когда мы разобрались с двигателем планет, мы можем рассмотреть изменённую форму объёма планеты. Изменённая форма объёма планеты определяется тем, что плоский установившийся поток насосного типа, или двигатель планеты <часть предложения неразборчива – ред.> образуется её радиальным гравитационным полем, или мы можем сказать, что силовое поле двигателя планеты имеет противоположное направление радиальному гравитационному полю планеты. По этой причине двигатель планеты уменьшает энергию планеты. Результатом уменьшения энергии планеты является её изменённая форма.

Сам принцип изменения формы планеты, или уменьшения её энергии, остаётся неизменным в данном случае. Это значит, что все принципы и выкладки для воздействия на планету гравитационного поля другой планеты, которые мы получили выше, остаются верными и для случая, когда силовое поле организуется внутри планеты. По этой причине главной фигурой изменившегося объёма планеты является фигура подобная эллипсу, но имеющая одинаковую величину ($h_{пр}$) в двух взаимоперпендикулярных направлениях. Эта фигура образует изменённый объём планеты.

При силовом воздействии одной планеты на другую эта фигура образовала объём планеты, имеющей плоскость симметрии. При внутреннем силовом воздействии образуется объём, имеющий и ось симметрии и плоскость симметрии. Ось симметрии совпадает с осью вращения планеты и служит началом отсчёта для силового воздействия планетного двигателя, а плоскость симметрии совпадает с экваториальной плоскостью планеты. Это значит, что используя уравнения (22) и (35) мы можем определить величину уменьшившегося давления и энергии в целом.

Экваториальный радиус Земли больше полярного на 21,4 км. По нашим расчётам эта разность должна быть равна двум высотам ($h_{пр}$). Отсюда мы можем вычислить высоту ($h_{пр}$). Она будет равна

$$h_{пр} = 21{,}4 : 2 = 10{,}7 \text{ (км)}$$

По высоте ($h_{пр}$) мы можем определить, что плотность расплавленной массы Земли в десять раз больше плотности воды. Тогда мы будем иметь давление равное 10700 кг/см$^2$. Если мы умножим величину этого давления на объём планеты, то получим величину уменьшившейся энергии планеты. Здесь имеется в виду объём расплавленной массы.

Энергия планеты уменьшается планетным двигателем. Из этого условия следует, что мы здесь должны будем воспользоваться законом сохранения энергии. Изменившаяся форма объёма планеты представляет собой застывшее движение с определённой величиной энергии, а планетный двигатель представляет собой реальное движение планетной массы. В то же время по закону сохранения энергии энергия застывшего движения должна быть равна энергии планетного двигателя. Как тут быть? Если здесь пользоваться голыми математическими канонами, то мы здесь не смогли бы ничего сделать, т.к. застывшее движение не определено во времени, а реальное движение определено по времени. По этой причине мы в данном случае должны будем отказаться от математических канонов, как мы делали это раньше, и руководствоваться законами природы. Чтобы совместить живую энергию двигателя с застывшей энергией планеты, мы должны будем поступить следующим образом: ввести в зависимости застывшего движения время. В данном конкретном случае мы должны будем величину уменьшившейся энергии планеты отнести к единице времени. Тогда мы получим застывшую мощность.

Выше мы определили, что в данной работе мы приняли техническую систему единиц (МКС). По этой причине мы должны будем взять за единицу времени одну секунду. Чтобы получить мощность застывшей энергии планеты, мы должны разделить уравнение (35) на одну секунду, получим:

$$N_{ум} = V\rho h_{пр} w^2 \text{ (1/сек)} \qquad (41).$$

Уравнением (41) мы записали мощность убывшей в планете энергии.

Мощность убывшей энергии планеты согласно закону сохранения энергии должна быть равна мощности двигателя планеты, или плоского установившегося потока насосного типа. Отсюда вы видите, что изменённая форма планеты позволяет определить величину мощности механической энергии планетного двигателя.

*Отметим сразу, чтобы не забыть такой момент. До настоящего времени люди ещё не могут определиться с тем, что собой представляет время. Например, ни у кого не вызывает никаких сомнений декартова система координат. Все хорошо знают, что она служит для общего обозначения пространства, которое помогает понять конкретность реально существующих объёмов. В общем, мы знаем, что любая система координат является общим, или отвлечённым, обозначением пространства, не имеющим ничего общего с реальным пространством. Всякая система координат является просто выдумкой людей – только и всего. Эта выдумка нам нужна лишь для того, чтобы познать реальное пространство, т.к. работа нашего мышления основана на относительном принципе сравнения. Коль мы с помощью времени совмещаем застывшее движение и реальное движение, то время, как и система координат, является выдумкой людей, которая предназначена для познания и изучения движения. Для природы не существует времени.*

*Для неё существует только движение как форма изменения материи. Материей мы обозначаем и массу, и энергию, и силовое поле. Следовательно, время является общим или отвлечённым обозначением движения. Время есть наша выдумка. По этой причине нельзя придавать какую-либо материальность, как системе координат, так и времени, т.к. они относятся к области абстракции, или нашей выдумки.*

*Мы, например, знаем, что в теории относительности Эйнштейна времени придаётся материальность. Это значит, что теория относительности является элементарной чепухой, которая не имеет ничего общего с наукой и со здравым человеческим смыслом.*

Продолжаем свою прерванную тему.

Мы выяснили, что изменённая форма планеты даёт нам возможность установить величину мощности планетного двигателя, естественно, относительно механической энергии. В связи с тем, что нам неизвестны точные величины геометрических изменений Земли, то мы не можем точно вычислить ни плотность расплавленной массы Земли, ни величины уменьшения давления и энергии в планете, ни мощность её двигателя. Без этих величин мы не можем установить точную геометрию плоских установившихся потоков планеты. Хотя мы имеем все необходимые зависимости для них в работе [2]. В нашем положении невозможно получить необходимые точные размеры геометрии Земли. Поэтому нам придётся отложить расчёт её внутренней геометрии до лучших времён. Как бы там ни было, мы всё равно получили большое количество ценных сведений.

Теперь уже не приходится сомневаться в том, что внутри планет существуют планетные печки, где происходит преобразовании массы в энергию. Объём этой печки зависит от величины критического давления.

Двигатель нашей планеты снижает давление внутри неё приблизительно на 10700 атмосфер. Это уже сравнительно солидная величина. Уменьшение давления приводит к уменьшению радиуса и объёма печи, что снижает её потребность в массе. Надо полагать, что если двигатель нашей планеты остановится по каким-либо причинам, то масса Земли разогреется настолько, что если даже материки и останутся после этого, они значительно уменьшатся и температура их поверхности станет 500° - 1000°К. Мировой океан испарится, и биологическая жизнь прекратит своё существование. Вот что значит для нас планетный двигатель.

Силовое поле планетного двигателя взаимодействует только с расплавленной массой Земли и не оказывает своего влияния ни на мировой океан, ни на атмосферу, т.к. и то и другое располагается ровным слоем над земной корой.

Расплавленная масса планеты имеет одинаковый средний состав по физическим и химическим характеристикам, т.к. она непрерывно циркулирует во всём своём объёме и переходит из массы в энергию, а затем наоборот. Поэтому плотность расплавленной массы мы можем считать постоянной величиной во всём её объёме.

При своём движении расплавленная масса Земли образует различные местные, или локальные, силовые поля, которые взаимодействуют и с мировым океаном, и с атмосферой нашей планеты, создавая в них различные движения в виде различных океанических течений. Притом эти локальные силовые поля являются качественно разными, коль они взаимодействуют и с мировым океаном, и с атмосферой. Ибо мы выше выяснили, что вода и воздух взаимодействуют с различными силовыми полями. Отсюда следует, что по движению вод мирового океана и по движению воздушных масс в атмосфере мы можем следить за циркуляцией расплавленной массы Земли.

Вот так выглядят все те моменты, которые непосредственно связаны с силовым взаимодействием внутри планеты.

## г) ОБОБЩЁННЫЙ РАЗБОР ПЛАНЕТ СОЛНЕЧНОЙ СИСТЕМЫ

По нашему плану следующим вопросом должен быть вопрос об увеличении энергии в планетах. Выше мы установили, что планеты шаровой формы имеют максимальную величину энергии. Увеличить их энергию можно лишь путём увеличения объёма при неизменности шарообразной формы. Естественно, что увеличение объёма можно производить путём ввода в него дополнительной массы. Здесь могут быть два способа добавления массы к планете: при первом способе масса прибавляется к планете сравнительно медленно и постепенно. Реальным примером такого способа является выпадение на поверхность нашей планеты метеорной пыли и метеоритов. Когда читаешь литературу по этому вопросу, то там приводятся цифры такого порядка, что на поверхность Земли в течение года выпадает несколько миллиардов тонн космической пыли и метеоритов. Если учесть, что Земля существует несколько миллиардов лет, то за это время она должна была бы догнать по своей массе, пожалуй, даже само Солнце. Но этого не происходит. Видите, с какими данными здесь приходится сталкиваться. Хотя это очень важный вопрос, который бы помог выявить, какое количество массы выгорает в планетной печке.

Вторым способом прибавки массы к планете является способ, когда происходит соударение массы планеты с космическим телом сравнительно большой массы. Этот вопрос пока для нас является очень сложным вопросом, т.к. здесь мы не имеем никаких данных и получить их пока невозможно по техническим причинам, независящих от людей. Поэтому нам приходится опускать данный вопрос и переходить к следующему параграфу. <…>

<…> Самой интересной с научной и практической точки зрения является планета Венера. С одной стороны, она представляет собой идеальную планету, а с другой стороны, она близка по массе к нашей планете Земля. К настоящему времени условия на её поверхности нам известны. Температура там достигает 500°С, давление имеет величину порядка 100 атм. Но самым интересными и нужными её характеристиками люди ещё не удосужились поинтересоваться. Планета Венера представляет собой идеальную планету в том плане, что она имеет и планетную печь и идеально шарообразный объём. Это значит, что она имеет максимум энергии, который не уменьшается ни силовым воздействием другой планеты, ни внутренним силовым воздействием, т.к. она не имеет планетного двигателя. Об этом свидетельствует форма её объёма и то, что она почти не вращается. В силу этих причин Венера должна иметь предельно большой диаметр своей печки. Естественно, что объём её печки надо рассматривать относительно её массы. Предельно большой диаметр планетной печки должен быть потому, что критическое давление, при котором масса полностью переходит в энергию, имеет вполне определённую величину. Коль масса Венеры не разгружается никаким силовым воздействием, то критическое давление будет образовываться на меньшей глубине, чем у тех планет, массы которых разгружаются и внешним и внутренним силовым воздействием.

Современные технические средства позволяют провести замеры диаметра планетной печки Венеры и определить уровень жидкой массы, а затем измерить её действительный диаметр. Ибо действительным диаметром планет является диаметр их жидкой расплавленной массы.

Зная действительный диаметр Венеры, как диаметр объёма расплавленной массы и диаметр планетной печки мы сможем определить действительную плотность расплавленной массы. Полная величина массы планеты Венера нам известна, т.к. её определили с помощью законов Ньютона. Вам также хорошо известно, что твёрдая плавающая планетная кора, или литосфера, как и плавающий лёд, не нарушают механических характеристик жидкости. Это так же означает, что литосфера не нарушает величины плотности расплавленной массы планеты. Тогда мы можем поступить следующим образом: определим величину действительного объёма планеты по диаметру объёма расплавленной массы планеты, затем из этого действительного объёма планеты вычтем объём планетной печки и тогда мы получим чистый объём, в котором размещается вся масса планеты. Общая величина массы планеты нам известна. Разделив общую массу планеты на её чистый объём, мы получим действительную величину плотности жидкой массы планеты. Хотя эта величина будет действительной величиной её плотности, но всё же она относится к разряду средних величин, т.к. определена без учёта температурного распределения внутри планеты. Ведь температуры к её центру растут, а к поверхности – убывают. Если бы ещё нам удалось определить величину плотности расплавленной массы Венеры по выталкивающей силе Архимеда, тогда бы мы могли в какой-то степени связать плотность с температурой.

Вы можете спросить, зачем нам цепляться за плотность? Видите ли, плотность является одной из немногих характеристик, с помощью которой мы можем в какой-то степени познать и изучить химическую и физическую структуру расплавленной массы планет. Пока общие признаки указывают на то, что расплавленная масса планет имеет химическое и физическое состояние, которое отличается от привычного для нас состояния жидкой и твёрдой массы. На эти особые признаки расплавленной массы указывают следующие обстоятельства.

Во-первых, окружающая нас твёрдая и жидкая масса при всём своём разнообразии имеет вполне определённый химический состав, а для расплавленной массы планеты подобное не допускается. Она должна иметь какой-то общий химический состав. Об общем химическом составе массы мы пока ничего не знаем и не имеем ни малейшего представления, т.е. продолжаем лезть с химическим делением повсюду.

Во-вторых, расплавленная масса планет непрерывно циркулирует, переходя полностью в энергию и снова превращается в массу. Хотя Венера не имеет планетного двигателя в виде плоского установившегося потока насосного типа, но она всё равно должна иметь два плоских установившихся потока турбинного типа, которые организуют приток массы к планетной печке. И она имеет обычный установившийся поток, который организует отток массы от планетной печки. Под обычным установившимся потоком мы имеем в виду такой поток, который вы можете наблюдать при движении жидкостей и газов в трубах, при течении реки, течения мирового океана и т.д. Потоки турбинного типа организуют медленное вращение Венеры вокруг своей оси. Хотя по нашим представлениям именно плоские установившиеся потоки турбинного типа являются двигателями, но в особых условиях энергетического состояния планет они не могут выполнять функцию двигателя, а только насоса, перекачивающего жидкость из области высокого давления в область низкого давления, т.е. вакуума.

В-третьих, извергнутая вулканами нашей планеты масса делится на магму и пепел. По нашим обычным представлениям всё, что находится под воздействием высоких температур, должно иметь физические особенности расплава, а вулканический пепел их не имеет. Фактически же получается, что он не выделяет тепло, а наоборот, поглощает его. Если бы это было не так, то горячее дыхание вулканов распространялось бы на многие километры и сжигало бы всё способное гореть. Этим занимается только расплавленная лава. Из труб электростанций, работающих на каменном угле, вылетают частички несгоревшего угля в виде пепла, а из ванн расплавленного металла и шлака вылетают только оплавленные частички, маленькие и большие, но всё равно несущие большое количество тепла. Несгоревшие частички угля вылетают в виде пепла в первую очередь потому, что они не успевают разогреться до температуры горения, а в расплавленной массе металла, так же как в жерле вулкана, вся масса имеет одинаковую температуру и находится в расплавленном состоянии.

Мы не можем пока изучить непосредственно расплавленную массу планет относительно их химического и физического состава. Плотность же позволяет нам косвенным путём составить определённое представление о физическом и химическом состоянии расплавленной массы планет хотя бы солнечной системы. Мы можем жить, как и все живые организмы, только на поверхности планет, и то, в том случае, когда физические и химические условия на поверхности планет соответствуют требованиям живых организмов. В свою очередь поверхностные условия планет являются лишь следствием физического состояния их расплавленной массы. Следовательно, воздействием на расплавленную массу планет можно изменять их поверхностные условия, а не наоборот. Но прежде чем заняться подобными преобразованиями планет, мы должны убедиться в том, что все планеты солнечной системы имеют одинаковый физический и химический состав расплавленной массы. На худой конец, мы должны установить, что такая одинаковость присуща, например, планетам земной группы, а для других планет она может быть другой. Если мы этого не сделаем, то труды наши по преобразованию планет могут быть напрасными. Естественно, что об одинаковости химического и физического состава расплавленной массы планет можно судить по химическому и физическому составу их атмосфер и твёрдой планетной коры и ещё по целому ряду признаков. Дело в том, что планеты солнечной системы расположены от нас на разном расстоянии и имеют, прежде всего, различное физическое состояние. По этим причинам мы можем изучать лишь различия на этих планетах, хотя они являются общими для них. В этом плане для нас большой интерес представляют все планеты, в том числе и само Солнце. Ведь оно тоже является планетой, но только имеет очень высокую температуру своей массы. Опять же, вы теперь должны твёрдо себе уяснить, что изучение планет ведётся не ради самого изучения, а ради практических интересов самого человечества. Просто мы одни научные исследования можем сразу использовать в своих интересах, а другие придётся оставлять другим поколениям.

Результаты исследований автоматическими станциями условий на поверхности Венеры и Марса подтверждают в определённой степени одинаковость химического и физического состава расплавленной массы этих планет, но если мы обратимся к их средней плотности, то увидим существенные различия, например, Земля имеет следующую плотность 5,5 г/см$^3$, Венера – 5,2 г/см$^3$, а Марс всего 4 г/см$^3$. Чего не должно быть. Такое различие по средней плотности должно быть результатом неправильной методики определения средней плотности. Вернее, результаты такой средней плотности планет никому не нужны, т.к. они не дают количественной оценки действительной плотности этих планет. Выше мы установили, каким путём определяется действительная плотность расплавленной массы планет. Вот этой методикой необходимо пользоваться. Когда мы воспользуемся нашей методикой для определения плотности планет, то может оказаться, что столь существенной разницы между плотностями Земли и Венеры не существует или она есть, но сравнительно небольшая. Такая поправка может быть внесена по той причине, что Венера должна иметь больший объём своей печки, чем Земля. Если после такой поправки плотность Венеры всё равно будет меньше плотности Земли, но сравнительно на небольшую величину, то эту разницу мы можем списать на разницу температур масс Земли и Венеры. Тогда мы получим одновременно некоторое представление зависимости плотности от температуры. Но всё же этих подтверждениё ещё недостаточно. Ведь Марс имеет ещё большеe различие по плотности. Это значит, что на Марсе с помощью автоматических станций необходимо произвести замеры  диаметров его планетной печки и объёма его расплавленной массы. Здесь поправка в марсианской плотности должна выразится в том, что твёрдая кора, или литосфера, Марса имеет намного большую плотность, чем плотность земной коры. Косвенным подтверждением этому служат довольно низкие температуры на поверхности Марса – минус 84°С.

Вы сразу спросите, почему – косвенные? Отвечу вам кратко. Материки нашей планеты имеют достаточно большую толщину, чтобы свести тепловую радиацию расплавленной массы до минимума. Об этом свидетельствуют льды Антарктиды и большие участки вечной мерзлоты. Причиной такого положения является то обстоятельство, что остальные площади материков имеют так называемое водяное и паровое отопление. Воды поднимают тепло с глубин и обогревают поверхности материков. Это значит, что материки похожи на пень, изъеденный древоточцами. По этим ходам циркулирует вода, подогреваясь и охлаждаясь.

Дно мирового океана составляют плотные массы, которые не допускают циркуляцию воды. По этой причине воды мирового океана имеют температуру равную плюс 5°С. Хотя толщина дна мирового океана сравнительно незначительная – всего 5-10 км. Если бы материки не имели водяного отопления, то они были бы сплошь покрыты вечными льдами и вечной мерзлотой.

По этой причине мы отнесли температуры в разряд косвенных характеристик. Если после уточнения плотности массы Марса окажется, что она равна плотности массы Земли, а здесь должно быть равенство, т.к. Марс имеет планетный двигатель, то в этом случае мы с полной уверенностью можем сделать вывод, что плотности расплавленной массы планет, хотя бы для их земной группы, одинаковы.

Если мы получим такие характеристики по плотности для планет, то мы можем заняться их преобразованием. Коль Венера и Земля имеют незначительную разницу по массе, то это будет означать, что Венера может иметь почти одинаковые с земными условия на своей поверхности.

Преобразование планеты Венера будет сводиться к снижению температуры её расплавленной массы, вернее, к уменьшению теплового потока её планетной печки. Снижение температуры мы можем произвести только путём образования планетного двигателя, или плоского установившегося потока насосного типа. Сразу начнём искать ответ на этот вопрос: сможем ли мы это сделать или нет? Для этого нам снова придётся посмотреть на все планеты солнечной системы. Мы увидим, что все те планеты, которые имеют планетный двигатель и вращаются вокруг своей оси, в обязательном порядке имеют спутники, один или несколько, но всё равно имеют. Это положение говорит о том, что на определённом этапе существования планет существовало несоответствие между энергией планетной печки и потенциальной энергией массы планеты, связанной с её радиальным гравитационным полем. Стабилизация, или уравновешивание, энергий в каждой такой планете производилась путём отстрела излишков массы и образования планетного двигателя. Венера же не имеет ни спутников, ни своего планетного двигателя. Как видите, она оказалась гордой и непорочной женщиной. Для нас же всё это означает, что в силу каких-то причин на Венере не образовалось несоответствия между энергией её планетной печки и энергией массы. Следовательно, чтобы организовать в ней планетный двигатель, мы должны искусственным путём создать в ней такое несоответствие. Пока нам известен только один путь создания такого несоответствия – это столкновение с Венерой довольно крупного космического тела. Таким космическим телом для Венеры может быть расположенная рядом с ней планета Меркурий. Меркурий – небольшая по массе планета, находится вблизи Солнца и поэтому она не представляет большого интереса для человечества.

В то же время мы можем быть только уверенными в успехе подобного мероприятия, но 100% гарантии дать не можем. Мы бы имели 100% гарантию лишь в том случае, если бы имели множество тел с различной величиной массы и могли бы обстреливать Венеру до интересующего нас результата. Мы же имеем в своём распоряжении только одну планету Меркурий с определённой величиной массы. Ещё мы знаем, что в таком виде она представляет собой бомбу, которая может взорваться при ударе. Имея такие условия мы дальше уже должны будем опираться на свои силёнки. Чтобы столкнуть Меркурий с Венерой потребуется максимум усилий, а если расколоть его на части, то потребуется меньше усилий, но и вероятность успеха тоже может быть меньшей. Хотя более мелким кускам планеты можно придать более высокую скорость. Пока мы можем варьировать только так, исходя из своих технических возможностей.

Как бы там ни было, но имея такие данные на руках, уже сейчас необходимо приступить к созданию проекта по преобразованию планеты Венера, хотя современные технические возможности не позволяют его реализовать и многих данных ещё не хватает для его полного уточнения. Но всё равно такой проект необходим для человечества как стимул, как цель, которая способна организовать стремления людей и систематизировать их усилия. С одной стороны подобный проект определил бы конкретность для технического совершенствования, а с другой стороны, он бы придал конкретность наблюдениям и исследованиям планет. Даже в таком виде конкретность всегда необходима людям. Ибо она указывает на нужность их деятельности и тем самым придаёт им жизненную уверенность, как реальная и достижимая для них цель, которая требует от них больших усилий, т.е. самых больших усилий. Ведь люди, в общем, всегда стремятся к самому большому.

В современных конкретных условиях данный проект будет отвергнут в любой стране и любыми учёными. В этом не приходится сомневаться. Ибо человечество потеряло голову и всякий контроль за своей деятельностью. Теперь все надеются на хитрость и силу, забыв о своём человеческом природном предназначении. По этой причине всё это стадо устремилось к своей погибели.

<…>

Чем интересен для нас Марс, мы рассмотрели совместно с планетой Венера выше. Главный, конечно, интерес к нему проявляется в том, что он имеет остывшую планетную кору и является нашим соседом. Когда-нибудь люди пожелают разместить на нём свои службы, которые необходимы будут для обеспечения космических полётов и для слежения за внутреннем состоянием Марса. Условия на поверхности Марса совершенно не приспособлены для жизни людей – низкое давлении, углекислотная атмосфера и низкие температуры. Поэтому самая малая оплошность будет стоить жизни, находящимся на Марсе людям. По теперешним нашим понятиям причиной таких условий является очень малая масса Марса, она составляет лишь 0,11 часть от массы Земли. Улучшить условия на его поверхности можно увеличением массы. Искусственным путём можно увеличить массу Марса примерно вдвое. Подобное увеличение массы можно сделать за счёт следующих ресурсов: два его спутника добавить к его массе и воспользоваться материалом пояса астероидов, т.к. орбита пояса астероидов располагается по соседству с орбитой Марса.

Подобное мероприятие может привести к тому, что увеличится атмосферное давление на поверхности Марса, увеличится количество воды, что может привести к замене углекислотной атмосферы на азотнокислородную. Хотя условия на поверхности Марса так и останутся непригодными для жизни людей, но они для них будут не столь гибельными.

Планеты-гиганты интересны тем, что их масса находится в жидком и газообразном состоянии. Они имеют мощные планетные двигатели и фактически представляют собой тёмные солнца. Прежде всего, они позволяют следить за движением масс газов в их атмосферах, Если вы сравните, например, карту течений в мировом океане нашей планеты с движениями или течениями газообразных масс этих планет, то вы найдёте большую схожесть. Эта схожесть не случайна, т.к. океанические течения отражают в определённой степени движение, или циркуляцию, расплавленной массы планеты Земля. И движения атмосферы планет-гигантов тоже во многой степени зависит от циркуляции жидкой массы этих планет. Движения в атмосфере нашей планеты Земля тоже в большой мере зависят от циркуляции жидкой массы планеты. Мы рассмотрели ярко выраженные особенности планет-гигантов. Но выше мы определили новые характеристики для планет. К ним относятся: критическое давление, при котором масса полностью переходит в энергию. Используя геометрию планет и их плотность в конечном итоге можно найти размеры планетных печек этих планет.

Также мы установили действительные размеры планет, которые определяются диаметром объёма расплавленной массы планет. Что, в конечном итоге, даст возможность установить действительную плотность этих планет.

Нам известны расчёты плоского установившегося потока, которые позволят связать геометрию потока с его расходом и мощностью.

Геометрия планет-гигантов очень изменена, т.е. имеет большие отклонения от шарообразной формы. Эти изменения в геометрии дадут нам возможность в конечном итоге определить мощность плоского установившегося потока, или планетного двигателя. Различие в скорости вращения полярных и экваториальных областей планет поможет нам уточнить геометрию плоского установившегося потока.

В целом же мы здесь должны стремиться к тому, чтобы установить и полную общность и различие, т.е. по всем планетным пунктам, между планетами солнечной системы.

Как мы выше установили, Солнце тоже является планетой. Единственное его отличие от планет мы пока видим в его высокой температуре. Также Солнце имеет шарообразную форму своего объёма при наличии планетного двигателя, который должен был бы вызвать изменение формы объёма. Естественно, что такие изменения существуют, но мы пока их не видим потому, что не можем увидеть действительного, или реального, диаметра Солнца, который должен определятся его расплавленной массой, если так можно выразится, вернее, его массы, которая обладает свойством жидкости. Ведь выше над этой массой идёт уже атмосфера, масса которой имеет свойства газов.

Мы знаем, что жидкости образуют пары, которые создают над их поверхностью парциальное давление. Так вот атмосферы планет являются парами жидкой массы планет, которые создают на поверхности планет парциальное давление, которое мы назвали атмосферным давлением, и которые состоят из газов. В этом выражается особенность атмосфер. Луна не имеет атмосферы потому, что её масса находится в твёрдом состоянии. Атмосфера является одним из главных показателей, которые определяют, имеет ли та или иная планета свою массу в расплавленном состоянии.

Если на планетах-гигантах их атмосферы копируют изменённую форму объёмов планет, то солнечная атмосфера сохраняет шарообразную форму, т.е. форму его гравитационного радиального поля. В тоже время расплавленная масса Солнца имеет изменённую форму. В этом не следует сомневаться, коль оно имеет свой двигатель, который вращает его массу вокруг оси.

Главная особенность Солнца выражается в том, что циркуляция её массы сопровождается образованием очень мощных локальных и разнообразных полей, наглядное действие которых мы можем наблюдать. Ведь подобные локальные силовые поля образуются на любой планете солнечной системы, где существует циркуляция расплавленной массы. Следовательно, Солнце в первую очередь для нас интересно тем, что мы в наглядной форме можем наблюдать действие силовых полей.

Солнце интересно ещё тем, что на его поверхности возникают протуберанцы. Протуберанцы в свою очередь интересны тем, что они должны определяться выходами энергии планетной печки Солнца на солнечную поверхность. В свою очередь эта энергия частично выносит массу Солнца, которую мы определяем как его расплавленную массу. В общем, образование протуберанцев на Солнце мы можем отнести к разряду его вулканической деятельности. Следовательно, по ним мы можем установить в какой-то степени состав самого Солнца. Ибо сейчас бытует мнение, что солнечную массу больше чем на 90% составляет водород. На самом же деле всё обстоит иначе. Масса Солнца тоже состоит из какого-то общехимического вещества и подвергается циркуляции, которая существует на основе перехода массы в энергию и наоборот, а его атмосферу, действительно, составляет в основном водород.

Вулканическую деятельность мы пока можем наблюдать только на своей планете Земля и на Солнце.

Как видите, само Солнце и планеты солнечной системы представляют большой интерес для наблюдений. Хотя в настоящее время, как бы по негласному мнению, считается, что всё это изучено визуально и что теперь осталось всё это изучить с помощью автоматических станций и ракет. Хотя, как вы теперь сами убедились, неизученной остаётся даже сама наша планета Земля. Коль люди не знают, что они должны изучить и исследовать, то все современные исследования планет с помощью автоматических станций и ракет носят бессмысленный, а потому никому не нужный характер, дорогой забавы. В то же время изучение планет стало насущной задачей, коль появилась космическая техника. На основе этого требования времени мы просто обязаны продолжать научное изучение планет солнечной системы. При этом изучении основным методом изучения планет и Солнца остаётся визуальный метод с помощью телескопов и приданной им другой подобной техники, а автоматические станции и ракеты должны служить необходимым придатком ко всей технике. Ибо космические автоматические станции и ракеты являются ещё и очень дорогим удовольствием. Но главное здесь заключается в том, чтобы вся эта техника, само руководство ею, планирование всех космических исследований проводились не какими-то дельцами и разными искателями приключений, которые представляют все подобные исследования в качестве циркового аттракциона, призванного позабавить публику, а непосредственно специалистами–учёными. В современных условиях повсеместно специалисты-учёные играют второстепенную роль. В их обязанности входит лишь подготовка космических аттракционов и обеспечение их безопасности, а всё остальное проделывают шуты гороховые, которые всё же боятся только одного, как бы им не ушибиться больно.

Теперь мы можем подвести итоги. Мы начали данную работу с расчётов, а закончили рассуждениями. По той простой причине, что столкнулись с большим недостатком данных непосредственно о самих планетах, многие из которых считаются вполне изученными. Отсюда мы делаем заключение, что изучение планет необходимо продолжить в таком плане: считать нашу планету Земля как бы экспериментальной установкой, на которой мы обязаны проверять все наблюдения за другими планетами, их спутниками и за самим Солнцем. Только такая система позволит детально изучить внутреннее строение планет, а также их физику и химию. Особо надо подчеркнуть, что мы почти ничего не знаем о силовых полях. Хотя в современных условиях существует ядерная физика, но о переходе массы в энергию, как о переходе массы в качественно иное состояние, мы ничего не знаем.

Теперь не вызывает сомнения, что все планеты солнечной системы были отстреляны в своё время самим Солнцем и потому являются его частью. Что спутники планет являются частями планет, т. к. они были отстреляны планетами. Это подтверждается тем, что с одной стороны, пылевая конденсация планет не могла бы происходить без взрыва, который бы вернул вещества в их исходное состояние, а с другой стороны, не всякая масса является носителем радиального гравитационного поля. Об этом свидетельствуют хотя бы астероиды. Притом везде мы наблюдаем соприкосновение космических тел, как малых, так и больших, как довольно сильные соударения. В силу этих причин о пылевой конденсации как об образовании планет не может быть и речи.

Коль Солнце отстрелило все планеты, то это обстоятельство указывает на то, что само Солнце тоже образовалось в определённый момент времени. Надо полагать, что оно отстрелило свои спутники на самом начальном этапе своего существования. Образование самого Солнца пока остаётся загадкой, а происхождение планет мы можем определить. Для этого необходимо лишь установить механизм отстрела спутников. В этом нам может помочь механика безынертной массы и дальнейшее изучение физической сущности планет.

Можно считать, что данная работа является необходимым этапом в подготовке изучения происхождения планет. Мы не будем возражать, если найдутся желающие и опередят нас в этом плане. Мы подобное можем только приветствовать. <…>

<…> Приведу вам такой пример. Все привыкли считать, что газы сжимаемы, и затем на этой основе теоретики строят все свои математические выкладки, не считаясь с законами природы, которые присущи газам. На этой основе все считают, что акустические волны в воздухе образуются подобно тому, как волны сжатия и разрежения образуются во всякой пружине при воздействии на неё импульсной силой. В общем, для воздуха считают, что акустическая волна состоит из двух участков: участка сжатия и участка расширения. Когда же мы нашли законы природы механики безынертной массы как механики жидкости и газа, то акустическая волна предстала перед нами в своём ином виде. Она не имеет участков сжатия и разрежения, просто жидкости и газы движутся в этой волне без всякого сжатия и разрежения сразу в двух взаимоперпендикулярных направлениях. На одном участке волны они движутся в одном направлении, а на другом её участке меняют направление своего движения на противоположное – только и всего.

Далее на основании сжимаемости воздуха считают, что летательные аппараты, летящие со сверхзвуковой скоростью, образуют, так называемый, скачок уплотнения, т.е. считают, что в лобовом потоке воздух сжимается и создаёт повышенное давление сопротивления полёту. В действительности всё происходит иначе. Действительную картину нам рисуют всё те же законы природы механики безынертной массы. В лобовом потоке при полёте летательного аппарата со сверхзвуковой скоростью не происходит уплотнения воздуха. Плотность его остаётся неизменной и равной плотности воздуха невозмущённого потока. Просто механическая энергия движения летательного аппарата полностью переходит в тепловую энергию. В результате разогрева воздуха, без изменения его плотности, происходит увеличение давления, которое создаёт дополнительное сопротивление полёту.

Когда мы знаем действительную картину интересующего нас явления, которую мы получили на основании законов природы, а не на основании досужих вымыслов, то мы можем обойтись без всяких теоретиков и сами посчитать все количественные характеристики этого явления с помощью соответствующих математических зависимостей, которые мы получили на основании законов природы. Ибо законы природы определяют нам в общем виде количественную зависимость в явлениях, а не просто так кем-то выдуманные математические выкладки. В этом случае мы в любом конкретном моменте получаем действительные количественные величины, без всяких искажений с желаемой для нас точностью, а в том теоретическом виде с искажённой картиной явления и выдуманными математическими зависимостями при всей их кажущейся реальности мы не можем получить действительных количественных величин. Поэтому получается так, что подобные теории существуют сами по себе, а практика – сама по себе, т.к. она вынуждена добиваться искомых результатов на основе экспериментов, а не на основе расчётов.

Естественно, что астрономов не устраивает пассивное наблюдение за структурами вселенной. Они стремятся объяснить их назначение. В то же время всё это не означает, что они в этом случае должны что-то выдумывать. Следовательно, в своих пояснениях и разъяснениях они должны исходить из уже известных людям законов природы, которые были найдены другими науками, и, исследуя вселенную с помощью этих законов, они могут видеть, что укладывается в эти законы, а что – нет. Выявленные астрономами явления, которые не укладываются в уже известные законы природы, представляют собой большой интерес, т.к. таким путём они указывают, что людям известно и что – неизвестно.

Притом при подобных исследованиях нельзя путать или смешивать известные людям законы природы с используемыми людьми явлениями природы. Примером таких явлений могут служить электрические явления. Хотя электричество и электромагнитные волны широко используются в практике людей, но законы природы для электрических явлений нам неизвестны. По этой причине на электрические явления нельзя опираться в объяснениях явлений вселенной. В настоящее время астрономы имеют на своём вооружении мощные радиотелескопы. С их помощью они сделали много интересных открытий во вселенной. Естественно, что они пытаются дать объяснения своим открытиям. Коль они не знают законов природы для электромагнитных колебаний, то они могут делать свои пояснения и разъяснения лишь с точки зрения устройства самого радиотелескопа. Вы сами хорошо понимаете, что в этом случае всякие разъяснения и пояснения могут оказаться лишь сплошной выдумкой, которая не имеет общего с действительными явлениями природы. Но в то же время сами наблюдения, сделанные с помощью радиотелескопов, представляют собой большой интерес.

Мы проводим свои исследования с помощью законов механики безынертной массы. Теперь мы посмотрим на вселенную и выясним, что укладывается в эти законы, а что не укладывается. Проведём анализ общего характера, не вдаваясь в детали.

Выше мы установили, что при очень больших давлениях масса переходит в энергию, т.е. в энергию светового, теплового и всякого другого излучения, электрическую энергию, энергию силовых полей. В общем, при больших давлениях масса полностью переходит в качественно иное состояние. Это качественно иное состояние массы характеризуется тем, что она в таком виде не может существовать в состоянии покоя, а совершает непрерывное движение в своей новой материальной форме. В связи с тем, что масса тоже обладает кинетической и потенциальной энергией, то определение перехода массы в энергию звучит не совсем точно. Это значит, что энергию мы должны определять относительно соответствующей материальности. Всякая материальность для нас существует в виде реального объёма. Следовательно, масса, как материальная объёмность, переходит в иную материальную объёмность, качественно новую. Для этой качественно новой материальности мы тоже должны дать своё определение, которым мы могли бы пользоваться. Воспользуемся для этой цели уже известным определением, которое уже существует как «эфир». Это значит, что при больших давлениях масса полностью переходит в эфир, т.е. в качественно новую для себя материальность, которую мы назвали эфиром. Объём, заполненный эфиром, по отношению к массе мы называем вакуумом, или просто пустотой. В то же время материальность определяется объёмом. Масса сама по себе не включает в своё определение объёмную конкретность. Объёмную конкретность массе придаёт иное понятие, которое мы называем плотностью. Общность для массы и плотности определяется механической силой. Для эфира этой общности, как механической силы, не существует. Хотя Лебедев установил экспериментально, что световое излучение создаёт на поверхности из массы давление, но в объёме свет не создаёт давления, а масса его создаёт. В общем, мы точно знаем, что энергия эфира преобразуется в механическую энергию только через массу.

Хотя механическая сила является качественным различием между материальностью массы и материальностью эфира, но всё равно, коль эфир представляет собой материальность, то он имеет объём. Само определение эфира даёт нам неопределённость относительно его объёма. Вот эта неопределённость пока ставит нас в тупик.

В то же время мы видим, что эфир представлен как бы элементарными составляющими массы, т.е. представляет массу, разложенную на свои простейшие составляющие. Также мы видим, что простейшие составляющие массы не могут находиться в состоянии покоя, а сразу начинают своё движение в пространстве. В этом случае мы можем представить себе массу по отношению к эфиру, как состояние покоя эфира, т.е. упакованный в объёме эфир представляет собой массу, а не упакованная в объёме масса представляет собой эфир.

Материальность окружающего нас мира предстаёт перед нами в объёмах массы и в объёмах эфира. В то же время мы не можем определить для себя чёткость или определённость границ между этими двумя материальными объёмами. Ибо объёмы массы не существует для нас в чистом виде, т.е. без присутствия в них тепловой, электрической энергии, энергии силовых полей. Но всё равно мы для себя должны навести такие границы, чтобы разобраться в материальности окружающего нас мира. Для этого нам придётся поделить энергию между двумя этими материальностями. Общим для энергии массы является сила. Сила проявляется при соприкосновении масс, и она же создаёт давление в объёме, занятым массой. Силу, распределённую по площади, мы называем давлением, а силу взаимодействия мы называем просто силой, или сосредоточенной силой, как силой, приложенной в точке соприкосновения масс.

Если для энергии массы мы получили чёткую границу в виде механической силы, то для энергии эфира мы пока не можем получить такой чёткой границы в силу многих причин. Например, из ядерной физики мы видим, что масса имеет очень сложную упаковку по отношению к эфиру. Мы научились преобразовывать тепловую энергию в механическую, электрическую – тоже в механическую, и энергию некоторых силовых полей – тоже в механическую. Теперь мы чётко должны знать, что под механической энергией мы имеем в виду, однозначно, энергию массы. Но далеко не всякую лучистую энергию мы можем преобразовать в механическую энергию. Возможно, мы не знаем ещё о многих других энергиях. В силу этих причин мы не можем найти чёткую и определённую границу для энергии эфира. По этой же причине мы можем допустить существование других материальностей помимо эфира и массы.

Исходя из наших знаний и незнаний на данном этапе нашего существования, мы должны будем остановиться на двух материальностях: на массе и эфире. Механическую энергию, граница которой определяется силой, мы будем относить к массе, а все остальные виды известной и неизвестной нам энергии мы будем относить к эфиру, т.к. мы пока останавливаемся на двух материальностях.

Конкретность объёма массы мы получаем через плотность, а конкретность объёма эфира мы пока не можем получить в силу вышеизложенных причин. В общем, про эфир мы очень ещё мало знаем. Относительно этих двух материальностей мы рассмотрим вселенную в общем плане.

Энергия объёма массы определяется произведением давления на объём. Для центрального объёма планет мы получили энергию в виде неопределённости, т.к. в этом случае давление стремится к бесконечности, а объём стремится к нулю. Вот эта энергетическая неопределённость объёма массы указала нам на существование эфирного объёма в центре планет. Конкретную величину объёма эфира в планетах мы установили непосредственно замерами величины этого объёма, которые позволили нам установить величину критического давления для массы как границу между массой и эфиром.

С меркой энергетического объёма массы мы можем подойти к объёму всей вселенной. Тогда мы тоже получим неопределённость относительно энергии объёма массы только обратного характера, т.е. здесь мы будем иметь дело с тем, что объём вселенной стремится к бесконечности, а давления – к нулю, т.е.

$$Э_{вс} = V_{вс} \cdot P_{вс} = \infty \cdot 0,$$

при $V_{вс} \to \infty$

$P_{вс} \to 0.$

Отсюда мы можем сделать лишь один вывод, что объём вселенной определяется материальностью эфира. Это значит, что когда мы установим объёмную конкретность для материальности эфира, тогда мы сможем установить конкретную величину объёма вселенной. Даже такая неопределённость относительно объёмной энергии массы может определять нам конкретные задачи для поисков размеров вселенной. Отсюда мы видим, что мы должны искать конкретно, чтобы определить размеры вселенной. Вот так законы природы определяют нам конкретность задач для дальнейших исследований и поисков.

Задачи же эти очень большие, т.к. нам придётся начинать с того, что мы должны будем перетрясти многое науки, которые касаются или связаны с энергией эфира. Хотя многие из этих наук считаются давно законченными и что там делать больше нечего. Например, к разряду таких наук относится термодинамика. Эту науку придётся строить заново, т.к. всё её здание висит в воздухе. Ибо основанием, или фундаментом, всякой науки являются законы природы. В то же время основная масса наук до настоящего времени продолжает оставаться без этого фундамента и при этом они считаются законченными.

В наше время только теоретиков не смущает все эти обстоятельства. Им достаточно того, что их есть кончик пера, смоченный чернилами, а на всё остальное им наплевать. Они сочинили, например, такое учение, что вселенная расширяется. Естественно, что подобная выдумка имеет много изъянов. Тогда они назвали свою сказку гипотезой и таким путём покрывают все изъяны. Ибо они твёрдо знают, что их сказки нельзя проверить опытным путём. Сейчас подобное нельзя проверить лишь потому, что люди не понимают действительного назначения законов природы как истин, которые находятся по свойствам окружающего нас мира. Если бы все даже в современных условиях опирались на уже известные законы природы, то этого было бы вполне достаточно, чтобы выявить и пресечь все нелепости, которыми загадили современную науку и превратили её таким способом в сточную канаву для помоев, от которой люди уже давно отвернулись.

Настоящие учёные занимаются поиском законов природы и часто остаются с пустыми руками, т.к. задача эта довольно трудная. Но в то же время каждый из них вносит определённый камень в здание построения науки. Так называемые теоретики поступают проще: они сочиняют сказку. Затем накручивают к ней умопомрачительные выкладки, в которых очень трудно разобраться, и выдают всё это за действительность. Под эти вымыслы они получают целые институты с многочисленным штатом, и живут себе припеваючи без всяких мучительных дум о науке. Ибо они занимаются в последующем лишь тем, что подавляют и уничтожают всех тех, кто не согласен с их вымыслом. Если ты учёный и тебе действительно повезло, и ты нашёл ранее неизвестные людям законы природы, то на тебя сразу же набрасываются так называемые теоретики, которые представляют собой откровенных мошенников, и стараются уничтожить и тебя и твоё открытие. В этом плане они не хуже любых, даже самых отпетых уголовников. Это значит, чтобы воскресить снова науку, надо избавиться в первую очередь от теоретиков как фальсификаторов науки.

Вот так выглядит вселенная вместе с так называемыми разумными существами. Так она представлена относительно эфирной материальности.

В объёме вселенной находится множество тел, которые по своей материальности относятся к массе. Эти тела для нас существуют в конкретных объёмах. Основными телами вселенной являются звёзды. Наблюдениями астрономов установлено, что все звёзды делятся на такие основные группы: красные гиганты, которые имеют очень большие размеры и малую плотность своей массы. Жёлтые звёзды типа нашего Солнца. На красные и белые карлики, которые имеют сравнительно небольшие объёмы при очень большой плотности. Это значит, что материальность всех звёзд представлена массой. Они все имеют шарообразную форму объёма. Внутри каждой звезды имеется планетная печка, или эфирный объём, где масса превращается в эфир с предельно большим выделением эфирной энергии. Мы также знаем, что при такой циркуляции масс звёзд их масса может быть представлена в виде определённого общехимического вещества. В тоже время мы видим, что основанием для деления звёзд на вышеуказанные группы является, в основном, величина их планетной массы. Отсюда мы можем сделать вывод, что при общехимическом составе масс звёзд их массы имеют различия в своей физической структуре, т.е. различие масс звёзд определяется различием физического построения структуры самих масс. Говоря другими словами, мы должны искать различия для масс звёзд не в их химическом составе, который мы делим на химические соединения и на отдельные элементы, составляющие их, а непосредственно в самой физической структуре массы, которая является в обычных условиях основой для образования всех химических элементов и их соединений.

В спектре звёзд астрономы не нашли каких-либо химических элементов, которые не были бы известны нам. Это говорит о том, что несмотря на физические различия масс звёзд, которые выражаются в различии их плотности, всё равно из этих масс образуются те химические элементы, которые определены таблицей Менделеева. Говоря другими словами, расплавленная масса планет и звёзд, которую мы определили как общехимическое вещество, несмотря на свои физические различия, преобразуется в одни и те же химические элементы и соответствующие им соединения. Вот эта особенность массы представляет для нас определённый интерес и представляет интерес заполучить её.

Звёзды белые карлики имеют предельно большую величину плотности своей расплавленной массы, а все остальные планеты и звёзды соответственно меньшую величину своей плотности. Физические условия для расплавленной массы планет мы пока определяем температурой и давлением. Расплавленная масса, циркулируя в объёме планет и звёзд, попадает в области высоких давлений и температур в центре планет и в области сравнительно низких давлений и температур, которые существуют на поверхности расплавленной массы планет и звёзд. Поверхности расплавленной массы определяют нам действительный диаметр и объём планет и звёзд. Поверхностные условия планет и звёзд определяют нам минимальные значения температур и давлений, при которых расплавленная масса определяется давлением паров, или давлением атмосферы. Для расплавленной массы Земли давление на поверхность её расплавленной массы фактически определяется толщиной мирового океана. Мировой океан образует над поверхностью расплавленной массы водяной столб высотой примерно в 10 километров. Такой столб воды создаёт давление примерно в 1000 атмосфер. Температура расплавленной массы, извергаемой вулканами, равна 2000 - 3000°К. Сейчас пытаются сделать скважину и добыть расплавленную массу Земли в чистом виде. Чтобы эта расплавленная масса могла сохранить неизменными свои физические и химические свойства, её необходимо хранить при давлении не ниже 100 атмосфер и температуре не менее 2000°К. Если мы пожелаем получить расплавленную массу Земли искусственным путём, то нам придётся начинать с этих минимальных для неё физических условий.

Белые карлики при очень большой плотности своей массы фактически не имеют атмосферы. Это значит, что расплавленная масса этих звёзд имеет на своей поверхности незначительное давление и очень высокие температуры порядка сотен тысяч градусов. Это значит, что при искусственном получении расплавленной массы планет и звёзд мы можем компенсировать давление температурой.

Вот такого рода практические руководства мы можем получать при обращении с расплавленными массами планет и звёзд. Звёзды от нас далеко, но они нам определяют конкретность перспектив наших земных дел. Когда человечество получит в своё распоряжение планетную и звёздную массу, то оно сможет получать любые химические элементы в нужном для себя количестве. Печи, созданные из расплавленных масс планет и звёзд, смогут поглощать все отходы промышленного производства и общественного существования людей, превращая их в энергию и в нужные нам химические элементы.

В настоящее время подобными проблемами занимается ядерная физика. Физики проводят свои работы в условиях вакуума и сравнительно небольших температур. Они добывают разнотипные частицы и ищут сверхтяжёлые элементы. Исходя из условий проведения подобных работ, мы теперь можем сказать, что они бесперспективны и желаемы результатов от них не получишь, т.к. для этого необходимы бОльшие давления и бОльшие температуры.

Все свои работы ядерщики проводят на разнотипных ускорителях, которые имеют длину от нескольких сотен метров до нескольких километров. В то же время ускорители представляют собой самое нелепейшее, самое бессмысленное сооружение, когда-либо созданное человечеством. Поясним это предложение.

Все ускорители создавались на основе законов механики Ньютона. Ибо физики ещё по настоящее время полагают, что их частицы и атомы являются в полном смысле частицами и подчиняются в своём движении законам Ньютона. По этой причине длина всех ускорителей является очень большой величиной - от нескольких сотен метров до нескольких километров. Согласно второго закона Ньютона, сила равна произведению массы умноженной на ускорение. Отсюда исходит название ускорителей как ускорители частиц и ядер.

Жидкости и газы, когда они образуют поток, уже не подчиняются второму закону Ньютона, а подчиняются второму закону механики безынертной массы, который гласит, что сила равна произведению расхода массы в единицу времени умноженной на скорость.

Отметим, что зависимость для второго закона безынертной массы, как механики жидкости и газа, была известна людям давно. Её получил ещё Ньютон. В то же время никто не понимал, что эта зависимость между силой и массой является такой же единственной зависимостью для массы жидкостей и газов, какой является зависимость между силой и массой для твёрдых тел. Эта зависимость применяется в практике людей для расчёта реактивной тяги, и известна людям как реактивная сила. Моя заслуга здесь заключается лишь в том, что я в своих поисках выяснил, что она является законом для жидкостей и газов. Своим успехом я обязан Ньютону как своему учителю. Ибо его работы помогли мне завершить свои работы. Я не стыжусь этого и выказываю ему свою признательность и благодарность.

Так вот, жидкости и газы не имеют ускорения, а имеют постоянную скорость, величина которой определяется силовым полем. Это значит, что может меняется расход массы, а скорость остаётся неизменной. Это также означает, что жидкости и газы не могут запасать кинетическую энергию, а служат лишь переносчиками энергии поля как промежуточное рабочее тело. По этой причине механика жидкости и газа получила своё название механики безынертной массы. Эта механика не противоречит механике Ньютона, а является самостоятельным дополнением, основанном на свойстве текучести массы, которое не учитываются законами Ньютона. Потоки атомов, ядер, различных прочих частиц, которые ускоряются в ускорителях, как и всякие потоки жидкости и газа, обладают свойством текучести. По этой причине они в обязательном порядке должны подчиняться законам механики безынертной массы. Коль это так, то потоки частиц в ускорителях не разгоняются, а движутся с постоянной скоростью, которая соответствует величине магнитного поля ускорителя. По этой причине не следовало делать ускорители большой длины. Достаточно было бы оставить одну или две секции от всего ускорителя, длинной в метр или два, чтобы создать устойчивый поток, а все эти сотни и даже километры можно просто выбросить на свалку. Ибо величина магнитного поля не зависит от длины ускорителя, а зависит от характеристик секций составляющих его. Вот до чего можно докатиться, если руководствоваться домыслами теоретиков, а не законами природы.

***

Когда вы ознакомитесь с данной работой и даже с другими работами в области механики безынертной массы, то вы в её теории, в смысле математических выкладок, не найдёте что-либо нового для себя. Даже новые для вас зависимости покажутся хорошо вам известными, как само собой разумеющееся. Вот это положение используют в первую очередь все те, кто стремится уничтожить мои работы. Они не допускают их к печати, а просто рассказывают всем и распускают клеветнические слухи, как будто я всё списал и передрал у кого-то и теперь выдаю за своё открытие. Когда вы перейдёте непосредственно к практическим ощущениям окружающего вас мира и к практическим делам, только тогда каждый из вас сможет оценить большую качественную разницу, которая выявляет механику безынертной массы как новое для вас, как ранее неизвестное вам. Главное же заключается для людей именно в этом, а не в самих математических выкладках.

Например, современные турбины и поршневые двигатели имеют коэффициент полезного действия не более 40%. Законы механики безынертной массы позволяют его увеличить вдвое. Следовательно, все выпускаемые такого рода двигатели во всём мире являются устаревшими и хороши они лишь на свалке.

Законы механики безынертной массы найдены были мною в 1969 году. Почти десять лет они не допускаются в практику людей и неизвестно, когда они будут допущены. Ибо противники этой механики не собираются их допускать в практику людей. Таким путём они фактически заставляют всё человечество работать на свалку, т.е. делать ненужные для себя двигатели как очень устаревшие. Подобной деятельностью по их милости человечество занимается почти десять лет и неизвестно, сколько будет заниматься ещё. Следовательно, враги механики безынертной массы являются прямыми врагами всего человечества.

Данное открытие загрузило бы полезной для человечества работой тяжёлую промышленность ведущих стран мира лет на пять и более. Это условие позволило бы ликвидировать на это время экономический кризис в этих странах. Это значит, что война, как бойня для людей мирового масштаба, была бы отодвинута лет на пять и более, а затем, может быть, уже навсегда.

Противники механики безынертной массы хорошо это знают в отличие от всех. Значит, они умышленно толкают человечество в эту бойню и готовят её для них. Так что мы здесь имеем дело не просто с противниками механики, а с озверелыми до самой последней степени зверьми, которые извлекают для себя выгоды из всемирного побоища людей.[2]

**Снова обращаемся к вселенной.**

Некоторые звёзды во вселенной образуют звёздные скопления, которые называются галактиками. В основном встречаются три типа галактик – это неправильные галактики, которые имеют форму бесформенных облаков, эллиптические галактики, которые имеют объём в виде эллипсоида, и спиральные галактики. Наша солнечная система находится в одной из спиральных галактик,

Прежде всего, наше внимание привлекают спиральные галактики. Достаточно взглянуть на них, чтобы убедиться, что составляющие такую галактику спирали представляют собой форму логарифмической спирали. Из теории механики безынертной массы нам известно, что логарифмическая спираль представляет собой линию тока плоского установившегося потока, который бывает насосного и турбинного типов. Такое положение явно указывает на то, что спирали галактики являются траекториями движения масс галактики к её центру. Хотя астрономы считают, что звёзды в таких галактиках совершают движение по окружностям относительно её центра. Траектория логарифмической спирали показывает, что звезда движется одновременно в двух направлениях: по окружности и в радиальном направлении к центру галактики. Если спиральные галактики представляют собой плоский установившийся поток, то скорость движения звёзд по окружностям должна оставаться неизменной независимо от того, находятся ли они на периферии галактики или у её центра. Радиальная же скорость звёзд должна возрастать по мере их приближения к центру галактики, т.е. на периферии галактик радиальная скорость для звёзд будет минимальной, а в непосредственной близости от её центра она будет максимальной. Тогда спиральные галактики можно будет представить в виде единого газового потока с вкраплениями твёрдых тел в виде звёзд.

Как нам кажется, подобное можно проверить, если фотографировать галактики одним и тем же телескопом не через десятки и даже через сотни лет, а через, может быть, тысячи лет. Сравнением этих фотографий можно выявить характер действительного движения звёзд в спиральных галактиках. Что из этого получилось бы, трудно сказать, но такие данные всё равно нужны.

Не плохо было бы, если бы человечество построило хотя бы одну наглядную модель вселенной на основании уже известных звёзд и галактик. Такая модель могла бы очень помочь в осмысливании построения вселенной.

Сейчас же мы пока можем сделать один вывод насчёт галактик, что они образованы на основе каких-то галактических силовых полей, которые нам неизвестны. Коль формы галактик различны, то галактические силовые поля тоже имеют свои определённые различия. В общем, всё это относится к разряду предположений, основанных на законах природы, вернее, на её свойствах. Нужны или не нужны такие предположения людям, мы предоставляем им возможность судить об этом самим. Например, выше мы установили бесперспективность ядерной физики, но это вовсе не означает, что на ней надо сразу поставить крест. Просто надо найти для неё соответствующие рамки и поставить её на своё место. Для практики ядерная физика дала атомный реактор и атомную бомбу, для науки она дала совсем немного, т.к. не найдены ещё законы природы для неё.

---

[2] Автор неправ только в одном, что эти субъекты были не только противниками механики, но и много чего ещё, от чего к настоящему времени камня на камне не осталось. Он просто выяснил, что пробуксовка механики, даже с точки зрения её предметного обсуждения, слово ему так и не дали, объясняется не состоянием науки и какой-то конкуренцией внутри этой сферы, а пагубной политикой, которая велась и в СССР и за рубежом. Теперь уже наглядно видны её результаты. Хотя – это только цветочки, то ли ещё будет, когда созреют ягодки. Поскольку автор стал кое о чём догадываться и не молчал, это усугубило его трудности. Удивительная прозорливость, хотя нам всем лучше было бы, если бы он ошибся. То, что он затронул политику, окончательно поставило крест на возможности продвинуть механику безынертной массы, несмотря на то, что проверка её положений является очень быстрым и недорогим делом и даст однозначный ответ. Если допустить, что положения механики оказались бы соответствующими действительности, то читатель хотя бы в первом приближении может представить, какие пагубные последствия повлекло то, что теория не была допущена в научный оборот. Они – катастрофические. Определённая область науки зашла в тупик. Если механика безынертной массы не является выходом из него, то, значит, выхода нет. Теперь автор не может помешать политикам в их трудах и заботах, поэтому редактор надеется, что шанс у работ есть и после проверки практикой механика войдёт в научный оборот. Ведь без неё и ветряк невозможно довести до ума. (2022)

В этом плане представляют большой интерес элементарные частички, для которых необходимо определить материальность между материальностью массы и материальностью эфира. Подобные рамки нужны для ядерной физики. В таком плане надо понимать призыв к самостоятельному мышлению. Иначе можно наломать много дров.

Второй, или даже третьей, массовой материальностью вселенной является космическая пыль и газы. Газы вселенной в основном представлены водородом. Химический состав космической пыли нам неизвестен. Мы знаем, что она представляет собой мельчайшие частички твёрдого тела. По этой причине она полностью подчиняется законам Ньютона, а газы вселенной подчиняются законам механики безынертной массы. Если космическая пыль движется так же, как все другие космические тела больших размеров, то газы представляют собой промежуточное, или рабочее, тело для различных силовых полей. Это значит, что по характеру распределения газов мы можем выяснить характер силовых полей вселенной на доступных для нашего наблюдения расстояниях. Силовые поля космического пространства представляют собой большой интерес.

С житейской точки зрения космическая пыль и газ представляет собой космический мусор. Всякий мусор является спутником определённой деятельности. Следовательно, пыль и газы вселенной тоже являются спутниками определённой деятельности во вселенной. Мы тоже видим, что наибольшие плотности космической пыли и газов присущи галактикам. Это значит, что галактики являются наиболее активными местами деятельности во вселенной. Коль космическим газом в основном является водород, а водород составляет атмосферы звёзд, то это значит, что приоритет в космической деятельности принадлежит звёздам. По каким-то причинам они время от времени взрываются и заполняют космическое пространство пылью и газом, но сама масса звёзд при этом не рассеивается в космическом пространстве, а снова после взрыва объединяется в массу звёзд и в таком виде продолжает своё существование. С этой точки зрения химический состав космической пыли представляет собой большой интерес, т.к. он должен быть в какой-то степени родственным вулканическому пеплу нашей планеты и протуберанцам Солнца. Космическую пыль мы пока не можем получить в её естественном виде, т.к. она сгорает в атмосфере нашей планеты. В этом случае неплохо было иметь хотя бы спектральный анализ космической пыли. В этой связи большой интерес представляет собой вулканический пепел нашей планеты, который необходимо изучить, как и по месту расположения вулканов, так и в историческом плане, т.е. его состав в прошлых геологических эпохах. Космическая пыль и газы могут рассказать нам об очень многом, когда ты знаешь их назначение. Само же назначение вещей в природе не выдумывается, а находится. В этом принципе выражается главная сущность очеловечивания людей. Поиск же назначения вещей требует от каждого очень больших усилий как умственных, так и физических. Только в этом выражаются трудности человеческого существования, от которых люди стремятся уклониться и попадают в трудности животного существования, т.е. превращаются в зверьё. Зверь есть зверь, и им лучше жить в джунглях, а не в городах со всеми удобствами.

В целом же вселенная представляет собой объект, который требует очень длительного изучения. Человечество сравнительно недавно получило на своё вооружение телескопы. Для людей это время кажется большим, а для процессов вселенной оно является лишь кратковременным мигом. Поэтому можно считать, что изучение вселенной только началось. Поэтому человечеству придётся набраться терпения и продолжить свои наблюдения в веках и даже тысячелетиях. Изучение космического пространства нам необходимо в первую очередь для того, чтобы шире смотреть на мир.

***

последнее редактирование: август-сентябрь 2022 г.

## Литература

1. Шкурченко И. З. «Механика жидкости и газа, или механика безынертной массы» (рукопись, написана в декабре 1971 года)

2. Шкурченко И. З. «Строение Солнца и планет солнечной системы с точки зрения механики безынертной массы» (рукопись, написана в марте 1974)

## ПРИЛОЖЕНИЕ

### *От редактора*

Редактор посчитал нужным дать описание Марса из текста «Строение Солнца и планет солнечной системы» в качестве Приложения к данной работе «Энергетическая структура Солнца и планет солнечной системы».

Вот изъятая часть текста из работы «Строение Солнца и планет солнечной системы с точки зрения механики безынертной массы»:

Следующей планетой на нашем пути будет планета Марс, которая тоже относится к разряду планет земной группы. Для этой планеты мы лишь отметим её особенности с нашей точки зрения. Всвязи с тем, что Марс имеет продолжительность своих суток почти одинаковую с земными, то мы отметим, что в гидросфере планеты есть плоский установившейся поток насосного типа. В центре планеты имеется источник тепла определённой мощности. Марс имеет большее сжатие у полюсов, что свидетельствует о повышенном тепловом энергосодержании объёмной единицы марсианского источника тепла по сравнению с земным источником.

Согласно учебника астрономии, преобладающим газом в атмосфере Марса является углекислый газ. Это значит, что строение атмосферы Марса схоже со строением атмосферы Венеры. Коль атмосфера Марса состоит в основном из углекислого газа, то для поддержания её теплового баланса солнечной тепловой энергии будет просто не хватать. Ведь Земля получает в два с лишним раза больше тепловой энергии Солнца, чем Марс. Земля почти не имеет в своей атмосфере углекислого газа, а Венера, поверхность которой находится при температуре в несколько сот градусов, имеет атмосферу, содержание которой почти полностью составляет углекислый газ. По этой причине мы можем судить, что поверхность Марса имеет повышенные плюсовые температуры, которые компенсируют утечки тепловой энергии из атмосферы и тем самым поддерживают её тепловой энергитический баланс. Повышенные положительные температуры на поверхности Марса образуются за счёт того, что марсианская кора на много тоньше земной коры. По этой причине на поверхностях Марса, типа наших материковых, температура достигает нескольких десятков градусов, а в пониженных местах, типа кратеров, расщелин и им подобным углублениям температура достигает нескольких сот градусов. Только за счёт подобного температурного режима Марс может содержать свою углекислотную атмосферу. Как пишут, что давление атмосферы на марсе составляет лишь 0,006 от давления земной атмосферы. Но если, действительно, атмосферу Марса составляет углекислота, то давление её должно превышать давление земной атмосферы в 5 - 10 раз.

Вот так должна выглядеть планета Марс с точки зрения механики безинертной массы в соответствии с теми данными, которые нам известны о ней.[3]

Действительно, в 1974 году о Марсе было известно только то, что было написано в учебнике астрономии.

Автор сделал описание Марса в «Строении Солнца и планет солнечной системы» согласно только двум фактам – скорости вращения Марса вокруг своей оси и преобладания углекислого газа в составе атмосферы.

Скорость вращения Марса вокруг своей оси и его сплюснутая форма прямо говорят, что Марс имеет внутренний источник энергии, а движение его недр организовано по типу плоского установившегося потока насосного типа, благодаря чему Марс может иметь углекислотную атмосферу. Поэтому автор приходит к выводу, что температуры и давления на его поверхности должны быть намного выше, но исходит из того, что у Марса тонкая кора, во всяком случае, она имеет хорошую теплопроводность. В работе «Энергетическая структура Солнца» автор приходит к выводу, что кора планеты может образовать объём в виде сферы, в котором заключён внутренний поток, от коры планеты зависит теплообмен с поверхностью, поэтому полагает, что Марс имеет очень плотную кору. Поэтому он имеет низкие температуры на поверхности, а также разреженную атмосферу. Ведь некогда и Луна имела атмосферу, будучи расплавленным телом, но её атмосфера ушла в кору полностью по мере остывания источника теплового расширения газов. На Марсе источник не остывал, если судить о нём  по осевому вращению, поэтому разрежение атмосферы можно будет объяснить только изучением эволюцией коры, на которую наложил отпечаток тот факт, что Марс - почти безводная планета. На нём практически затухла вулканическая активность. Автор считает, что мощность источника энергии Марса можно увеличить путём увеличения массы Марса. Из всех планет земного типа только Земля не имеет монолитной коры благодаря огромному количеству воды, образовавшейся во время её формирования после того, как первоначальный сгусток вещества стал спутником Солнца. Всё это требует исследования, поэтому редактор поместил описание Марса как приложение к последней работе автора по теме механики безынертной массы.

[3] В то время было очень трудно купить копировальную бумагу и ленту для пишущей машинки. Поэтому автор был вынужден использовать синюю или фиолетовую копирку, когда не было  чёрной.

Вернёмся к теме.

В 1973 году были произведены запуски четырёх советских автоматических станций на Марс. 15 марта 1974 сообщили, как пишет автор в рукописи «Строение Солнца и планет солнечной системы»:

Отвлечёмся. Сегодня 15 марта 1974 года по радио было передано сообщение, что советские автоматические станции достигли поверхности Марса и произвели натурные замеры его атмосферы и температуры поверхности. Следовательно,мы скоро сможем проверить поверхностные условия Марса, которые мы получили для него в своём путешествии с помощью механики безинертной массы.

Но никакой информации автор не дождался. Поэтому оставил без изменения данное им описание Марса.

Как следует из современной существующей информации об этих исследованиях, например,
http://sovams.narod.ru/Mars/1973/bse.html
http://martiantime.narod.ru/History/lant2.htm (всё о полётах на Марс),
не удалось замерить ничего. Удалось получить лишь небольшое количество снимков. Пишут:

*« «Марс-4» в 1974 году пролетает мимо планеты из-за отказа корректирующей двигательной установки, «Марс-5» работает на орбите всего две недели, передав всего чуть больше 40 снимков, «Марс-6» тоже сквозит мимо, а его спускаемый аппарат прекращает связь с Землёй в самой близости от поверхности. «Марс-7» в том же 1974 году (четыре станции запустили в одно «окно»!) тоже пролетает мимо планеты, и её спускаемый аппарат — тоже»* (С) https://tsargrad.tv/articles/pervoe-preodolenie-prokljatja-marsa-i-poslednee_228496

Как я поняла, «полистав» интернет, практически вся информация о Марсе – американского происхождения. Пишут:

*«Однако даже первые пролётные миссии передавали снимки слишком низкого качества, чтобы приоткрыть завесу тайны над этим вопросом. Но уже Маринеру-9 (впервые среди земных зондов вышедшему на марсианскую орбиту и проработавшем на ней с 14 ноября 1971 года по 27 октября 1972-го) удалось передать более 7 тысяч снимков приличного качества с разрешением в 100-1000 метров на пиксель»*(с) https://habr.com/ru/post/431766/

После серии наших неудач, у американцев дела пошли в гору:

*«В 1976 году им (американцам) удаётся мягко посадить на планету «Викинг-1» (Viking-1) и «Викинг-2» (Viking-2). И эти машинки работали одна два, другая четыре года, передав на Землю громадное количество крайне интересной информации. Естественно, при этом они попытались дать ответ на извечный вопрос: есть ли жизнь на Марсе?»* (с) https://tsargrad.tv/articles/pervoe-preodolenie-prokljatja-marsa-i-poslednee_228496

И этот успех продолжается до сего дня, как и наши неудачи. У американцев на Марсе полно марсоходов, а также спутников на его орбите:

*«Mars Observer в 1993 году. Но в целом у них наряду с неудачами получались и прорывы: Mars Global Surveyor, работавший на орбите с 1996 по 2006 год, Mars Pathfinder с первым работавшим на поверхности планеты марсоходом Sojourner, марсоходы Spirit, Opportunity (проработавший с 2004 года до середины 2018 года, когда его накрыла пылевая буря!), спутник Mars Reconnaisance Orbiter, который в состоянии разглядеть на поверхности Марса объекты размером до 30 см (и, кстати, разглядевший наш «Марс-3» на месте его посадки). И так далее — вплоть до ныне работающих на Красной планете посадочной платформы InSight и марсохода Curiosity.»* https://tsargrad.tv/articles/pervoe-preodolenie-prokljatja-marsa-i-poslednee_228496

Все данные подтверждают низкое давление и температуры на Марсе.

Однако утверждается, что:

*«По оценкам изначальная атмосфера Марса имела давление в 6 раз больше текущего земного, но в результате поздней тяжёлой бомбардировки астероидами и кометами (случившейся 3,8 миллиарда лет назад) Марс потерял большую её часть сохранив давление в 0,5-1 земную атмосферу (500-1000 мбар). Но сейчас мы наблюдаем среднее давление у марсианской поверхности всего лишь в 6 мбар — куда же делось оставшееся?»* (С) https://habr.com/ru/post/431766/

Получается, что условия, которые даны автором, когда-то были на Марсе. Они были, пока кора Марса не стала монолитом.

В работе [2] автор проанализировал данные, полученные нашей станцией о Венере. Данные были помещены в журнале «Наука и жизнь». Насчёт Марса таких данных в журналах не появилось. Данные были только американские и не в таком конкретном виде. Поэтому автору и впоследствии, в частности в работе «Энергетическая структура Солнца», нечего было анализировать, и нечем было дополнить работу [2] .

В работе «Энергетическая структура Солнца» автор объясняет условия, которые создают климатические различия планет. Факт, что причины изменения климата Марса – существуют. Зная об их существовании, кто-то другой сможет разобраться во всём.

Редактор выскажет лишь своё мнение. Он считает, что есть два варианта объяснения истории с исследованием Марса.

1 вариант: американские данные соответствуют действительности, т.к. действительно получены при помощи межпланетных станций и марсоходов. Поэтому нужно искать причину изменения климата Марса. Где её искать и что можно найти, автор сказал. Поэтому редактор поместил изъятый из работы [2] отрывок как Приложение к тексту «Энергетическая структура Солнца». Сейчас в этой области идут сплошные гаданья на кофейной гуще.

2 вариант: американские исследования Марса – инсценировка.

Автор пишет, что сообщение о том, что советские автоматические станции достигли поверхности Марса и передают информацию, было передано по радио 15 марта. В СССР старались скрывать неудачи или как-то оправдывать их, поэтому сообщения всегда появлялись позже события. «Марс – 6» начал посадку на Марс 12 марта.

Теперь пишут:

*«12 марта 1974 года Советская межпланетная автоматическая станция Марс-6 вошла в атмосферу Марса. ……Из четырех станций (Марс – 4,5,6,7) спускаемыми аппаратами обладали две последние, а коснуться поверхности Красной планеты удалось лишь Марсу-6.»*

После чего связь прервалась.

В общем, получается, что к 15 марта было известно всем, кому «положено», что этот эксперимент по исследованию Марса закончился полным провалом. Но объявили о полном успехе. Такого просто не могло быть.

Это в США могли раструбить, что лунный грунт в неимоверном количестве был доставлен на Землю, после чего его никто не видел. В СССР, если что-то не получалось, то по возможности либо делали вид, что ничего не было, либо, если не удавалось замолчать, то придумывали обтекаемую причину неудачи, и старались побыстрее забыть. А если что-то получалось, то, естественно, о достижениях принято было трубить во все трубы. И результат был налицо – его можно было предъявить, например, лунный грунт, данные о Венере. В общем, старались говорить «гоп», когда перепрыгнули. А в данном случае раструбили об успехе, но ничего не предъявили? После чего планы исследования Марса стали изменяться, отходить на второй план, как и программы изучения Луны и Венеры, не говоря уже о других небесных телах.

С учётом того, что исследования Луны и Венеры в эти годы уже как бы саботировались под благовидными предлогами, и, таким образом, были медленно, но верно прекращены к настоящему времени, а в те времена для бывшего СССР оставили только МКС, да и то – обновлённую, а собственную утопили в мировом океане, то редактор обосновано полагает, что достижения американцев в отношении Марса – блеф. Теперь на игрушечной МКС космонавты поздравляют с орбиты с Новым годом, рассказывают, как они соблюдают гигиену и испражняются в невесомости, а также охотно делятся прочими житейскими подробностями. Хотя в последнее время и это «животрепещущее» потеряло интерес у публики, поэтому аттракцион с МКС закругляют. С исследовательской работой всё время было как-то туговато. Вклад в науку и технику не просматривается.

В общем, настоящие космические программы примерно с семидесятых годов стали сворачиваться под разным соусом и профанироваться повсеместно (если было, что сворачивать, в других странах они даже не разворачивались). Поэтому наступившую беспрецедентную деградацию нельзя объяснить даже конкуренцией. США потеряли не меньше, чем СССР и РФ.

В таком случае все имеющиеся на сегодняшний день данные о Марсе – косвенные, и подгоняются под одну версию, начало которой положила информация «от имени» «Викингов». Возможно, местами данные даже соответствуют действительности, но только местами. Дёшево и сердито. «Утопить» в переносном смысле могли и результаты советских исследований Марса. О них только успели объявить, что они получены. Разве в других вопросах у нас не стали «подыгрывать» США? До сих пор не разобрались с «Аполлонами» и т.н. «лунной эпопеей». Ведь если это блеф, то удивляет, прежде всего, слаженность участников постановки с обеих сторон. Почему же «игра» должна была ограничиться лишь двумя эпизодами? Единственное, что не смогли «утопить» – результаты исследования Венеры, т.к. они успели быть обнародованы. И в США, возможно (я даже надеюсь), есть что-то настоящее, например, «Вояджеры». Пока сердце кровью обливается от очередного аттракциона, когда узнаёшь, что Маск одной левой «запустил» автомобиль на орбиту Марса с манекеном в кабине. По личному мнению редактора

«обнаружение» на Марсе русл рек и др. признаков воды является доказательством того, что данные выдумываются, и на самом деле исследования не проводятся. Нет внятных результатов даже о таинственных полярных шапках, нет ответа на вопрос – откуда на Марсе пыль (пылевые бури – факт)? Например, пыль могла образоваться в результате мощной вулканической деятельности во времена формирования коры Марса. Недоумение у редактора вызывают равнины Марса, усеянные камнями. Откуда камни? Они не скатились с каких-то горных образований, т.к. их нет поблизости, и они явно - не выходы скальных пород. Зато непременно пытаются обнаружить «жизнь», т.к. обывателя только этот вопрос и интересует. Поиски «жизни» являются главными «научными исследованиями». Убери их, и в явном виде образуется вакуум исследований не только у редактора.

Научные подвиги не повторяют, а ведь электроника и новые материалы позволяет теперь творить чудеса, о которых даже не могли мечтать люди шестидесятых годов, проводить практически любые исследования планет и спутников. По крайней мере, Марс должен уже кишеть разного рода аппаратурой и Луна – тоже. Нет, вместо этого носятся с неким ИИ, который «учат» «распознавать лица» и намерены автоматизировать всю бюрократическую работу. Нет, чтобы упразднить бюрократию и контроль, - вот был бы прогресс! Бюрократию увековечивают, автоматизировав её. «Распознавать лица» - главное (какой же бездарный пшик получился из невиданного прогресса! Таков конец, который всему делу - венец). А космос – кому он нужен? Бессмысленное и дорогое удовольствие (не то, что компьютерная графика). Какая разница, что творится на планетах и спутниках? Ведь теперь абсолютно ясно, что человек привязан к Земле. Вот и будем «распознавать лица» и выдавать ему, человеку, справки и разрешения. А любопытство (*curiosity*) публики можно удовлетворять полностью легко и дёшево через монитор.

Редактор не знает, какая версия соответствует действительности, может быть, всё объясняется как-то иначе, - это надо выяснять. Но нет возможности. Редактор высказывает оба соображения **лишь для того**, чтобы прочитав то, что автор говорит о Марсе, читатель теперь среагировал правильно, а не шаблонно, прочитав это в работе [2], мол, вот настоящее пятно, прямое доказательство, что вся работа автора и о планетах, и даже о «безынертной массе» - полная чушь. Чего-чего, а Марс-то мы знаем, как облупленный благодаря американцам. Если бы не это обстоятельство, редактор удержал бы своё мнение при себе.

Редактор считает, что сбылись самые мрачные догадки автора о будущем, поскольку на своём опыте он понял, к чему всё придёт. Направление так и не было изменено. Пришли туда, куда шли. Однако редактор на всякий случай сократил не все соображения автора, как можно будет выбраться из этого гиблого места и вернуться к норме человеческого существования.

редакция: август - сентябрь 2022 года

---

[i] Оригинал работы «Энергетическая структура Солнца и планет солнечной системы с точки зрения механики безынертной массы» представляет собой не машинописный, а рукописный текст. Привожу образец.

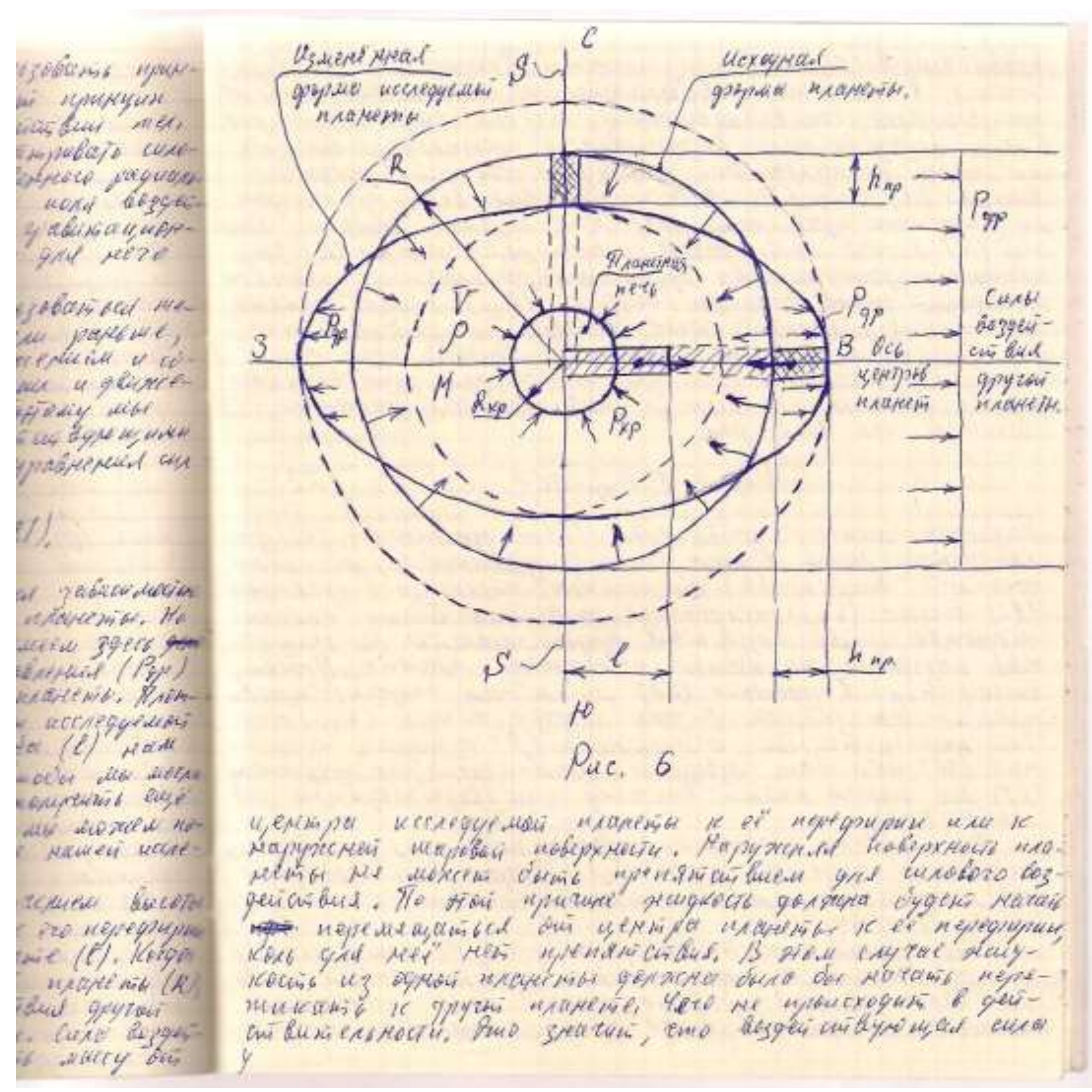

Рис. 6

центра исследуемой планеты к её периферии или к наружной шаровой поверхности. Наружная поверхность планеты не может быть препятствием для силового воздействия. По этой причине жидкость должна будет начать перемещаться от центра планеты к её периферии, коль для неё нет препятствия. В этом случае жидкость из одной планеты должна было бы начать перетекать к другой планете. Чего не происходит в действительности. Это значит, что воздействующая сила у

*образец оригинального текста*

***

По всем вопросам можно обращаться к редактору по адресу: arodionova1@rambler.ru